\documentclass[pdflatex,sn-mathphys-num,oneside]{sn-jnl}

\usepackage[T1]{fontenc}
\usepackage[utf8]{inputenc}
\usepackage{tracefnt}
\usepackage{float}
\usepackage{graphicx}
\usepackage{multirow}
\usepackage{amsmath,amssymb,amsfonts}
\usepackage{amsthm}
\usepackage[scr=boondox]{mathalpha} 
\usepackage[title]{appendix}
\usepackage{xcolor}
\usepackage{textcomp}
\usepackage{manyfoot}
\usepackage{booktabs}
\usepackage{caption}
\usepackage{subcaption} 
\usepackage{algorithm}
\usepackage{algorithmicx}
\usepackage{algpseudocode}
\usepackage{listings}
\usepackage[version=4]{mhchem}
\usepackage[normalem]{ulem}
\usepackage{gensymb}

\begin{document}

\title[propSAPT2]{Direct calculation of dipole moment surfaces through second-order symmetry-adapted perturbation theory for properties}



\author[1]{\fnm{Bartosz} \sur{Tyrcha}}\email{btyrcha@doktorant.umk.pl}

\author[1]{\fnm{Marcin} \sur{Stachowiak}}\email{mstach@umk.pl}

\author[2]{\fnm{Konrad  } \sur{Patkowski}}\email{patkowsk@auburn.edu}

\author*[1]{\fnm{Piotr S. } \sur{\.Zuchowski}}\email{pzuch@fizyka.umk.pl}

\affil[1]{\orgdiv{Institute of Physics, Faculty of Physics, Astronomy and Applied Informatics}, \orgname{Nicolaus Copernicus University in Torun}, \orgaddress{\street{Grudziadzka}, \city{Torun}, \postcode{87-100}, \country{Poland}}}

\affil[2]{\orgdiv{Department of Chemistry and Biochemistry}, \orgname{Auburn University}, \orgaddress{\city{Auburn}, \postcode{36849, Alabama}, \country{USA}}}

\abstract{
Interaction-induced dipole moment surfaces are vital for calculating line intensities and collision-induced absorption coefficients for van der Waals complexes. We introduce the perturbative corrections, of second order with respect to the interaction operator, to the dipole moment (or, in general, any one-electron property) change due to noncovalent interactions. These are obtained in the spirit of the symmetry-adapted perturbation theory for properties (propSAPT) method, developed recently by some of us [B. Tyrcha et al., JCTC \textbf{21}(9), 4562–4578 (2025)]. When orbital relaxation effects due to the external property field are properly introduced, these can be understood as analytical derivatives of the second-order SAPT interaction energies. We apply the newly introduced formalism to calculations of dipole moment surfaces for three light van der Waals complexes: \ce{He\bond{...}Ar}, \ce{He\bond{...}H2} and the \ce{H2}~dimer. The propSAPT method is compared against finite-field SAPT calculations and finite-field supermolecular CCSD(T) as the reference.
}


\maketitle

\section{Introduction}
\label{sec:introduction}

Potential energy surfaces (PESs) play a central role in the simulation of liquids, molecular spectroscopy, and collision dynamics. Modern \textit{ab initio} electronic-structure methods can provide PESs for systems spanning a remarkably broad range of sizes and target accuracies. At one end of this spectrum, local-correlation approaches can achieve near-CCSD(T) accuracy for systems containing more than one hundred atoms~\cite{nagy_local_2024}.
At the other end, interaction energies of small complexes can be calculated with uncertainties well below those conventionally associated with the ``gold standard'' of quantum chemistry.
Notable examples include \ce{He\bond{...}H2}, for which highly accurate interaction potentials enable the reproduction of spectroscopic line-shape parameters~\cite{bakr2013Highly,thibault2017Rovibrational,slowinski2020H2He,wcislo2021first,slowinski2022Collisional}, and \ce{H2\bond{...}CO}, which has become a benchmark system for stringent comparisons between experimentally measured and theoretically predicted rovibrational bound states~\cite{jankowski2005new,jankowski2012Theory,jankowski2013comprehensive,stachowiak2024Theory}.
Recent low-energy collision experiments have also provided an opportunity to study positions of scattering resonances, which are very sensitive to the details of interactions
~\cite{lavert-ofir2014Observation,klein2017Directly,horn2024Improving,horn2025Feshbach}. 

In many applications, however, the knowledge of the interaction energy alone is insufficient. Molecular properties and their variation with the intermolecular geometry are also required. A particularly important example is the dipole-moment surface (DMS), which determines spectroscopic transition intensities, or -- for nonpolar molecules in the gas phase -- is responsible for phenomena such as collision-induced absorption (CIA)~\cite{karman_update_2019,karman_o2-o2_2018}. Accurate descriptions of these effects are essential in atmospheric chemistry, remote sensing, planetary science, and astrochemistry. 

The calculation of interaction-induced properties is generally more demanding than that of interaction energies because such properties are energy derivatives and often represent very small differences between the corresponding properties of the complex and its isolated monomers. 
The studies of induced dipoles in weakly-bound complexes were initiated by the Hartree--Fock calculations of a dipole moment of a supermolecule~\cite{wormer1979initio}. 
Later on, Meyer and Frommhold noticed the key importance of electron correlation for CIA in systems containing helium and molecular hydrogen~\cite{frommhold1987Collisioninduced,frommhold1990Erratum,meyer1986Collisioninduceda,meyer1986Collisioninduced,meyer1986initio,meyer1989Absorption,meyer1989Rototranslational,meyer1994Longrange,frommhold1993Collisioninduced}.
A crucial theoretical advance was made by Hunt, who used perturbation theory and the Hellmann--Feynman theorem to derive the dispersion-induced polarization and to provide the first general proof of Feynman's conjecture: at long range, the leading dispersion force originates from the attraction of the nuclei in each molecule to the dispersion-induced distortion of its own electronic charge density~\cite{hunt_dispersion_1990}. 
Most recent, very accurate DMSs obtained for astrochemistry and planetary atmosphere applications include \ce{He\bond{...}H2}, the \ce{H2} dimer, the \ce{N2} dimer, and oxygen-nitrogen mixtures~\cite{li_interaction-induced_2012,karman_o2-o2_2018,karman_quantum_2015}.

One of the most widely used and physically transparent approaches to intermolecular interactions is symmetry-adapted perturbation theory (SAPT)~\cite{jeziorski_perturbation_1994}. 
The theoretical foundations of this method, particularly its modern many-electron formulation, owe much to the pioneering work of Bogumi{\l} Jeziorski. 
By combining intermolecular perturbation theory with the machinery of many-body quantum chemistry, Jeziorski and his collaborators established a framework in which the interaction energy is calculated directly and resolved into physically meaningful contributions. This work provided the conceptual and mathematical foundations of SAPT, which has since become an indispensable tool for analysing the origins of noncovalent interactions.
In particular, the seminal studies published in 1976 established rigorous symmetry-adapted expressions for the first-order exchange interaction~\cite{jeziorski1976FirstOrder,jeziorski1976Variationperturbation}. 
These developments were followed by a symmetry-adapted double-perturbation formulation in which intramonomer electron correlation was introduced through a Møller--Plesset-like expansion~\cite{szalewicz_symmetry-adapted_1979}, eventually leading towards many-electron applications~\cite{rybak1991Manybody,moszynski_second_1995,moszynski1994Manybodya,williams_dispersion_1995,rybak_intraatomic_1987}. 
For property surfaces, it is possible to combine the finite-field method with SAPT interaction energies. Heijmen et al. were able to calculate the resulting induced dipoles and polarizabilities for \ce{He\bond{...}H2} and \ce{He2} very accurately, and the derivatives of SAPT components are seemingly more stable than the supermolecular ones~\cite{heijmen_symmetry-adapted_1996}.

Jeziorski later extended the philosophy of direct perturbative calculations beyond the interaction energy itself. He pioneered, together with Piszczatowski and Łach, a general method for the direct evaluation of interaction-induced molecular properties~\cite{piszczatowski2008Direct}. Rather than calculating the property of the dimer and subtracting the corresponding monomer values, their formulation targeted the interaction-induced increment directly. The approach was demonstrated for the relativistic mass--velocity and Darwin contributions to the interaction energy of two hydrogen atoms. It is also important to mention here the perturbation approach of Hunt and her group, which was applied to the model \ce{He\bond{...}H} system~\cite{juanositimoneda1986Labelfree}. 

The advantages of a direct treatment of interaction-induced properties are especially pronounced for dipole-moment surfaces. The interaction-induced dipole moment is an inherently small quantity: for the systems considered here, it is of the order of $10^{-3}$~a.u. near the van der Waals minimum, and much smaller in the long-range region that governs the far wings of the CIA profile. 
In a supermolecular calculation, this quantity has to be recovered as a vanishingly small residue of the total dimer moment -- or, in a finite-field approach, of the field derivative of the total dimer energy --  relative to the very similar sum of monomer contributions. 
The numerical noise of the supermolecular difference -- orbital and amplitude convergence thresholds, density-fitting errors, or the finite-field step size -- is essentially independent of the intermolecular separation $R$, so that the attainable signal-to-noise ratio deteriorates precisely where the surface is needed the most. 
This is one practical reason why the available induced-dipole surfaces are least reliable at long separations~\cite{li2010CollisionInduced}. 
Even more complications arise due to the basis-set superposition error (BSSE) and the counterpoise correction~\cite{boys1970calculation} used to mitigate it, as the dimer basis must be used throughout the calculations of an interaction-induced property. 
A theory formulated directly for the interaction-induced increment avoids both problems: the increment is obtained as a sum of terms built exclusively from monomer quantities, so it is free of basis-set superposition error by construction, and each term carries its own well-defined asymptotic behaviour.

Bogumi{\l} Jeziorski's idea provided the starting point for the recently introduced SAPT framework for interaction-induced properties, termed propSAPT~\cite{tyrcha2025Analytical}. In propSAPT, an interaction-induced property is formulated through a double perturbation expansion with respect to the intermolecular interaction operator and an external perturbation associated with the property of interest. The first-order formulation yields analytical expressions constructed exclusively from monomer quantities and includes the effects of intermolecular exchange. It therefore permits not only the direct calculation of an interaction-induced property, but also its
decomposition into physically interpretable contributions.

In this paper, we extend propSAPT to second order in the intermolecular interaction, where the induction and dispersion contributions to the interaction-induced property appear. We derive the corresponding expressions for the interaction-induced dipole moment and test their accuracy on three light van der Waals complexes that are important in the context of CIA spectroscopy: \ce{He\bond{...}Ar}, \ce{He\bond{...}H2}, and the \ce{H2} dimer. 
The results are compared with supermolecular CCSD(T) dipole moments obtained by the finite-field method, with particular attention paid to the long-range region, where the supermolecular approach is least reliable. We expect that the new method will make the dipole-moment surfaces easier to calculate, free of BSSE, and directly interpretable in terms of physical contributions.

\section{Theory}
\label{sec:theory}

The starting point for the theory is the expression derived by Piszczatowski et al. \cite{piszczatowski2008Direct} for the change of property $X$ of dimer \ce{A\bond{...}B} resulting from the interaction between the molecules.
Let us denote by $h_A$ and $h_B$ the  Hamiltonians of monomers A and B, respectively, and by $V$ the interaction between A and B.  Let also $e_A$ and $e_B$ denote the appropriate eigenvalues of $h_A$ and $h_B$,  $\phi_A$ and $\phi_B$ the corresponding eigenfunctions, and $e_\text{int}$ will be the interaction energy. 

Consider a one-electron perturbation with the formal expansion parameter $\zeta$, $\zeta X = \zeta (X_A + X_B)$, added to the Hamiltonians of the interacting and non-interacting dimers:
\begin{align}
    (h_A + h_B + V + \zeta X) \Psi(\zeta) & = E(\zeta) \Psi(\zeta)\ ,                 \\
    (h_A + h_B + \zeta X) \Phi(\zeta)     & = (E_A(\zeta) + E_B(\zeta)) \Phi(\zeta)\ ,
    \label{eq:shrodinger_nonint_with_X}
\end{align}
where $E_A(\zeta)$, $E_B(\zeta)$, and $E(\zeta)$  represent the monomer and dimer energies perturbed by the $X$ operator, respectively.
One can project the above equations onto the non-interacting wavefunction $\Phi(\zeta)$ (see Ref.~\cite{piszczatowski2008Direct}) to obtain the effect of the $(X_A + X_B)$ perturbation on the energy difference between the dimer and the monomers:
\begin{equation}
    E(\zeta) - E_A(\zeta)- E_B(\zeta) =
    \frac{\langle \Phi(\zeta) | V \Psi(\zeta) \rangle}
    {\langle \Phi(\zeta) | \Psi(\zeta) \rangle}\ .
    \label{eq:Ediff}
\end{equation}
By taking the first derivative of the above equation with respect to $\zeta$ and setting $\zeta = 0$, one gets a formula for the interaction-induced {\em change   } of property $\Delta X$  via the exact dimer wavefunction $\psi$:  
\begin{equation}
    \Delta X = \frac{1}{ \langle \phi_{A} \phi_{B} | \psi \rangle }
    \left[ \langle \Phi^{[1]} | ( V - e_{\text{int} } )\psi \rangle
        + \langle \phi_{A} \phi_{ B} | ( V- e_{\text{int} } )\Psi^{[1]} \rangle \right]\ .
    \label{eq:Xdiff}
\end{equation}
We use $[1]$ to indicate linear (first order) with respect to the $X$ operator, so that $\Psi^{[1]}$ and $\Phi^{[1]}$ represent the linear part of the wavefunctions $\Psi(\zeta)$ and $\Phi(\zeta)$, respectively.
The function $\Psi^{[1]}$ is the solution of the following linear equation:
\begin{equation}
    \big[ h_A + h_B + V - (e_A + e_B + e_{\text{int}} )\big] \Psi^{[1]} =
    \big[ \langle \psi | X \psi \rangle - X_A - X_B \big] \psi\ ,
    \label{eq:psi1eq}
\end{equation}
and for $\Phi^{[1]}$ the solution can be separated into the $A$ and $B$ parts, $\phi^{[1]}_A$ and $\phi^{[1]}_B$, defined as
\begin{equation}
    \Phi^{[1]} = \phi^{[1]}_A \phi_B
    + \phi_A \phi^{[1]}_B
\end{equation}
for which the equations read
\begin{align}
    \big[ h_A  - e_A \big] \phi^{[1]}_{A} & =
    \big[ \langle \phi_A | X_A \phi_A  \rangle - X_A \big] \phi_A\ ,
    \label{eq:phi1}                           \\
    \big[ h_B - e_B \big] \phi^{[1]}_{B}  & =
    \big[ \langle \phi_B | X_B \phi_B  \rangle - X_B \big] \phi_B\ .
    \label{eq:phi2}
\end{align}

However, in order to apply the method to many-electron systems, it is necessary to approximate the used wavefunctions reasonably.
Therefore, from now on we will assume that $\phi_A$ and $\phi_B$ are the Hartree--Fock (or Kohn--Sham) determinants of the isolated monomers.
As a result, the $\Phi^{[1]}$ wavefunction can be obtained through the solution of the coupled-perturbed Hartree--Fock (CPHF) equations with the perturbations $X_A$ and $X_B$.
To make an approximation to the unknown $\psi$, the wavefunction of the interacting subsystems in the absence of the $X$ perturbation, we substitute $\psi \to \mathscr{A} \psi_\text{pol}$, where $\psi_\text{pol}$ is the polarisation wavefunction from the symmetrised Rayleigh-Schr{\"o}dinger perturbation theory (SRS)~\cite{jeziorski_perturbation_1994, rybak1991Manybody, moszynski1994Manybody, moszynski1994Manybodya}  and $\mathscr{A}$ is the full antisymmetriser of the dimer given as
\begin{equation}
    \mathscr{A} = \frac{N_A! N_B!}{(N_A + N_B)!} \: \mathscr{A}_A \: \mathscr{A}_B \left(1 + P \right)\ ,
    \label{eq:antisymmetrizer}
\end{equation}
where $N_A$ and $N_B$ are the numbers of electrons of monomers $A$ and $B$, $\mathscr{A}_A$ and $\mathscr{A}_B$ are the antisymmetrisers of the isolated monomers, and $P$ is the operator collecting all the permutations interchanging the indices of electrons between the monomers.
Similarly, we substitute $\Psi^{[1]} \to \mathscr{A} \Psi^{[1]}_\text{pol}$, where 
\begin{equation}
    \Psi^{[1]}_\text{pol}(\lambda) = \sum_{n = 0}^{\infty} \lambda^{n} \Psi^{(n), [1]}_\text{pol}
\end{equation}
is the part of the solution of the dimer equation that is first order in the external perturbation,
\begin{equation}
    \left( h_A + h_B + \lambda V + \zeta X \right) \Psi_\text{pol}(\lambda, \zeta) = 
    E(\lambda, \zeta) \Psi_\text{pol}(\lambda, \zeta)\ ,
    \label{eq:schrodinger_dimer_pol}
\end{equation}
with the exact solution being
\begin{equation}
    \Psi_\text{pol}(\lambda, \zeta) = \sum_{n = 0}^{\infty} \sum_{m = 0}^{\infty} \lambda^{n} \zeta^{m} \Psi^{(n), [m]}_\text{pol}\ .
    \label{eq:double_perturbed_dimer}
\end{equation}
Note that while the square brackets, e.g. $\Psi^{[1]}$, denote the order in the property operator $X$, we will use parentheses, e.g. $\Psi^{(1)}$, to denote the order in the intermolecular interaction operator $V$.
It is important to mention that due to the substitution $\Psi^{[1]} \to \mathscr{A} \Psi^{[1]}_\text{pol}$, with a perturbational wavefunction, we lose the description of orbital relaxation effects.
Therefore, they will have to be accounted for later in our derivations.
Moreover, in order to avoid mixing coupled and uncoupled terms in this part of the derivations, we further simplify $\Phi^{[1]} \to \Phi^{[1]}_\text{unc} = \Psi^{(0), [1]}_\text{pol}$ to be taken from the simple perturbation theory expansion.
Using these approximations in Eq.~(\ref{eq:Xdiff}), we obtain the general expression for a one-electron interaction-induced property (IIP) as
\begin{equation}
    \Delta X = 
    \frac{1}{ \langle \phi_{A} \phi_{B} | \mathscr{A} \psi_\text{pol} \rangle }
    \Big[ \langle \Phi^{[1]}_\text{unc} | ( V - e_\text{SRS} ) \mathscr{A} \psi_\text{pol} \rangle 
    + \langle \phi_{A} \phi_{B} | ( V - e_\text{SRS} ) \mathscr{A} \Psi^{[1]}_\text{pol} \rangle \Big]\ ,
    \label{eq:Delta_X_general}
\end{equation}
so that $e_\text{int} \to e_\text{SRS}$ becomes the SRS interaction energy.

In the first-order propSAPT theory of Ref.~\cite{tyrcha2025Analytical}, it was sufficient to know only the product of the monomer wavefunctions $\Psi^{(0), [0]}_\text{pol} = \phi_A \phi_B$ and the first-order correction with respect to the external field of the property, namely $\Psi^{(0),[1]}_\text{pol} = \Phi^{[1]}_\text{unc}$.
However, the induction and dispersion contributions to the change of a given property $X$ are of second order in the intermolecular operator $V$, and to obtain them one needs also the first-order polarisation wavefunction $\Psi^{(1), [0]}_\text{pol} = \psi^{(1)}_\text{pol}$, as well as the term bilinear in both $V$ and $X$, the $\Psi^{(1), [1]}_\text{pol}$ wavefunction, obtained from the double perturbation theory applied to Eq.~(\ref{eq:schrodinger_dimer_pol}).
The wavefunctions $\psi^{(1)}_\text{pol}$ and $\Phi^{[1]}_\text{unc}$ have the well-known forms
\begin{align}
    \label{eq:psi_pol_1}
    \psi^{(1)}_\text{pol} &= \mathcal{R} V \phi_A \phi_B\ , \\
    \label{eq:phi_1_unc}
    \Phi^{[1]}_\text{unc} &= \mathcal{R} X \phi_A \phi_B\ ,
\end{align}
where $\mathcal{R}$ is the resolvent operator defined as
\begin{equation}
    \mathcal{R} = \left( \langle F_A \rangle + \langle F_B \rangle - F_A - F_B + | \phi_A \phi_B \rangle \langle \phi_A \phi_B | \right)^{-1}
        \left( 1 - | \phi_A \phi_B \rangle \langle \phi_A \phi_B | \right)\ ,
\end{equation}
where $F_A$ and $F_B$ are Fock operators
of the noninteracting monomers $A$ and $B$ respectively.
From the double-perturbation theory we obtain the bilinear term in Eq.~(\ref{eq:double_perturbed_dimer}) as
\begin{equation}
    \Psi^{(1), [1]}_\text{pol} =
    \big(
      \mathcal{R} \left( V - \langle V \rangle \right) \mathcal{R} X 
    + \mathcal{R} \left( X - \langle X \rangle \right) \mathcal{R} V 
    \big)
    \phi_A \phi_B\ ,
    \label{eq:Psi_11} 
\end{equation}
which can be rewritten using the cluster-expansion amplitudes
\begin{equation}
    \Psi^{(1),[1]}_\text{pol} = \left( T^{(1),[1]} + T^{[1]} T^{(1)} \right) \phi_A \phi_B\ ,
    \label{eq:Psi11_with_T_operators}
\end{equation}
where $T^{(1)} = \mathcal{R} V$, $T^{[1]} = \mathcal{R} X$, and 
\begin{equation}
    T^{(1),[1]} = 
    \mathcal{R} [V, T^{[1]}] +
    \mathcal{R} [X, T^{(1)}]
    \label{eq:T11_definition}
\end{equation}
is the connected part of the bilinear excitation operator.

The second-order correction to the IIP in propSAPT is obtained by performing an order-by-order analysis of Eq.~(\ref{eq:Delta_X_general}) and keeping all the terms of second order in the interaction operator:
\begin{multline}
    \Delta X^{(2)} = 
    \frac{1}{ \langle \phi_{A} \phi_{B} | \mathscr{A} \phi_{A} \phi_{B} \rangle}
    \Big[
    \langle \Phi^{[1]}_\text{unc} | ( V - e^{(1)}_\text{SRS} ) \mathscr{A} \psi^{(1)}_\text{pol} \rangle \\
    + \langle \phi_{A} \phi_{B} | ( V - e^{(1)}_\text{SRS} ) \mathscr{A} \Psi^{(1), [1]}_\text{pol} \rangle 
    - \Delta X^{(1)} \langle \phi_{A} \phi_{B} | \mathscr{A} \psi^{(1)}_\text{pol} \rangle
    \Big]\ .
    \label{eq:Delta_X_2}
\end{multline}
By collecting the terms of Eq.~(\ref{eq:Delta_X_2}) without any intermonomer permutations, that is, those arising from the trivial part of the antisymmetriser, it is possible to write the polarisation part of the second-order IIP as
\begin{equation}
    \Delta X^{(2)}_\text{pol} =
      \langle \Phi^{[1]}_\text{unc} | V \psi^{(1)}_\text{pol} \rangle
    - \langle V \rangle \langle \Phi^{[1]}_\text{unc} | \psi^{(1)}_\text{pol} \rangle
    + \langle \phi_A \phi_B | V \Psi^{(1), [1]}_\text{pol} \rangle\ .
\end{equation}
Inserting the cluster-operator forms of the $\psi^{(1)}_\text{pol}$, $\Phi^{[1]}_\text{unc}$, and $\Psi^{(1), [1]}_\text{pol}$ wavefunctions (Eqs.~(\ref{eq:psi_pol_1}), (\ref{eq:phi_1_unc}), and~(\ref{eq:Psi_11}) respectively), and using Eq.~(\ref{eq:T11_definition}), the second-order polarisation part of IIP can be rewritten as
\begin{equation}
\begin{split}
    \Delta X^{(2)}_\text{pol} &=
              2 \operatorname{Re} \langle T^{[1]} | V T^{(1)} \rangle
            - 2 \operatorname{Re} \langle V \rangle \langle T^{[1]} | T^{(1)} \rangle \\
    &\quad + \langle T^{(1)} | X T^{(1)} \rangle 
            - \langle X \rangle \langle T^{(1)} | T^{(1)} \rangle\ .
    \label{eq:Delta_X_pol_2_with_cluster_operators}
\end{split}
\end{equation}
In the above expression, the shorthand notation $\langle Y \phi_A \phi_B | Z \phi_A \phi_B \rangle = \langle Y | Z \rangle$ was introduced.
One may show that the role of terms 
$- 2 \operatorname{Re} \langle V \rangle \langle T^{[1]} | T^{(1)} \rangle$ 
and 
$- \langle X \rangle \langle T^{(1)} | T^{(1)} \rangle$
is merely to cancel out the unlinked diagram contributions of the main terms.
Hence, one may rewrite Eq.~(\ref{eq:Delta_X_pol_2_with_cluster_operators}) as
\begin{equation}
    \Delta X^{(2)}_\text{pol} =
    2 \operatorname{Re} \langle T^{[1]} | V T^{(1)} \rangle_L
    + \langle T^{(1)} | X T^{(1)} \rangle_L\ ,
\end{equation}
where the subscript $L$ denotes the linked diagram contributions of a given term only.

Similarly to the standard SAPT, the second-order polarisation term decomposes into induction and dispersion contributions
\begin{equation}
    \Delta X^{(2)}_\text{pol} = 
          \Delta X^{(2)}_\text{ind}(B \to A) + \Delta X^{(2)}_\text{ind}(A \to B) 
        + \Delta X^{(2)}_\text{disp}\ .
    \label{eq:eq:Delta_X_pol_2_with_cluster_operators_linked}
\end{equation}
However, in order to obtain this decomposition, it is convenient to introduce the resolvent operator decomposed into a series of $\mathcal{R}_{nm}$ operators corresponding to $n$-tuple excitations on monomer $A$ and $m$-tuple excitations on monomer $B$:
\begin{equation}
    \mathcal{R} = \sum_{n = 0}^{N_A} \sum_{m = 0}^{N_B} \mathcal{R}_{n m} \, \text{ for } \,  n + m > 0\ .
    \label{eq:resolvent_series}
\end{equation}
To define the $\mathcal{R}_{nm}$ operator, we first need to introduce the second-quantised normal-ordered form of the spinorbital substitution operators in their usual definitions~\cite{moszynski1994Manybody,jeziorski_perturbation_1994,paldus_clifford_1988}
\begin{align}
    a^{\kappa_1 \cdots \kappa_n}_{\lambda_1 \cdots \lambda_n} 
    &= a^{\dagger}_{\kappa_1} \cdots a^\dagger_{\kappa_n} a_{\lambda_n} \cdots a_{\lambda_1}\ , \\ 
    b^{\mu_1 \cdots \mu_m}_{\nu_1 \cdots \nu_m} 
    &= b^{\dagger}_{\mu_1} \cdots b^\dagger_{\mu_m} b_{\nu_m} \cdots b_{\nu_1}\ ,
\end{align}
with $a^{\dagger}_{\kappa}$, $a_{\kappa}$ and $b^{\dagger}_{\mu}$, $b_{\mu}$ being creation and annihilation operators corresponding to monomer $A$ and $B$ spinorbitals, respectively. 
In the general expressions of this subsection, the spinorbitals are labelled $\alpha,\beta$ for the orbitals occupied in $A$ and $B$, and $\rho,\sigma$ for the virtual orbitals of $A$ and $B$, respectively.
The Einstein summation convention (a repeating upper and lower index implies summation over it) is used throughout this work.
Moreover, $\kappa,\lambda$ and $\mu, \nu$ will denote both occupied and virtual molecular orbitals of monomer $A$ and $B$, respectively.
The second-quantized form of $\mathcal{R}_{nm}$ is
\begin{equation}
    \mathcal{R}_{n m} = \left( \frac{1}{n! m!} \right)^2
    \frac{| a^{\rho_1 \cdots \rho_n}_{\alpha_1 \cdots \alpha_n} b^{\sigma_1 \cdots \sigma_m}_{\beta_1 \cdots \beta_m} \phi_A \phi_B \rangle \langle a^{\rho_1 \cdots \rho_n}_{\alpha_1 \cdots \alpha_n} b^{\sigma_1 \cdots \sigma_m}_{\beta_1 \cdots \beta_m} \phi_A \phi_B |}
    {\varepsilon_{\rho_1 \cdots \rho_n}^{\alpha_1 \cdots \alpha_n} + \varepsilon_{\sigma_1 \cdots \sigma_m}^{\beta_1 \cdots \beta_m}}\ ,
\end{equation}
where $\varepsilon_{\rho_1 \cdots \rho_n}^{\alpha_1 \cdots \alpha_n}$, $\varepsilon_{\sigma_1 \cdots \sigma_m}^{\beta_1 \cdots \beta_m}$ denote the differences of occupied and virtual spinorbital energies,
\begin{equation}
    \varepsilon_{\rho_1 \cdots \rho_n}^{\alpha_1 \cdots \alpha_n} = \varepsilon_{\alpha_1} + \cdots + \varepsilon_{\alpha_n} - \varepsilon_{\rho_1} - \cdots - \varepsilon_{\rho_n}\ .
\end{equation}
The second-quantised definition of the interaction operator $V$ reads
\begin{equation}
    V = 
    v^{\kappa \mu}_{\lambda \nu} a^{\lambda}_{\kappa} b^{\nu}_{\mu}
    + (v_A)^{\mu}_{\nu} b^{\nu}_{\mu}
    + (v_B)^{\kappa}_{\lambda} a^{\lambda}_{\kappa}
    + V_\text{nuc-nuc}\ ,
\end{equation}
where $v^{\kappa \mu}_{\lambda \nu} = ( \kappa \lambda | \mu \nu )$ is a standard two-electron repulsion integral (ERI), $(v_A)^{\mu}_{\nu}$ and $(v_B)^{\kappa}_{\lambda}$ describe the electrostatic interaction with the nuclei of monomer $A$ and $B$, respectively, and $V_\text{nuc-nuc}$ is the additive nuclei-nuclei repulsion term.
Additionally, we are introducing the normal-ordered version of the intermolecular operator $V$ as
\begin{equation}
    V = (V_\text{el-el})_N + (\Omega_A)_N + (\Omega_B)_N + \langle V \rangle\ .
    \label{eq:V_normal_ordered}
\end{equation}
Using the normal-ordered product of the creation/annihilation operators with respect to the Fermi vacuum, denoted  $\{ \cdot \}$, operators from the right-hand side of Eq.~(\ref{eq:V_normal_ordered}) read
\begin{align}
    (V_\text{el-el})_N &= \sum_{\kappa \lambda \mu \nu} v^{\kappa \mu}_{\lambda \nu} \{ a^{\dagger}_{\lambda} a_{\kappa} \} \{ b^{\dagger}_{\nu} b_{\mu} \}\ , \\
    (\Omega_A)_N &= \sum_{\mu \nu} (\omega_A)^{\mu}_{\nu} \{ b^{\dagger}_{\nu} b_{\mu} \}\ , \\
    (\Omega_B)_N &= \sum_{\kappa \lambda} (\omega_B)^{\kappa}_{\lambda} \{ a^{\dagger}_{\lambda} a_{\kappa} \}\ , 
\end{align}
with $\Omega_A$ and $\Omega_B$ being electrostatic potential operators of unperturbed monomers, with their matrix elements defined as
\begin{align}
    (\omega_A)^{\mu}_{\nu} &= (v_A)^{\mu}_{\nu} + v^{\alpha \mu}_{\alpha \nu}\ , \\
    (\omega_B)^{\kappa}_{\lambda} &= (v_B)^{\kappa}_{\lambda} + v^{\kappa \beta}_{\lambda \beta}\ .
\end{align}

Applying Eq.~(\ref{eq:resolvent_series}) to the $T^{(1)}$ and $T^{[1]}$ operators, we get their splitting with respect to the number of excitations on each monomer
\begin{align}
    &
    \begin{array}{rl}
        T^{(1)} &= T^{(1)}_{10} + T^{(1)}_{01} + T^{(1)}_{11} \\
                &= \mathcal{R}_{10} (\Omega_A)_N + \mathcal{R}_{01} (\Omega_B)_N + \mathcal{R}_{11} (V_\text{el-el})_N\ , \\
    \end{array}
    \\ &
    \begin{array}{rl}
        T^{[1]} &= T^{[1]}_{10} + T^{[1]}_{01} \\
                &= \mathcal{R}_{10} X_A + \mathcal{R}_{01} X_B\ .
    \end{array}
\end{align}
The $nm$-tuple excitation operator expressions are defined in Eqs.~(\ref{eq:T(1)_10})-(\ref{eq:T[1]_01}), together with their corresponding amplitudes:
\begin{align}
    \label{eq:T(1)_10}
    T^{(1)}_{10} &= t^{\alpha}_{\rho} a^{\rho}_{\alpha}\ ,
    & t^{\alpha}_{\rho} &= \frac{(\omega_B)^{\alpha}_{\rho}}{\varepsilon_{\alpha} - \varepsilon_{\rho}}\ , \\
    \label{eq:T(1)_01}
    T^{(1)}_{01} &= t^{\beta}_{\sigma} b^{\sigma}_{\beta}\ , 
    & t^{\beta}_{\sigma} &= \frac{(\omega_A)^{\beta}_{\sigma}}{\varepsilon_{\beta} - \varepsilon_{\sigma}}\ , \\
    \label{eq:T(1)_11}
    T^{(1)}_{11} &= t^{\alpha \beta}_{\rho \sigma} a^{\rho}_{\alpha} b^{\sigma}_{\beta}\ , 
    & t^{\alpha \beta}_{\rho \sigma} &= \frac{v^{\alpha \beta}_{\rho \sigma}}{\varepsilon_{\alpha} + \varepsilon_{\beta} - \varepsilon_{\rho} - \varepsilon_{\sigma}}\ , \\
    \label{eq:T[1]_10}
    T^{[1]}_{10} &= (\chi_A)^{\alpha}_{\rho} a^{\rho}_{\alpha}\ , 
    & (\chi_A)^{\alpha}_{\rho} &= \frac{(x_A)^{\alpha}_{\rho}}{\varepsilon_{\alpha} - \varepsilon_{\rho}}, \\
    \label{eq:T[1]_01}
    T^{[1]}_{01} &= (\chi_B)^{\beta}_{\sigma} b^{\sigma}_{\beta}\ , 
    & (\chi_B)^{\beta}_{\sigma} &= \frac{(x_B)^{\beta}_{\sigma}}{\varepsilon_{\beta} - \varepsilon_{\sigma}}\ .
\end{align}
With this notation, we obtain
\begin{equation}
\begin{split}
    \Delta X^{(2)}_\text{pol} &=
      2 \operatorname{Re} \langle T^{[1]}_{10} | (\Omega_B)_N T^{(1)}_{10} \rangle 
    + 2 \operatorname{Re} \langle T^{[1]}_{01} | (\Omega_A)_N T^{(1)}_{01} \rangle 
    \\ &\quad
    + 2 \operatorname{Re} \underline{\langle T^{[1]}_{10} | (\Omega_A)_N T^{(1)}_{11} \rangle} 
    + 2 \operatorname{Re} \underline{\langle T^{[1]}_{01} | (\Omega_B)_N T^{(1)}_{11} \rangle} 
    \\ &\quad
    + 2 \operatorname{Re} \langle T^{[1]}_{10} | (V_\text{el-el})_N T^{(1)}_{01} \rangle 
    + 2 \operatorname{Re} \langle T^{[1]}_{01} | (V_\text{el-el})_N T^{(1)}_{10} \rangle 
    \\ &\quad
    + 2 \operatorname{Re} \langle T^{[1]}_{10} | (V_\text{el-el})_N T^{(1)}_{11} \rangle
    + 2 \operatorname{Re} \langle T^{[1]}_{01} | (V_\text{el-el})_N T^{(1)}_{11} \rangle 
    \\ &\quad
    + \langle T^{(1)}_{10} | (X_A)_N T^{(1)}_{10} \rangle 
    + \langle T^{(1)}_{01} | (X_B)_N T^{(1)}_{01} \rangle 
    \\ &\quad
    + 2 \operatorname{Re} \underline{\langle T^{(1)}_{11} | (X_A)_N T^{(1)}_{01} \rangle}
    + 2 \operatorname{Re} \underline{\langle T^{(1)}_{11} | (X_B)_N T^{(1)}_{10} \rangle}
    \\ &\quad
    + \langle T^{(1)}_{11} | (X_A)_N T^{(1)}_{11} \rangle
    + \langle T^{(1)}_{11} | (X_B)_N T^{(1)}_{11} \rangle\ ,
    \label{eq:Delta_X_2_pol_expanded}
\end{split}
\end{equation}
where $(X_A)_N = X_A - \langle X_A \rangle$.
Typically, in SAPT, terms that contain $T^{(1)}_{10}$ excitations only are assigned to the $B \to A$ induction contribution, terms containing only $T^{(1)}_{01}$ to the $A \to B$ induction, and those containing only $T^{(1)}_{11}$ excitation operators to dispersion.
This assignment works for second-order interaction energy corrections, but becomes misleading for higher-order terms~\cite{patkowski2006Thirdorder}.
The interaction-induced correction $\Delta X^{(2)}_\text{pol}$ is of second order in the intermolecular operator $V$ and of first order in the property $X$; therefore, we should be careful when relying on such an assignment.
Additionally, in Eq.~(\ref{eq:Delta_X_2_pol_expanded}) we observe that two (of the underlined) terms contain simultaneously either $T^{(1)}_{10}$ or $T^{(1)}_{01}$, and $T^{(1)}_{11}$.
Investigating closely the underlined terms of Eq.~(\ref{eq:Delta_X_2_pol_expanded}), we can rewrite them as
\begin{align}
    \begin{split}
        \langle T^{(1)}_{11} | (X_A)_N T^{(1)}_{01} \rangle 
               &= \langle \mathcal{R}_{11} (V_\text{el-el})_N | (X_A)_N T^{(1)}_{01} \rangle 
            \\ &= \langle (V_\text{el-el})_N \mathcal{R}_{11} \big( (X_A)_N T^{(1)}_{01} \big) \rangle\ ,
    \end{split}
    \\ 
    \begin{split}
        \langle T^{(1)}_{11} | (X_B)_N T^{(1)}_{10} \rangle 
               &= \langle \mathcal{R}_{11} (V_\text{el-el})_N | (X_B)_N T^{(1)}_{10} \rangle 
            \\ &= \langle (V_\text{el-el})_N \mathcal{R}_{11} \big( (X_B)_N T^{(1)}_{10} \big) \rangle\ ,
    \end{split}
\end{align}
and
\begin{align}
    \begin{split}
        \operatorname{Re} \langle T^{[1]}_{10} | (\Omega_A)_N T^{(1)}_{11} \rangle  
               &= \operatorname{Re} \langle T^{(1)}_{11} | (\Omega_A)_N T^{[1]}_{10} \rangle
            \\ &= \operatorname{Re} \langle (V_\text{el-el})_N \mathcal{R}_{11} \big( (\Omega_A)_N T^{[1]}_{10} \big) \rangle\ ,
    \end{split}
    \\
    \begin{split}
        \operatorname{Re} \langle T^{[1]}_{01} | (\Omega_B)_N T^{(1)}_{11} \rangle 
               &= \operatorname{Re} \langle T^{(1)}_{11} | (\Omega_B)_N T^{[1]}_{01} \rangle
            \\ &= \operatorname{Re} \langle (V_\text{el-el})_N \mathcal{R}_{11} \big( (\Omega_B)_N T^{[1]}_{01} \big) \rangle\ .
    \end{split}
\end{align}
Then we observe that
\begin{multline}
    \mathcal{R}_{11} \big( (X_A)_N T^{(1)}_{01} \big) 
    + \mathcal{R}_{11} \big( (\Omega_A)_N T^{[1]}_{10} \big) 
    = \\ =
    \frac{(x_A)^{\alpha}_{\rho} (\omega_A)^{\beta}_{\sigma}}
        {(\varepsilon_{\beta} - \varepsilon_{\sigma})
        (\varepsilon_{\alpha} + \varepsilon_{\beta} - \varepsilon_{\rho} - \varepsilon_{\sigma})} 
        a^{\rho}_{\alpha} b^{\sigma}_{\beta}
    + 
    \frac{(x_A)^{\alpha}_{\rho} (\omega_A)^{\beta}_{\sigma}}
        {(\varepsilon_{\alpha} - \varepsilon_{\rho}) 
        (\varepsilon_{\alpha} + \varepsilon_{\beta} - \varepsilon_{\rho} - \varepsilon_{\sigma})}
        a^{\rho}_{\alpha} b^{\sigma}_{\beta}
    = \\ = 
    \frac{(x_A)^{\alpha}_{\rho} (\omega_A)^{\beta}_{\sigma}}
        {\varepsilon_{\alpha} + \varepsilon_{\beta} - \varepsilon_{\rho} - \varepsilon_{\sigma}}
    \left[ 
          \frac{\varepsilon_{\beta} - \varepsilon_{\sigma}}
            {(\varepsilon_{\alpha} - \varepsilon_{\rho})(\varepsilon_{\beta} - \varepsilon_{\sigma})}
        + \frac{\varepsilon_{\alpha} - \varepsilon_{\rho}}
            {(\varepsilon_{\alpha} - \varepsilon_{\rho})(\varepsilon_{\beta} - \varepsilon_{\sigma})} 
    \right] a^{\rho}_{\alpha} b^{\sigma}_{\beta}
    = \\ = 
    \frac{(x_A)^{\alpha}_{\rho} (\omega_A)^{\beta}_{\sigma}}
        {(\varepsilon_{\alpha} - \varepsilon_{\rho}) (\varepsilon_{\beta} - \varepsilon_{\sigma})} 
        a^{\rho}_{\alpha} b^{\sigma}_{\beta}
    = T^{[1]}_{10} T^{(1)}_{01}\ ,
\end{multline}
and similarly
\begin{equation}
    \mathcal{R}_{11} \big( (X_B)_N T^{(1)}_{10} \big) 
    + \mathcal{R}_{11} \big( (\Omega_B)_N T^{[1]}_{01} \big) 
    = T^{[1]}_{01} T^{(1)}_{10}\ .
\end{equation}
Therefore, the underlined terms of Eq.~(\ref{eq:Delta_X_2_pol_expanded}) can be combined to a significantly simplified form:
\begin{align}
    \label{eq:simplified_A_to_B}
    \langle T^{(1)}_{11} | (X_A)_N T^{(1)}_{01} \rangle 
    + \langle T^{(1)}_{11} | (\Omega_A)_N T^{[1]}_{10} \rangle 
    &= \langle (V_\text{el-el})_N T^{[1]}_{10} T^{(1)}_{01} \rangle\ ,
    \\
    \label{eq:simplified_B_to_A}
    \langle T^{(1)}_{11} | (X_B)_N T^{(1)}_{10} \rangle
    + \langle T^{(1)}_{11} | (\Omega_B)_N T^{[1]}_{01} \rangle
    &= \langle (V_\text{el-el})_N T^{[1]}_{01} T^{(1)}_{10} \rangle\ .
\end{align}
Based on this simplification, we assign the terms of Eq.~(\ref{eq:simplified_A_to_B}) to the $\Delta X^{(2)}_\text{ind}(A \to B)$ contribution and the terms of Eq.~(\ref{eq:simplified_B_to_A}) as the part of $\Delta X^{(2)}_\text{ind}(B \to A)$ contribution.
Therefore, the second-order induction contribution to the induced property change can be written as
\begin{equation}
\begin{split}
    \Delta X^{(2)}_\text{ind}(B \to A) &=
      2 \operatorname{Re} \langle T^{[1]}_{10} | (\Omega_B)_N T^{(1)}_{10} \rangle 
    + 2 \operatorname{Re} \langle T^{[1]}_{01} | (V_\text{el-el})_N T^{(1)}_{10} \rangle \\ &\quad
    + 2 \operatorname{Re} \langle (V_\text{el-el})_N T^{[1]}_{01} T^{(1)}_{10} \rangle
    + \langle T^{(1)}_{10} | (X_A)_N T^{(1)}_{10} \rangle\ .
    \label{eq:Delta_X_2_ind_B_to_A}
\end{split}
\end{equation}
The molecular orbital contractions describing the $\Delta X^{(2)}_\text{ind}(B \to A)$ correction, assuming real orbitals and the restricted Hartree--Fock (RHF) reference, read
\begin{equation}
\begin{split}
    \Delta X^{(2)}_\text{ind}(B \to A) &=
      16 v^{rs}_{ab} (\chi_B)^{b}_{s} t^{a}_{r} 
    + 4 (\chi_A)^{r'}_{a} (\omega_B)^{r}_{r'} t^{a}_{r}        
    \\&\quad 
    - 4 (\chi_A)^{r}_{a'} (\omega_B)^{a'}_{a} t^{a}_{r} 
    + 2 t^{r'}_{a} (x_A)^{r}_{r'} t^{a}_{r}
    \\&\quad
    - 2 t^{r}_{a'} (x_A)^{a'}_{a} t^{a}_{r}\ , 
    \label{eq:Delta_X_2_ind_B_to_A_contractions}
\end{split}
\end{equation}
where the indices $a, b$ label the occupied orbitals of monomers $A$ and $B$, while $r, s$ label the virtual orbitals of monomers $A$ and $B$, respectively.
Similarly, by performing the monomer substitution of the used quantities, we can obtain the $\Delta X^{(2)}_\text{ind}(A \to B)$ correction from Eq.~(\ref{eq:Delta_X_2_ind_B_to_A}) and its molecular orbital contractions from Eq.~(\ref{eq:Delta_X_2_ind_B_to_A_contractions}).

Then, the remaining terms of Eq.~(\ref{eq:Delta_X_2_pol_expanded}) are assigned as the dispersion contribution
\begin{equation}
\begin{split}
    \Delta X^{(2)}_\text{disp} &=
      2 \operatorname{Re} \langle T^{[1]}_{10} | (V_\text{el-el})_N T^{(1)}_{11} \rangle
    + 2 \operatorname{Re} \langle T^{[1]}_{01} | (V_\text{el-el})_N T^{(1)}_{11} \rangle 
    \\ &\quad
    + \langle T^{(1)}_{11} | (X_A)_N T^{(1)}_{11} \rangle
    + \langle T^{(1)}_{11} | (X_B)_N T^{(1)}_{11} \rangle\ ,
    \label{eq:X_2_disp_operators_non-resp}
\end{split}
\end{equation}
and the corresponding molecular orbital expression, in case of RHF reference and real orbitals, reads
\begin{equation}
\begin{split}
    \Delta X^{(2)}_\text{disp} &=
              8 (\chi_A)^{r'}_{a} v^{rs}_{r'b} t^{ab}_{rs}
            + 8 (\chi_B)^{s'}_{b} v^{rs}_{as'} t^{ab}_{rs} \\
    &\quad - 8 (\chi_A)^{r}_{a'} v^{a's}_{ab} t^{ab}_{rs} 
            - 8 (\chi_B)^{s}_{b'} v^{rb'}_{ab} t^{ab}_{rs} \\
    &\quad + 4 t^{r's}_{ab} (x_A)^{r}_{r'} t^{ab}_{rs}
            + 4 t^{rs'}_{ab} (x_B)^{s}_{s'} t^{ab}_{rs} \\
    &\quad - 4 t^{rs}_{a'b} (x_A)^{a'}_{a} t^{ab}_{rs}
            - 4 t^{rs}_{ab'} (x_B)^{b'}_{b} t^{ab}_{rs}\ .
\end{split}
\end{equation}

\subsection{Second-order exchange corrections}
\label{subsec:symmetry_forcing}

In order to account for exchange contributions to the interaction-induced property, the weak symmetry forcing procedure was used in Eq.~(\ref{eq:Delta_X_general}), which can be understood as projecting the polarization wavefunction obtained with the double-perturbation expansion of Eq.~(\ref{eq:double_perturbed_dimer}) onto the subspace of antisymmetrized wavefunctions.
The operator of such a projection is the antisymmetrizer $\mathscr{A}$ of the whole system.
Typically in the second-quantised formalism of SAPT, the permutation operator $P$ from  it is written in the form of a series of $P_{2n}$ operators, each of them interchanging exactly $n$ pairs of electron indices between the fragments: 
\begin{equation}
    P = \sum_{n=1}^{\infty} P_{2n}.
\end{equation}
In the derivations of the exchange corrections in SAPT and its extensions, the so-called \textit{single-exchange} approximation is often utilised, where the permutation expansion of the given correction is limited to the terms quadratic in the intermolecular overlap integrals $S^{\lambda}_{\mu} = \langle \varphi_{\mu} | \varphi_{\lambda} \rangle$, and it is sufficient to consider only the first term $P_2$ of the series expansion of the permutation operator~\cite{murrell1965theory,murrell1967Intermolecular} 
\begin{equation}
    P_2 = - S^{\lambda}_{\mu} S^{\nu}_{\kappa} a^{\kappa}_{\lambda} b^{\mu}_{\nu}\ .
\end{equation}
The single-exchange approximation was also utilised in this work to obtain the exchange corrections for the interaction-induced property change.
Collecting all the terms of Eq.~(\ref{eq:Delta_X_2}) quadratic in the overlap matrix elements gives 
\begin{equation}
\begin{split}
    \Delta X^{(2)}_\text{exch}(S^2) 
    &= 
             \langle \Phi^{[1]} | V P_2 \psi^{(1)}_\text{pol} \rangle 
           + \langle \phi_{A} \phi_{B} | V P_2 \Psi^{(1), [1]}_\text{pol} \rangle \\
    &\quad - e^{(1)}_\text{pol} \langle \Phi^{[1]} | P_2 \psi^{(1)}_\text{pol} \rangle
           - e^{(1)}_\text{pol} \langle \phi_{A} \phi_{B} | P_2 \Psi^{(1), [1]}_\text{pol} \rangle \\
    &\quad - e^{(1)}_\text{exch}(S^2) \langle \Phi^{[1]} | \psi^{(1)}_\text{pol} \rangle 
           - \Delta X^{(1)}_\text{pol} \langle \phi_{A} \phi_{B} | P_2 \psi^{(1)}_\text{pol} \rangle \\
    &\quad - \Delta X^{(2)}_\text{pol} \langle \phi_{A} \phi_{B} | P_2 \phi_{A} \phi_{B} \rangle\ , 
\end{split}
\label{eq:Delta_X_2_exch}
\end{equation}
where $e^{(1)}_\text{pol}$ and $e^{(1)}_\text{exch}(S^2)$ are the first-order electrostatic and exchange energy corrections from the SRS method~\cite{jeziorski_perturbation_1994, rybak1991Manybody, moszynski1994Manybody, moszynski1994Manybodya}, which are defined as
\begin{align}
    e^{(1)}_\text{pol} &= \langle \phi_A \phi_B | V \phi_A \phi_B \rangle\ , \\
    e^{(1)}_\text{exch}(S^2) &= \langle \phi_A \phi_B | V P_2 \phi_A \phi_B \rangle - \langle \phi_A \phi_B | V \phi_A \phi_B \rangle \langle \phi_A \phi_B | P_2 \phi_A \phi_B \rangle\ ,
\end{align}
and $\Delta X^{(1)}_\text{pol}$ is the first-order polarisation term of the interaction-induced property change, derived in Ref.~\cite{tyrcha2025Analytical},
\begin{equation}
    \Delta X^{(1)}_\text{pol} = 2 \operatorname{Re} \langle \phi_A \phi_B | V \mathcal{R} X \phi_A \phi_B \rangle\ .
\end{equation}
Similarly as with the $\Delta X^{(2)}_\text{pol}$ contribution, the first two terms in Eq.~(\ref{eq:Delta_X_2_exch}) are the main ones and the other terms are cancelling out the unlinked contributions of the main terms.
Therefore, we can write the second-order exchange correction in the single-exchange approximation as 
\begin{equation}
    \Delta X^{(2)}_\text{exch}(S^2) = 
      \langle \Phi^{[1]} | V P_2 \psi^{(1)}_\text{pol} \rangle_L  
    + \langle \phi_{A} \phi_{B} | V P_2 \Psi^{(1), [1]}_\text{pol} \rangle_L\ .
    \label{eq:Delta_X_2_exch_S2_linked}
\end{equation}
Inserting $\psi^{(1)}_{\rm pol}$ from Eq.~(\ref{eq:psi_pol_1}) and $\Psi^{(1),[1]}_\text{pol}$ from Eq.~(\ref{eq:Psi11_with_T_operators}) into Eq.~(\ref{eq:Delta_X_2_exch_S2_linked}), we obtain the general form
\begin{equation}
    \Delta X^{(2)}_\text{exch}(S^2) = 
      \langle T^{[1]} | V P_2 T^{(1)} \rangle_L 
    + \langle V P_2 T^{(1),[1]} \rangle_L
    + \langle V P_2 T^{[1]} T^{(1)} \rangle_L\ .
\end{equation}

Contrary to the second-order polarization correction $\Delta X^{(2)}_\text{pol}$, the explicit use of the bilinear $T^{(1),[1]}$ operator cannot be avoided.
Similarly as for the $T^{(1)}$ and $T^{[1]}$ excitation operators, we split it into the $nm$-tuple excitations contributions
\begin{equation}
    T^{(1),[1]} = T^{(1),[1]}_{10} + T^{(1),[1]}_{01} + T^{(1),[1]}_{11}\ .
\end{equation}
Those can be written into the form of
\begin{align}
    T^{(1),[1]}_{10} &= u^{\alpha}_{\rho} a^{\rho}_{\alpha}\ , \\ 
    T^{(1),[1]}_{01} &= u^{\beta}_{\sigma} b^{\sigma}_{\beta}\ , \\ 
    T^{(1),[1]}_{11} &= u^{\alpha \beta}_{\rho \sigma} a^{\rho}_{\alpha} b^{\sigma}_{\beta}\ .
\end{align}
by introducing the bilinear amplitudes $u$ as
\begin{equation}
    \begin{split}
        u^{\alpha}_{\rho} &= \Big(
          (x_A)^{\rho'}_{\rho} t^{\alpha}_{\rho'} 
        + (\chi_A)^{\alpha}_{\rho'} (\omega_B)^{\rho'}_{\rho} 
        \\ &\quad
        - (x_A)^{\alpha}_{\alpha'} t^{\alpha'}_{\rho} 
        - (\chi_A)^{\alpha'}_{\rho} (\omega_B)^{\alpha}_{\alpha'}
        \\ &\quad
        + (x_B)^{\sigma}_{\beta} t^{\alpha\beta}_{\rho\sigma} 
        + (\chi_B)^{\beta}_{\sigma} v^{\alpha\sigma}_{\rho\beta}
        \Big) / (\varepsilon_{\alpha} - \varepsilon_{\rho})\ ,
    \end{split}
\end{equation}
\begin{equation}
    \begin{split}
        u^{\beta}_{\sigma} &= \Big(
          (x_B)^{\sigma'}_{\sigma} t^{\beta}_{\sigma'} 
        + (\chi_B)^{\beta}_{\sigma'} (\omega_A)^{\sigma'}_{\sigma}
        \\ &\quad
        - (x_B)^{\beta}_{\beta'} t^{\beta'}_{\sigma} 
        - (\chi_B)^{\beta'}_{\sigma} (\omega_A)^{\beta}_{\beta'} 
        \\ &\quad
        + (x_A)^{\rho}_{\alpha} t^{\alpha\beta}_{\rho\sigma} 
        + (\chi_A)^{\alpha}_{\rho} v^{\rho\beta}_{\alpha\sigma}
        \Big) / (\varepsilon_{\beta} - \varepsilon_{\sigma})\ ,
    \end{split}
\end{equation}
and
\begin{equation}
    \begin{split}
        u^{\alpha\beta}_{\rho\sigma} &= \Big(
          (x_A)^{\rho'}_{\rho} t^{\alpha\beta}_{\rho'\sigma}  
        - (x_A)^{\alpha}_{\alpha'} t^{\alpha'\beta}_{\rho\sigma} 
        \\ &\quad
        + (x_B)^{\sigma'}_{\sigma} t^{\alpha\beta}_{\rho\sigma'}  
        - (x_B)^{\beta}_{\beta'} t^{\alpha\beta'}_{\rho\sigma} 
        \\ &\quad
        + (\chi_A)^{\alpha}_{\rho'} v^{\rho'\beta}_{\rho\sigma} 
        - (\chi_A)^{\alpha'}_{\rho} v^{\alpha\beta}_{\alpha'\sigma} 
        \\ &\quad
        + (\chi_B)^{\beta}_{\sigma'} v^{\alpha\sigma'}_{\rho\sigma}
        - (\chi_B)^{\beta'}_{\sigma} v^{\alpha\beta}_{\rho\beta'}
        \Big) / (\varepsilon_{\alpha} + \varepsilon_{\beta} - \varepsilon_{\rho} - \varepsilon_{\sigma}) 
    \end{split}
\end{equation}
respectively.
Applying it to Eq.~(\ref{eq:Delta_X_2_exch_S2_linked}) we obtain
\begin{equation}
\begin{split}
    \Delta X^{(2)}_\text{exch}(S^2) &= 
          \langle T^{[1]} | V P_2 T^{(1)}_{10} \rangle_L 
        + \langle T^{[1]} | V P_2 T^{(1)}_{01} \rangle_L 
        + \underline{\langle T^{[1]} | V P_2 T^{(1)}_{11} \rangle_L} 
        \\ &\quad 
        + \langle V P_2 T^{(1),[1]}_{10} \rangle_L
        + \langle V P_2 T^{(1),[1]}_{01} \rangle_L 
        + \langle V P_2 T^{(1),[1]}_{11}\rangle_L
        \\ &\quad
        + \langle V P_2 T^{[1]} T^{(1)}_{10} \rangle_L 
        + \langle V P_2 T^{[1]} T^{(1)}_{01} \rangle_L
        + \langle V P_2 T^{[1]} T^{(1)}_{11} \rangle_L\ .
        \label{eq:Delta_X_2_exch_S2_linked_expanded}
\end{split}
\end{equation}
In order to properly assign terms of Eq.~(\ref{eq:Delta_X_2_exch_S2_linked_expanded}) to induction- and dispersion-like contributions, we need to observe that the underlined term can be represented as
\begin{equation}
    \langle T^{[1]} | V P_2 T^{(1)}_{11} \rangle_L = 
      \langle [ (T^{[1]})^\dagger, V P_2] T^{(1)}_{11} \rangle_L
    + \langle V P_2 (T^{[1]})^\dagger T^{(1)}_{11} \rangle_L\ ,
\end{equation}
where the first term on the right-hand side is the contraction of two explicitly two-body quantities, and therefore constitutes a truly dispersive interaction, while in the second term the $(T^{[1]})^\dagger T^{(1)}_{11} | \phi_A \phi_B \rangle$ part can be understood as effectively one-electron excitations of the bilinear type, and therefore corresponds to an induction-like part of the induced property.
Hence, we rewrite the one-electron parts of the bilinear excitation operator $T^{(1),[1]}$ as
\begin{align}
    T^{(1),[1]}_{\text{ind},10}  
    &= T^{(1),[1]}_{10} + \langle a^{\alpha}_{\rho} (T^{[1]}_{01})^\dagger T^{(1)}_{11} \rangle a^{\rho}_{\alpha}
    = (u_\text{ind})^{\alpha}_{\rho} a^{\rho}_{\alpha}\ , \\ 
    T^{(1),[1]}_{\text{ind},01}
    &= T^{(1),[1]}_{01} + \langle b^{\beta}_{\sigma} (T^{[1]}_{10})^\dagger T^{(1)}_{11} \rangle b^{\sigma}_{\beta}
    = (u_\text{ind})^{\beta}_{\sigma} b^{\sigma}_{\beta}\ , 
\end{align}
where the modified bilinear induction amplitudes $(u_\text{ind})^{\alpha}_{\rho}$, $(u_\text{ind})^{\beta}_{\sigma}$ are given as 
\begin{equation}
    \begin{split}
        (u_\text{ind})^{\alpha}_{\rho} &= \Big(
          (x_A)^{\rho'}_{\rho} t^{\alpha}_{\rho'} 
        + (\chi_A)^{\alpha}_{\rho'} (\omega_B)^{\rho'}_{\rho} 
        \\ &\quad
        - (x_A)^{\alpha}_{\alpha'} t^{\alpha'}_{\rho} 
        - (\chi_A)^{\alpha'}_{\rho} (\omega_B)^{\alpha}_{\alpha'} 
        \\ &\quad
        + (\chi_B)^{\sigma}_{\beta} v^{\alpha\beta}_{\rho\sigma} 
        + (\chi_B)^{\beta}_{\sigma} v^{\alpha\sigma}_{\rho\beta}
        \Big) / (\varepsilon_{\alpha} - \varepsilon_{\rho})\ ,
    \end{split}
\end{equation}
and
\begin{equation}
    \begin{split}
        (u_\text{ind})^{\beta}_{\sigma} &= \Big(
          (x_B)^{\sigma'}_{\sigma} t^{\beta}_{\sigma'} 
        + (\chi_B)^{\beta}_{\sigma'} (\omega_A)^{\sigma'}_{\sigma}
        \\ &\quad
        - (x_B)^{\beta}_{\beta'} t^{\beta'}_{\sigma} 
        - (\chi_B)^{\beta'}_{\sigma} (\omega_A)^{\beta}_{\beta'}
        \\ &\quad
        + (\chi_A)^{\rho}_{\alpha} v^{\alpha\beta}_{\rho\sigma} 
        + (\chi_A)^{\alpha}_{\rho} v^{\rho\beta}_{\alpha\sigma}
        \Big) / (\varepsilon_{\beta} - \varepsilon_{\sigma}) 
    \end{split}
\end{equation}
respectively.
With this notation, we define the second-order exchange-induction and exchange-dispersion induced property change, in the single-exchange approximation, as
\begin{equation}
\begin{split}
    \Delta X^{(2)}_\text{exch-ind}(S^2) &= 
          \langle T^{[1]} | V P_2 T^{(1)}_{10} \rangle_L 
        + \langle T^{[1]} | V P_2 T^{(1)}_{01} \rangle_L 
        \\ &\quad 
        + \langle V P_2 T^{(1),[1]}_{\text{ind},10} \rangle_L
        + \langle V P_2 T^{(1),[1]}_{\text{ind},01} \rangle_L 
        \\ &\quad
        + \langle V P_2 T^{[1]} T^{(1)}_{10} \rangle_L 
        + \langle V P_2 T^{[1]} T^{(1)}_{01} \rangle_L\ , 
    \label{eq:Delta_X_2_exch-ind_S2_operators}
\end{split}
\end{equation}
and
\begin{equation}
\begin{split}
    \Delta X^{(2)}_\text{exch-disp}(S^2) &= 
        \langle T^{[1]} | V P_2 T^{(1)}_{11} \rangle_L 
        - \langle V P_2 (T^{[1]})^\dagger T^{(1)}_{11} \rangle_L 
        \\&\quad
        + \langle V P_2 T^{(1),[1]}_{11}\rangle_L
        + \langle V P_2 T^{[1]} T^{(1)}_{11} \rangle_L
\end{split}
\end{equation}
respectively.
However, due to the extensive form of the molecular orbital expressions for the second-order exchange corrections, we limit ourselves to presenting only the field-relaxed versions of these quantities.
Those are derived in Section~\ref{subsec:exchange_part_relaxation}, and their molecular orbital contractions are presented in the Appendices~\ref{sec:appendix_ind} and~\ref{sec:appendix_disp}.

\section{Including orbital relaxation effects}
\label{sec:orbital_relaxation}

The sources of orbital relaxation effects entering the second-order propSAPT interaction-induced property contributions are twofold.
Firstly, they arise through the relaxation with respect to the external property field $X$.
Secondly, they arise through the relaxation with respect to the electrostatic field of the other monomer.
The importance of including the latter is well-known in the perturbational theory for intermolecular interactions, and including it is a standard practice for the second-order induction and exchange-induction~\cite{moszynski1994Manybodyb} as well as third-order induction~\cite{patkowski2010Orbital} terms in SAPT.

\subsection{Relaxation of polarization contributions}
\label{subsec:polarisation_part_relaxation}

The functional for the orbital-relaxed second-order induction energy can be obtained from expanding the energy expression
\begin{equation}
    E_\text{ind,resp} = \langle e^{-\kappa_A(\lambda) - \kappa_B(\lambda)} (h_A + h_B + \lambda V) e^{\kappa_A(\lambda) + \kappa_B(\lambda)} \rangle\ ,
    \label{eq:induction_functional}
\end{equation}
to the second-order with respect to the parameter $\lambda$~\cite{patkowski2010Orbital,sadlej1980Long}.
The operators $\kappa_{A}(\lambda)$ and $\kappa_{B}(\lambda)$ from Eq.~(\ref{eq:induction_functional}) are defined as
\begin{align}
    \kappa_{A}(\lambda) &= \sum_n \lambda^n \left( C^{(n)}_A - (C^{(n)}_A)^\dagger \right)\ , \\
    \kappa_{B}(\lambda) &= \sum_n \lambda^n \left( C^{(n)}_B - (C^{(n)}_B)^\dagger \right)\ ,
\end{align}
and $C^{(n)}_A$, $C^{(n)}_B$ are one-electron excitation operators acting on monomer $A$ and $B$, respectively,
Note that the operators $\kappa_A$ and $\kappa_B$ are anti-Hermitian and are elements of the Lie algebras of spinorbital rotations of the corresponding monomers~\cite{rowe1980Manybody}.
Then, the second-order induction energy with orbital relaxation is written as
\begin{equation}
    E_\text{ind,resp}^{(2)} = 
      \langle [V, \kappa^{(1)}_A + \kappa^{(1)}_B] \rangle
    + \frac{1}{2} \langle \big[ [h_A + h_B, \kappa^{(1)}_A + \kappa^{(1)}_B], \kappa^{(1)}_A + \kappa^{(1)}_B \big] \rangle\ ,
\end{equation}
and by generalising its form to be a functional of arbitrary operators 
$\tilde{\kappa}_A = \tilde{C}_A - (\tilde{C}_A)^\dagger$, 
$\tilde{\kappa}_B = \tilde{C}_B - (\tilde{C}_B)^\dagger$, 
we obtain a Hylleraas-like functional
\begin{equation}
\begin{split}
    J[\tilde{\kappa}_A, \tilde{\kappa}_B] &=
      \langle [V, \tilde{\kappa}_A + \tilde{\kappa}_B] \rangle
    + \frac{1}{2} \langle \big[ [h_A + h_B, \tilde{\kappa}_A + \tilde{\kappa}_B], \tilde{\kappa}_A + \tilde{\kappa}_B \big] \rangle\ ,
\end{split}
\end{equation}
which decomposes naturally into 
$J[\tilde{\kappa}_A, \tilde{\kappa}_B] = J_{A}[\tilde{\kappa}_A] + J_{B}[\tilde{\kappa}_B]$,
where
\begin{equation}
\begin{split}
    J_{A}[\tilde{\kappa}_A] &=
      \langle [V, \tilde{\kappa}_A] \rangle
    + \frac{1}{2} \langle \big[ [h_A, \tilde{\kappa}_A], \tilde{\kappa}_A \big] \rangle\ ,
    \label{eq:JA_functional}
\end{split}
\end{equation}
and similarly for $J_{B}[\tilde{\kappa}_B]$.
The functional of Eq.~(\ref{eq:JA_functional}) satisfies the stationary condition at $\kappa^{(1)}_A = C^{(1)}_A - (C^{(1)}_A)^\dagger$, where the $C^{(1)}_A$ operator is the solution of the CPHF equation
\begin{equation}
    \langle \delta C_A | [h_A, C^{(1)}_A - (C^{(1)}_A)^\dagger] + \Omega_B\rangle = 0\ .
\end{equation}
When evaluated at the stationary point, it yields the orbital-relaxed second-order induction energy, i.e.~\cite{moszynski1994Manybodyb,jeziorski1993SAPT,patkowski2010Orbital}
\begin{equation}
    J_{A}[\kappa^{(1)}_A]  = E^{(2)}_{\text{ind,resp}}(B \to A)\ .
\end{equation}
In the presence of the external perturbation $\zeta X$, the Hylleraas functional can be rewritten with the transformed operators
\begin{equation}
    J_{A}[\tilde{\kappa}_A] =
      \langle [\mathscr{V}(\zeta), \tilde{\kappa}_A] \rangle
    + \frac{1}{2} \langle \big[ [\mathscr{h}_A(\zeta), \tilde{\kappa}_A], \tilde{\kappa}_A \big] \rangle\ ,
\end{equation}
where
\begin{equation}
    \mathscr{h}_A(\zeta) = e^{-\kappa_A(\zeta)} (h_A + \zeta X_A) e^{\kappa_A(\zeta)}\ ,
\end{equation}
is the field-transformed monomer Hamiltonian and
\begin{equation}
    \mathscr{V}(\zeta) = e^{-\kappa_A(\zeta) - \kappa_B(\zeta)} V e^{\kappa_A(\zeta) + \kappa_B(\zeta)}
\end{equation}
is the transformed interaction operator.
The anti-Hermitian operators $\kappa_{A}(\zeta)$ and $\kappa_{B}(\zeta)$ transform the reference state $\phi_A \phi_B$ into the solution of Eq.~(\ref{eq:shrodinger_nonint_with_X}), i.e. $\Phi(\zeta) = e^{\kappa_{A}(\zeta) + \kappa_{B}(\zeta)} \phi_A \phi_B$, and
\begin{align}
    \kappa_{A}(\zeta) &= \sum_n \zeta^n \left( C^{[n]}_A - (C^{[n]}_A)^\dagger \right)\ , \\
    \kappa_{B}(\zeta) &= \sum_n \zeta^n \left( C^{[n]}_B - (C^{[n]}_B)^\dagger \right)\ ,
\end{align}
with $C^{[n]}_A$ and $ C^{[n]}_B$ being single-excitation operators in the spaces of monomers $A$ and $B$, respectively.

The total derivative of $J_{A}[{\tilde{\kappa}}_A]$ with respect to $\zeta$ at the field value zero can be written as
\begin{equation}
    \frac{\operatorname{d} J_{A}}{\operatorname{d} \zeta} \bigg|_{\zeta = 0} = 
      \frac{\partial J_A}{\partial \zeta} \bigg|_{\zeta = 0}
    + \sum_{\kappa \lambda} \frac{\partial J_A}{\partial (\tilde{\kappa}_A)^{\kappa}_{\lambda}} \frac{\partial (\tilde{\kappa}_A)^{\kappa}_{\lambda}}{\partial \zeta} \bigg|_{\zeta = 0}\ ,
    \label{eq:JA_derivative}
\end{equation}
where $(\tilde{\kappa}_A)^{\kappa}_{\lambda}$ are the matrix elements of the operator $\tilde{\kappa}_A$.
Evaluating the functional $\frac{\operatorname{d} J_{A}}{\operatorname{d} \zeta} \big|_{\zeta = 0}$ at its stationary point $\kappa^{(1)}_A$, we obtain the second-order orbital-relaxed induction correction to the property change as
\begin{equation}
    \Delta X^{(2)}_\text{ind,resp}(B \to A) = 
    \left( \frac{\partial J_A}{\partial \zeta} \bigg|_{\zeta = 0} \right) [\kappa^{(1)}_A]\ ,
    \label{eq:Delta_X_2_ind_resp_B_to_A_def}
\end{equation}
since the second term in Eq.~(\ref{eq:JA_derivative}) vanishes when the stationarity condition is fulfilled.
Therefore, by evaluating the right-hand side of Eq.~(\ref{eq:Delta_X_2_ind_resp_B_to_A_def}), the second-order induction correction to the property change can take the form
\begin{equation}
\begin{split}
     \Delta X^{(2)}_\text{ind,resp}(B \to A) &=
      \langle \big[ [\Omega_B, \kappa^{[1]}_A], \kappa^{(1)}_A \big] \rangle
    + \langle \big[ [V, \kappa^{[1]}_B], \kappa^{(1)}_A \big] \rangle
    \\ &\quad
    + \frac{1}{2} \langle \big[ [X_A, \kappa^{(1)}_A], \kappa^{(1)}_A \big] \rangle
    \\ &\quad
    + \frac{1}{2} \langle \Big[ \big[ [W_A, \kappa^{[1]}_A], \kappa^{(1)}_A \big], \kappa^{(1)}_A \Big] \rangle\ ,
    \label{eq:Delta_X_2_ind_resp_B_to_A_with_kappa}
\end{split}
\end{equation}
where $W_A$ comes from the M{\o}ller-Plesset partitioning of the Hamiltonian, given as $h_A = F_A + W_A$, with $F_A$ being the Fock operator.
The Fock operator is diagonal when the canonical orbital basis is used, hence only the part containing $W_A$ survives in the last term of Eq.~(\ref{eq:Delta_X_2_ind_resp_B_to_A_with_kappa}).
Evaluating the commutators in Eq.~(\ref{eq:Delta_X_2_ind_resp_B_to_A_with_kappa}) together with
$\kappa^{(1)}_A = C^{(1)}_A - (C^{(1)}_A)^\dagger$ and 
$\kappa^{[1]}_A = C^{[1]}_A - (C^{[1]}_A)^\dagger$, 
after some tedious algebra, yields
\begin{equation}
\begin{split}
    \Delta X^{(2)}_\text{ind,resp}(B \to A) &=
      2 \operatorname{Re} \langle C^{[1]}_A | (\Omega_B)_N C^{(1)}_A \rangle
    + 2 \operatorname{Re} \langle C^{[1]}_B | (V_\text{el-el})_N C^{(1)}_A \rangle
    \\ &\quad
    + 2 \operatorname{Re} \langle (V_\text{el-el})_N C^{[1]}_B C^{(1)}_A \rangle 
    + \langle C^{(1)}_A | (X_A)_N C^{(1)}_A \rangle
    \\ &\quad
    + 2 \operatorname{Re} \langle C^{(1)}_A | W_A C^{[1]}_A C^{(1)}_A \rangle
    +   \operatorname{Re} \langle (C^{(1)}_A)^2 | W_A C^{[1]}_A \rangle\ .
    \label{eq:Delta_X_2_ind_resp_B_to_A_operators_final}
\end{split}
\end{equation}
The molecular orbital representation of the $\Delta X^{(2)}_\text{ind,resp}(B \to A)$ contribution, assuming the RHF reference and the real orbitals, reads
\begin{equation}
\begin{split}
    \Delta X^{(2)}_\text{ind,resp}(B \to A) &=
      16 v^{rs}_{ab} (C^{[1]}_B)^{b}_{s} (C^{(1)}_A)^{a}_{r} 
    \\&\quad
    + 4 (C^{[1]}_A)^{r'}_{a} (\omega_B)^{r}_{r'} (C^{(1)}_A)^{a}_{r} 
    - 4 (C^{[1]}_A)^{r}_{a'} (\omega_B)^{a'}_{a} (C^{(1)}_A)^{a}_{r} 
    \\&\quad 
    + 2 (C^{(1)}_A)^{r'}_{a} (x_A)^{r}_{r'} (C^{(1)}_A)^{a}_{r}
    - 2 (C^{(1)}_A)^{r}_{a'} (x_A)^{a'}_{a} (C^{(1)}_A)^{a}_{r} 
    \\&\quad 
    + 4 (C^{[1]}_A)^{a}_{r} (C^{(1)}_A)^{r'}_{a'} (C^{(1)}_A)^{a'}_{r''} \theta^{rr''}_{ar'}
    - 4 (C^{[1]}_A)^{a}_{r} (C^{(1)}_A)^{r'}_{a'} (C^{(1)}_A)^{a''}_{r'} \theta^{ra'}_{aa''} 
    \\&\quad 
    + 4 (C^{[1]}_A)^{a}_{r} (C^{(1)}_A)^{r'}_{a} (C^{(1)}_A)^{a'}_{r''} \theta^{rr''}_{r'a'} 
    - 4 (C^{[1]}_A)^{a}_{r} (C^{(1)}_A)^{r}_{a'} (C^{(1)}_A)^{a''}_{r'} \theta^{r'a'}_{a''a} 
    \\&\quad 
    + 4 (C^{[1]}_A)^{a}_{r} (C^{(1)}_A)^{r'}_{a} (C^{(1)}_A)^{r''}_{a'} \theta^{ra'}_{r'r''} 
    - 4 (C^{[1]}_A)^{a}_{r} (C^{(1)}_A)^{r}_{a'} (C^{(1)}_A)^{r'}_{a''} \theta^{a'a''}_{ar'}\ ,
    \label{eq:Delta_X_2_ind_resp_B_to_A_contractions}
\end{split}
\end{equation}
where the antisymmetrised spin-adapted two-electron integral 
$\theta^{kk'}_{ll'} = 2 v^{kk'}_{ll'} - v^{k'k}_{ll'}$
is used.
The contribution $\Delta X^{(2)}_\text{ind,resp}(A \to B)$ and its corresponding molecular orbital contractions can be obtained by taking the monomer reflection of Eqs.~(\ref{eq:Delta_X_2_ind_resp_B_to_A_operators_final}) and (\ref{eq:Delta_X_2_ind_resp_B_to_A_contractions}), respectively.

The second-order propSAPT dispersion contribution to the induced property $\Delta X^{(2)}_\text{disp}$, in its simplest form, includes only the effects of relaxation due to the property field $X$.
Including also the effects of the relaxation due to the interaction with the partner molecule would correspond to the type of effects captured in the so-called coupled dispersion energy $E^{(2)}_{\text{disp,cp}}$~\cite{jeziorski_perturbation_1994}.
Here, we restrict ourselves to derive only the uncoupled version of the dispersion contribution in propSAPT.
To achieve this, we write the Hylleraas functional for the second-order uncoupled dispersion energy
\begin{equation}
\begin{split}
    J_{AB}[\tilde{T}_{11}] &= 
      \langle \tilde{T}_{11} | \mathscr{V}(\zeta) \rangle 
    + \langle \mathscr{V}(\zeta) | \tilde{T}_{11} \rangle
    \\ &\quad
    + \langle \tilde{T}_{11} | \left( \mathscr{F}_A(\zeta) \right)_N | \tilde{T}_{11} \rangle
    + \langle \tilde{T}_{11} | \left( \mathscr{F}_B(\zeta) \right)_N | \tilde{T}_{11} \rangle\ .
    \label{eq:J_AB}
\end{split}
\end{equation}
In this context, 
$\tilde{T}_{11} = \tilde{t}^{\alpha \beta}_{\rho \sigma} a^{\rho}_{\alpha} b^{\sigma}_{\beta}$
is a double-excitation operator of 11-type being varied, and $\mathscr{F}_A(\zeta)$, $\mathscr{F}_B(\zeta)$ are the field-transformed Fock operators of monomers $A$ and $B$, respectively.
The full field-transformed monomer Hamiltonian separates into the Fock and fluctuation parts
$\mathscr{h}_{A}(\zeta) = \mathscr{F}_A(\zeta) + \mathscr{W}_A(\zeta)$,
where
\begin{align}
    \mathscr{F}_A(\zeta) &= F_A + \zeta X_A + \zeta [F_A, \kappa^{[1]}_A] + \zeta [W_A, \kappa^{[1]}_A]_{1} + \mathcal{O}(\zeta^2), \\
    \mathscr{W}_A(\zeta) &= W_A + \zeta [W_A, \kappa^{[1]}_A]_{2} + \mathcal{O}(\zeta^2)\ ,
\end{align}
the subscripts $1$ and $2$ denote one- and two-electron parts of the commutator, and 
$\mathcal{O}(\zeta^2)$
denotes terms of second- and higher-order in the external field parameter $\zeta$.
The commutator $\zeta [F_A, \kappa^{[1]}_A]$ contains only occupied-virtual and virtual-occupied blocks; hence, it does not contribute to Eq.~(\ref{eq:J_AB}).
The one-electron part of the commutator $[W_A, Y]$ can be defined in terms of a superoperator $\mathcal{G}_A(Y)$ understood as
\begin{equation}
    \mathcal{G}_A(Y) = \mathcal{Q}_{A}\left([W_A, Y]\right)\ ,
\end{equation}
where $\mathcal{Q}_{A}(Y)$ is a superoperator projecting the operator $Y$ onto the space of one-electron operators acting on monomer $A$, and can be defined as
\begin{equation}
    \mathcal{Q}_A(Y) = \sum_{\kappa \lambda} \langle \{ [ Y, a_{\lambda}], a^{\dagger}_{\kappa} \} \rangle 
    a^{\dagger}_{\lambda} a_{\kappa}\ .
\end{equation}
For example, acting on the hamiltonian of monomer $A$, i.e. $\mathcal{Q}_A(h_A) = F_A$, we simply obtain the Fock operator of monomer $A$~\cite{rowe1980Manybody}.
A similar superoperator can be defined for monomer $B$.
Additionally, let us denote the matrix elements of the operators
$\mathcal{G}_A(\kappa^{[1]}_A)$ and 
$\mathcal{G}_B(\kappa^{[1]}_B)$ 
by $(g_A)^{\kappa}_{\lambda}$ and $(g_B)^{\mu}_{\nu}$, respectively.
Their spinorbital formulas can then be written as
\begin{align}
    (g_A)^{\kappa}_{\lambda} &= 
      (C^{[1]}_A)^{\alpha}_{\rho} v^{\rho \kappa}_{\alpha \lambda}
    - (C^{[1]}_A)^{\alpha}_{\rho} v^{\kappa \rho}_{\alpha \lambda}
    + (C^{[1]}_A)^{\rho}_{\alpha} v^{\alpha \kappa}_{\rho \lambda}
    - (C^{[1]}_A)^{\rho}_{\alpha} v^{\alpha \kappa}_{\lambda \rho}\ ,
    \\
    (g_B)^{\mu}_{\nu} &= 
      (C^{[1]}_B)^{\beta}_{\sigma} v^{\sigma \mu}_{\beta \nu}
    - (C^{[1]}_B)^{\beta}_{\sigma} v^{\mu \sigma}_{\beta \nu}
    + (C^{[1]}_B)^{\sigma}_{\beta} v^{\beta \mu}_{\sigma \nu}
    - (C^{[1]}_B)^{\sigma}_{\beta} v^{\beta \mu}_{\nu \sigma}\ .
\end{align}

The stationary point of $J_{AB}[\tilde{T}_{11}]$ is obtained at $T^{(1)}_{11}$ and gives the second-order uncoupled dispersion contribution from SRS,
\begin{equation}
    J_{AB}[T^{(1)}_{11}] = E^{(2)}_\text{disp}\ .
\end{equation}
Taking the total derivative of the functional $J_{AB}$ at $\zeta = 0$, similarly as for Eq.~(\ref{eq:JA_derivative}), we obtain
\begin{equation}
    \frac{\operatorname{d} J_{AB}}{\operatorname{d} \zeta} \bigg|_{\zeta=0} = 
      \frac{\partial J_{AB}}{\partial \zeta} \bigg|_{\zeta=0}
    + \sum_{\alpha \beta \rho \sigma} \frac{\partial J_{AB}}{\partial \tilde{t}^{\alpha \beta}_{\rho \sigma}} \frac{\partial \tilde{t}^{\alpha \beta}_{\rho \sigma}}{\partial \zeta} \bigg|_{\zeta=0}\ .
    \label{eq:J_AB_derivative}
\end{equation}
Then, by evaluating it at the stationary point, the second-order dispersion correction to the interaction-induced property change with the external field relaxation reads
\begin{equation}
\begin{split}
    \Delta X^{(2)}_\text{disp} &=
      2 \operatorname{Re} \langle C^{[1]}_A | (V_\text{el-el})_N T^{(1)}_{11} \rangle
    + 2 \operatorname{Re} \langle C^{[1]}_B | (V_\text{el-el})_N T^{(1)}_{11} \rangle 
    \\ &\quad
    + \langle T^{(1)}_{11} | (X_A)_N T^{(1)}_{11} \rangle
    + \langle T^{(1)}_{11} | \big( \mathcal{G}_{A}(\kappa^{[1]}_A) \big)_N T^{(1)}_{11} \rangle
    \\ &\quad
    + \langle T^{(1)}_{11} | (X_B)_N T^{(1)}_{11} \rangle
    + \langle T^{(1)}_{11} | \big( \mathcal{G}_{B}(\kappa^{[1]}_B) \big)_N T^{(1)}_{11} \rangle\ ,
\end{split}
\end{equation}
and its molecular orbital form in the RHF reference assuming real orbitals is given as
\begin{equation}
\begin{split}
    \Delta X^{(2)}_\text{disp} &=
       8 (C^{[1]}_A)^{r'}_{a} v^{rs}_{r'b} t^{ab}_{rs}
     + 8 (C^{[1]}_B)^{s'}_{b} v^{rs}_{as'} t^{ab}_{rs} 
    \\&\quad
    - 8 (C^{[1]}_A)^{r}_{a'} v^{a's}_{ab} t^{ab}_{rs} 
    - 8 (C^{[1]}_B)^{s}_{b'} v^{rb'}_{ab} t^{ab}_{rs} 
    \\&\quad 
    + 4 t^{r's}_{ab} (x_A)^{r}_{r'} t^{ab}_{rs}
    + 4 t^{rs'}_{ab} (x_B)^{s}_{s'} t^{ab}_{rs} 
    \\&\quad 
    - 4 t^{rs}_{a'b} (x_A)^{a'}_{a} t^{ab}_{rs}
    - 4 t^{rs}_{ab'} (x_B)^{b'}_{b} t^{ab}_{rs}
    \\&\quad 
    + 8 t^{r's}_{a'b} \theta^{r''a}_{r'r} (C^{[1]}_A)^{r}_{a} t^{a'b}_{r''s} 
    + 8 t^{rs'}_{ab'} \theta^{s''b}_{s's} (C^{[1]}_B)^{s}_{b} t^{ab'}_{rs''} 
    \\&\quad 
    - 8 t^{r's}_{a'b} \theta^{aa'}_{ra''} (C^{[1]}_A)^{r}_{a} t^{a''b}_{r's} 
    - 8 t^{rs'}_{ab'} \theta^{bb'}_{sb''} (C^{[1]}_B)^{s}_{b} t^{ab''}_{rs'}\ .
\end{split}
\end{equation}

Alternatively, it is possible to obtain the same expression for $\Delta X^{(2)}_\text{disp}$ by introducing the carefully chosen substitutions into Eq.~(\ref{eq:X_2_disp_operators_non-resp}).
For this purpose, it is sufficient to substitute~\cite{korona2002Electrostatic}
\begin{align}
    \label{eq:X_A_to_X_A_+_commutator}
    X_A &\to X_A + \mathcal{G}_{A}(\kappa^{[1]}_A)\ , \\
    \label{eq:X_B_to_X_B_+_commutator}
    X_B &\to X_B + \mathcal{G}_{B}(\kappa^{[1]}_B)\ .
\end{align}
together with the substitution replacing excitation operators that are first-order in $X$ with their relaxed CPHF counterparts, i.e.
$T^{[1]}_{10} \to C^{[1]}_A$ and
$T^{[1]}_{01} \to C^{[1]}_B$.

\subsection{Relaxation of exchange contributions}
\label{subsec:exchange_part_relaxation}

In order to include the relaxation effects in the exchange counterparts of the second-order interaction-induced property change contributions, it is necessary to reformulate the theory so that all additional relaxation effects are accounted for in the excitation operators, and therefore can be placed inside the modified $u$ amplitudes, rather than in the form of additional terms.

The coupled versions of the bilinear amplitudes
$(u_\text{ind,cp})^{\alpha}_{\rho}$, 
$(u_\text{ind,cp})^{\beta}_{\sigma}$
can be obtained by solving the following coupled-perturbed type equations
\begin{align}
    \label{eq:CPHF_U_A}
    \langle \delta U_A | [h_A, U^{(1),[1]}_A - (U^{(1),[1]}_A)^\dagger] + {\Xi}_A \rangle &= 0\ , \\
    \label{eq:CPHF_U_B}
    \langle \delta U_B | [h_B, U^{(1),[1]}_B - (U^{(1),[1]}_B)^\dagger] + {\Xi}_B \rangle &= 0\ ,
\end{align}
where
\begin{align}
    U^{(1),[1]}_A &= (u_\text{ind,cp})^{\alpha}_{\rho} a^{\rho}_{\alpha}\ , \\
    U^{(1),[1]}_B &= (u_\text{ind,cp})^{\beta}_{\sigma} b^{\sigma}_{\beta}\ ,
\end{align}
and the perturbing operator ${\Xi}_A$ is defined as 
\begin{equation}
    {\Xi}_A = \mathcal{Q}_A \left(
      [\Omega_B, \kappa^{[1]}_A]
    + [V_\text{el-el}, \kappa^{[1]}_B]
    + [X_A, \kappa^{(1)}_A] 
    + \big[ [W_A, \kappa^{[1]}_A], \kappa^{(1)}_A \big] 
    \right)\ ,
\end{equation}
and the ${\Xi}_B$ operator is defined similarly with the corresponding monomer $B$ equivalents.
The last term includes both $\mathcal{G}_{A}( \kappa^{[1]}_A )$ and something that can be interpreted as the first derivative of the Hessian matrix over the external field parameter $\zeta$, evaluated at $\zeta = 0$.
To solve Eqs.~(\ref{eq:CPHF_U_A}) and~(\ref{eq:CPHF_U_B}) numerically, it is enough to know only the occupied-virtual matrix elements of operators ${\Xi}_A$ and ${\Xi}_B$; these read
\begin{equation}
    \begin{split}
        ({\Xi}_A)^{\alpha}_{\rho} &= 
          (C^{[1]}_A)^{\alpha}_{\rho'} (\omega_B)^{\rho'}_{\rho} 
        - (C^{[1]}_A)^{\alpha'}_{\rho} (\omega_B)^{\alpha}_{\alpha'} 
        \\&\quad 
        + (C^{[1]}_B)^{\sigma}_{\beta} v^{\alpha\beta}_{\rho\sigma} 
        + (C^{[1]}_B)^{\beta}_{\sigma} v^{\alpha\sigma}_{\rho\beta}
        \\&\quad 
        + (x_A)^{\rho'}_{\rho} (C^{(1)}_A)^{\alpha}_{\rho'} 
        - (x_A)^{\alpha}_{\alpha'} (C^{(1)}_A)^{\alpha'}_{\rho} 
        \\&\quad 
        + (g_A)^{\rho'}_{\rho} (C^{(1)}_A)^{\alpha}_{\rho'} 
        - (g_A)^{\alpha}_{\alpha'} (C^{(1)}_A)^{\alpha'}_{\rho} 
        \\&\quad
        + (C^{[1]}_A)^{\alpha}_{\rho''} \theta^{\rho'' \rho'}_{\rho \alpha'} (C^{(1)}_A)^{\alpha'}_{\rho'}
        + (C^{[1]}_A)^{\rho'}_{\alpha''} \theta^{\alpha'' \alpha}_{\rho \alpha'} (C^{(1)}_A)^{\alpha'}_{\rho'}
        \\&\quad 
        + (C^{[1]}_A)^{\alpha''}_{\rho} \theta^{\alpha \rho'}_{\alpha'' \alpha'} (C^{(1)}_A)^{\alpha'}_{\rho'}
        + (C^{[1]}_A)^{\rho''}_{\alpha'} \theta^{\rho' \alpha}_{\rho \rho''} (C^{(1)}_A)^{\alpha'}_{\rho'}
        \\&\quad 
        + (C^{[1]}_A)^{\alpha}_{\rho''} \theta^{\rho'' \alpha'}_{\rho \rho'} (C^{(1)}_A)^{\rho'}_{\alpha'}
        + (C^{[1]}_A)^{\alpha'}_{\rho''} \theta^{\alpha \rho''}_{\rho \rho'} (C^{(1)}_A)^{\rho'}_{\alpha'}
        \\&\quad 
        + (C^{[1]}_A)^{\alpha''}_{\rho} \theta^{\alpha \alpha'}_{\alpha'' \rho'} (C^{(1)}_A)^{\rho'}_{\alpha'}
        + (C^{[1]}_A)^{\alpha''}_{\rho'} \theta^{\alpha \alpha'}_{\rho \alpha''} (C^{(1)}_A)^{\rho'}_{\alpha'}\ ,
    \end{split}
\end{equation}
and
\begin{equation}
    \begin{split}
        ({\Xi}_B)^{\beta}_{\sigma} &=
          (C^{[1]}_B)^{\beta}_{\sigma'} (\omega_A)^{\sigma'}_{\sigma} 
        - (C^{[1]}_B)^{\beta'}_{\sigma} (\omega_A)^{\beta}_{\beta'} 
        \\&\quad 
        + (C^{[1]}_A)^{\rho}_{\alpha} v^{\alpha\beta}_{\rho\sigma} 
        + (C^{[1]}_A)^{\alpha}_{\rho} v^{\rho\beta}_{\alpha\sigma}
        \\&\quad 
        + (x_B)^{\sigma'}_{\sigma} (C^{(1)}_B)^{\beta}_{\sigma'} 
        - (x_B)^{\beta}_{\beta'} (C^{(1)}_B)^{\beta'}_{\sigma}
        \\&\quad 
        + (g_B)^{\sigma'}_{\sigma} (C^{(1)}_B)^{\beta}_{\sigma'} 
        - (g_B)^{\beta}_{\beta'} (C^{(1)}_B)^{\beta'}_{\sigma} 
        \\&\quad 
        + (C^{[1]}_B)^{\beta}_{\sigma''} \theta^{\sigma'' \sigma'}_{\sigma \beta'} (C^{(1)}_B)^{\beta'}_{\sigma'}
        + (C^{[1]}_B)^{\sigma'}_{\beta''} \theta^{\beta'' \beta}_{\sigma \beta'} (C^{(1)}_B)^{\beta'}_{\sigma'}
        \\&\quad 
        + (C^{[1]}_B)^{\beta''}_{\sigma} \theta^{\beta \sigma'}_{\beta'' \beta'} (C^{(1)}_B)^{\beta'}_{\sigma'}
        + (C^{[1]}_B)^{\sigma''}_{\beta'} \theta^{\sigma' \beta}_{\sigma \sigma''} (C^{(1)}_B)^{\beta'}_{\sigma'}
        \\&\quad 
        + (C^{[1]}_B)^{\beta}_{\sigma''} \theta^{\sigma'' \beta'}_{\sigma \sigma'} (C^{(1)}_B)^{\sigma'}_{\beta'}
        + (C^{[1]}_B)^{\beta'}_{\sigma''} \theta^{\beta \sigma''}_{\sigma \sigma'} (C^{(1)}_B)^{\sigma'}_{\beta'}
        \\&\quad 
        + (C^{[1]}_B)^{\beta''}_{\sigma} \theta^{\beta \beta'}_{\beta'' \sigma'} (C^{(1)}_B)^{\sigma'}_{\beta'}
        + (C^{[1]}_B)^{\beta''}_{\sigma'} \theta^{\beta \beta'}_{\sigma \beta''} (C^{(1)}_B)^{\sigma'}_{\beta'}
    \end{split}
\end{equation}
respectively.
Additionally, as previously mentioned for the $\Delta X^{(2)}_\text{disp}$ contribution, the excitation operators that are first-order in $X$, should be substituted with their relaxed CPHF counterparts
$T^{[1]}_{10} \to C^{[1]}_A$ and
$T^{[1]}_{01} \to C^{[1]}_B$.
A similar substitution is performed for the one-electron excitation operators that are first-order in $V$, i.e. we substitute
$T^{(1)}_{10} \to C^{(1)}_A$ and
$T^{(1)}_{01} \to C^{(1)}_B$.
By performing all the aforementioned substitutions in Eq.~(\ref{eq:Delta_X_2_exch-ind_S2_operators}), we obtain the orbital-relaxed exchange-induction contribution to the property change as
\begin{equation}
\begin{split}
    \Delta X^{(2)}_\text{exch-ind,resp}(S^2) &= 
          \langle C^{[1]}_A | V P_2 C^{(1)}_A \rangle_L 
        + \langle C^{[1]}_A | V P_2 C^{(1)}_B \rangle_L 
        \\&\quad 
        + \langle C^{[1]}_B | V P_2 C^{(1)}_A \rangle_L 
        + \langle C^{[1]}_B | V P_2 C^{(1)}_B \rangle_L 
        \\&\quad 
        + \langle V P_2 U^{(1),[1]}_A \rangle_L
        + \langle V P_2 U^{(1),[1]}_B \rangle_L 
        \\&\quad
        + \langle V P_2 C^{[1]}_A C^{(1)}_A \rangle_L 
        + \langle V P_2 C^{[1]}_A C^{(1)}_B \rangle_L
        \\&\quad
        + \langle V P_2 C^{[1]}_B C^{(1)}_A \rangle_L 
        + \langle V P_2 C^{[1]}_B C^{(1)}_B \rangle_L\ .
\end{split}
\end{equation}
The molecular orbital contractions for the $\Delta X^{(2)}_\text{exch-ind,resp}(S^2)$ expression, due to its lengthy form, are presented in the Appendix~\ref{sec:appendix_ind}.

In case of the second-order exchange-dispersion correction to the induced property, it is sufficient to modify the bilinear double-excitation operator $T^{(1),[1]}_{11}$, to include the relaxed first order excitations $C^{(1)}_A$ and $C^{(1)}_B$ in place of $T^{(1)}_{10}$ and $T^{(1)}_{01}$, respectively, as well as to use the substitution of Eqs.~(\ref{eq:X_A_to_X_A_+_commutator}) and~(\ref{eq:X_B_to_X_B_+_commutator}). This gives us a form of a bilinear double-excitation operator 
$T^{(1),[1]}_{\text{cp},11} = (u_\text{cp})^{\alpha\beta}_{\rho\sigma} a^{\rho}_{\alpha} b^{\sigma}_{\beta}$,
that includes the orbital-relaxation effects with respect to the external operator $X$. It can be obtained from the equation
\begin{equation}
\begin{split}
    T^{(1),[1]}_{\text{cp},11} &= \mathcal{R}_{11} \Big( 
      [V_\text{el-el}, C^{[1]}_A + C^{[1]}_B]
    + [X_A + X_B, T^{(1)}_{11}] 
    \\ &\quad \quad
    + [\mathcal{G}_{A}(\kappa^{[1]}_A) + \mathcal{G}_{B}(\kappa^{[1]}_B), T^{(1)}_{11}] 
    \Big)\ ,
    \label{eq:T11_cp_definition}
\end{split}
\end{equation}
and the amplitudes $(u_\text{cp})^{\alpha\beta}_{\rho\sigma}$ are given as
\begin{equation}
    \begin{split}
        (u_\text{cp})^{\alpha\beta}_{\rho\sigma} &= \big(
          (x_{A})^{\rho'}_{\rho} t^{\alpha\beta}_{\rho'\sigma}  
        - (x_{A})^{\alpha}_{\alpha'} t^{\alpha'\beta}_{\rho\sigma} 
        \\&\quad
        + (x_{B})^{\sigma'}_{\sigma} t^{\alpha\beta}_{\rho\sigma'}  
        - (x_{B})^{\beta}_{\beta'} t^{\alpha\beta'}_{\rho\sigma}
        \\&\quad
        + (g_{A})^{\rho'}_{\rho} t^{\alpha\beta}_{\rho'\sigma}  
        - (g_{A})^{\alpha}_{\alpha'} t^{\alpha'\beta}_{\rho\sigma} 
        \\&\quad
        + (g_{B})^{\sigma'}_{\sigma} t^{\alpha\beta}_{\rho\sigma'}  
        - (g_{B})^{\beta}_{\beta'} t^{\alpha\beta'}_{\rho\sigma} 
        \\&\quad
        + (C^{[1]}_A)^{\alpha}_{\rho'} v^{\rho'\beta}_{\rho\sigma} 
        - (C^{[1]}_A)^{\alpha'}_{\rho} v^{\alpha\beta}_{\alpha'\sigma} 
        \\&\quad
        + (C^{[1]}_B)^{\beta}_{\sigma'} v^{\alpha\sigma'}_{\rho\sigma}
        - (C^{[1]}_B)^{\beta'}_{\sigma} v^{\alpha\beta}_{\rho\beta'}
        \big) / (\varepsilon_{\alpha} + \varepsilon_{\beta} - \varepsilon_{\rho} - \varepsilon_{\sigma})\ .
    \end{split}
\end{equation}
Using the $T^{(1),[1]}_{\text{cp},11}$ operator and the other previously utilised substitutions, we obtain the orbital-relaxed version of the second-order exchange-dispersion change of induced property as
\begin{equation}
\begin{split}
    \Delta X^{(2)}_\text{exch-disp}(S^2) &= 
          \langle C^{[1]}_A | V P_2 T^{(1)}_{11} \rangle_L 
        + \langle C^{[1]}_B | V P_2 T^{(1)}_{11} \rangle_L 
        \\&\quad
        - \langle V P_2 (C^{[1]}_A)^\dagger T^{(1)}_{11} \rangle_L 
        - \langle V P_2 (C^{[1]}_B)^\dagger T^{(1)}_{11} \rangle_L 
        \\&\quad
        + \langle V P_2 C^{[1]}_A T^{(1)}_{11} \rangle_L
        + \langle V P_2 C^{[1]}_B T^{(1)}_{11} \rangle_L
        \\&\quad
        + \langle V P_2 T^{(1),[1]}_{\text{cp},11}\rangle_L\ .
\end{split}
\end{equation}
The corresponding molecular orbital expression for $\Delta X^{(2)}_\text{exch-disp}(S^2)$ is given in the Appendix~\ref{sec:appendix_disp}.

\section{Computational details}
\label{sec:computational_details}

To test the new theory, we selected several systems expected to exhibit a strong second-order dispersion contribution to the induced dipole moment. In particular, for two interacting atoms, the dispersion-induced dipole is the dominant long-range contribution, motivating our choice of the \ce{He\bond{...}Ar} complex~\cite{chandran2025Complete}. Given the recent progress in high-precision spectroscopy of \ce{H2}~\cite{stankiewicz2026Cavityenhanced}, we complemented our test set with the \ce{He\bond{...}H2} and \ce{H2\bond{...}H2} dimers.

As reference values, we computed CCSD(T) dipole moments using the finite-field approach with a field increment of $10^{-3}\ E_h / e\,a_0$ and a central finite-difference formula.

For each system, we also performed a detailed analysis of the asymptotic behaviour of the interaction-induced dipole moment in order to determine how rapidly it approaches its asymptotic power-law regime. To this end, calculations were carried out at intermolecular separations extending up to $25\ a_0$. We evaluated the local (effective) decay exponent
\begin{equation}
n(R) = - \frac{d \log |\Delta \vec{\mu}(R)|}{d \log R}\ ,
\label{eq:exponent}
\end{equation}
which should approach a certain constant at long range.
This quantity allows us not only to verify the decay rate, but also to determine the intermolecular distance at which the leading asymptotic power law becomes dominant and compare it with the rigorous asymptotic theory~\cite{li2010CollisionInduced}.

The propSAPT calculations were performed using the propSAPT Python module~\cite{tyrcha2026propSAPT} interfaced with the Psi4 quantum chemistry package~\cite{smith_psi4_2020,smith_psi4numpy_2018}. 
Molecular orbital contractions and their numerical implementation were generated through the symbolic Python package for second-quantised SAPT derivations developed by our group~\cite{tyrcha2026sym2quantizedsapt}.
The aug-cc-pVTZ~\cite{dunning1989Gaussian,kendall1992Electron,woon1994Gaussian,woon1993Gaussian} orbital basis set was used throughout, together with the corresponding aug-cc-pVTZ-RI~\cite{weigend2002Efficient} and aug-cc-pVTZ-JKFIT~\cite{weigend2002fully} auxiliary basis sets for density fitting. Apart from propSAPT based on the Hartree--Fock reference, we have also used Kohn--Sham orbitals obtained with the PBE0 functional~\cite{adamo_toward_1999} with the GRAC asymptotic correction~\cite{heselmann_density-functional_2004, misquitta_intermolecular_2002}, with the $I_p-\epsilon_{\rm HOMO}$ shift parameters for He, Ar, and \ce{H2} of $0.234$, $0.162$, and $0.139\ E_h$, respectively.

\section{Results and discussion}
\label{sec:results}

\subsection{Comparison with the finite-field SAPT }
 
In our previous work~\cite{tyrcha2025Analytical}, we demonstrated that the first-order propSAPT expressions, evaluated within the Hartree–Fock reference, reproduce the corresponding finite-field SAPT (FF-SAPT) values to 4 significant digits — an agreement which is essentially exact. 
This is not the case for the propSAPT(DFT) implementation: the asymptotic correction applied to the Kohn–Sham potential used to construct the unperturbed reference density is not consistent with the coupled-perturbed Kohn–Sham (CPKS) response kernel used in the second-order Hessian evaluation. 
Hence, it is of interest to assess the extent to which propSAPT can be regarded as a fully relaxed theory.  

\begin{figure}
    \centering
    \includegraphics[width=\linewidth]{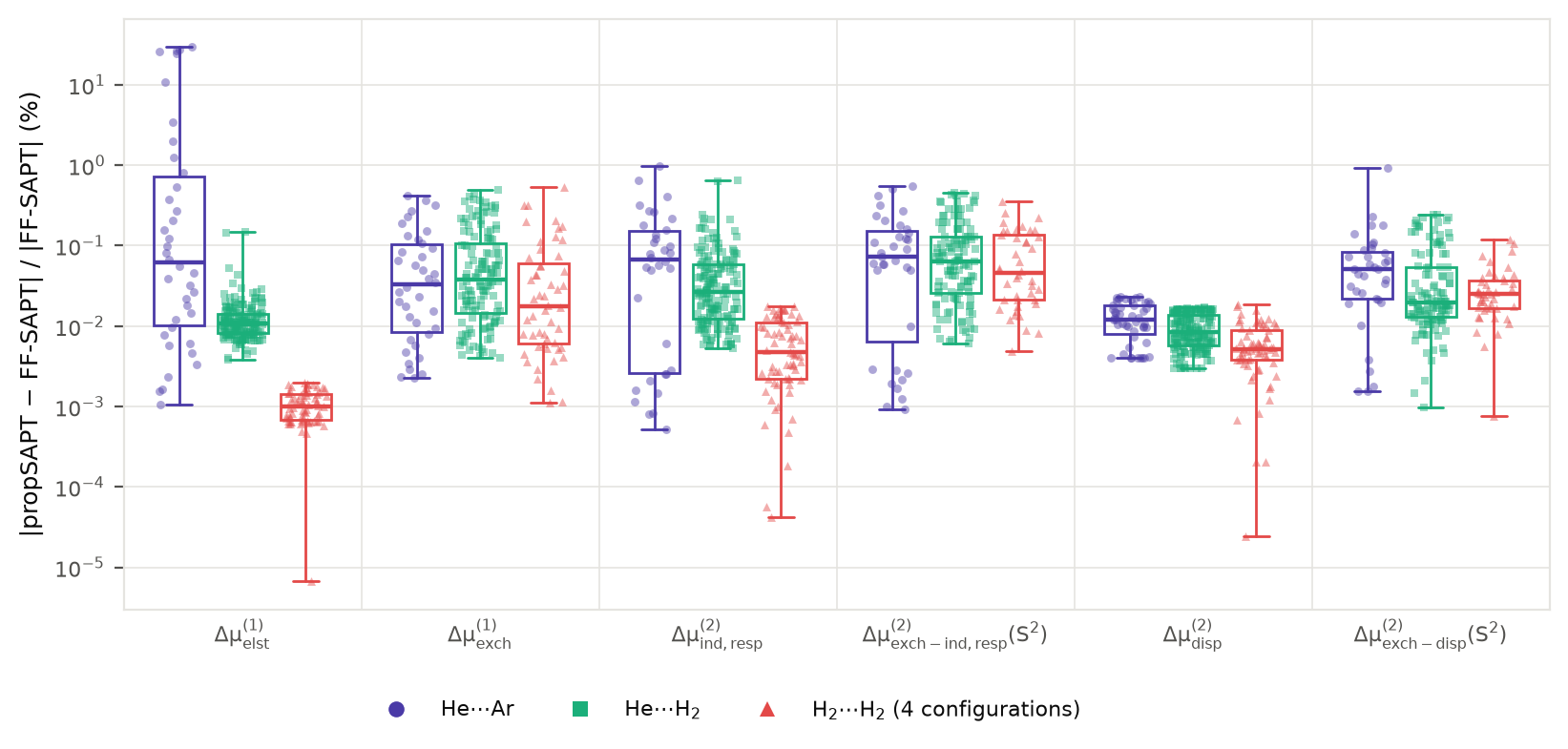}
    \caption{Relative error of the propSAPT corrections to the induced dipole moment with respect to FF-SAPT, evaluated from the difference between the corresponding dipole-moment vectors. Intermolecular distances range from 4 to 10 \AA. For \ce{He\bond{...}H2}, both T-shaped and collinear geometries were considered, while for the \ce{H2} dimer, T-shaped and planar tilted geometries with a 45$^\circ$ angle between the molecular axes were used. Points for which $|\Delta \vec{\mu}| < 10^{-8}\ e\,a_0$ were excluded.}
    \label{fig:correspondence_boxplot}
\end{figure}

In Fig.~\ref{fig:correspondence_boxplot} we compare relative errors for each correction in the second-order propSAPT with finite-field SAPT(HF) for representative series of distances (up to 10 \AA) and angular orientations. 
The agreement between the two approaches is striking: each correction from analytical propSAPT reproduces finite-field results with a median relative error of better than 0.1\%, and, in the case of the \ce{H2} dimer, the electrostatic, induction, and dispersion corrections are all below 0.01\%. The largest discrepancies are found for the first-order polarisation contribution in the \ce{He\bond{...}Ar} system, where $|\Delta \vec{\mu}^{(1)}|$ decays exponentially with intermolecular distance and is particularly sensitive to the choice of the RI auxiliary basis set. In all other cases, even the largest outliers remain below 1\%. Such a level of agreement between numerical FF-SAPT and analytical derivatives is convincing enough to consider that the latter formulation exactly reproduces the orbital relaxation of the SAPT0 theory.

 \subsection{Long-range propSAPT and FF-SAPT}

\begin{figure}
    \centering
    \includegraphics[width=0.6\linewidth]{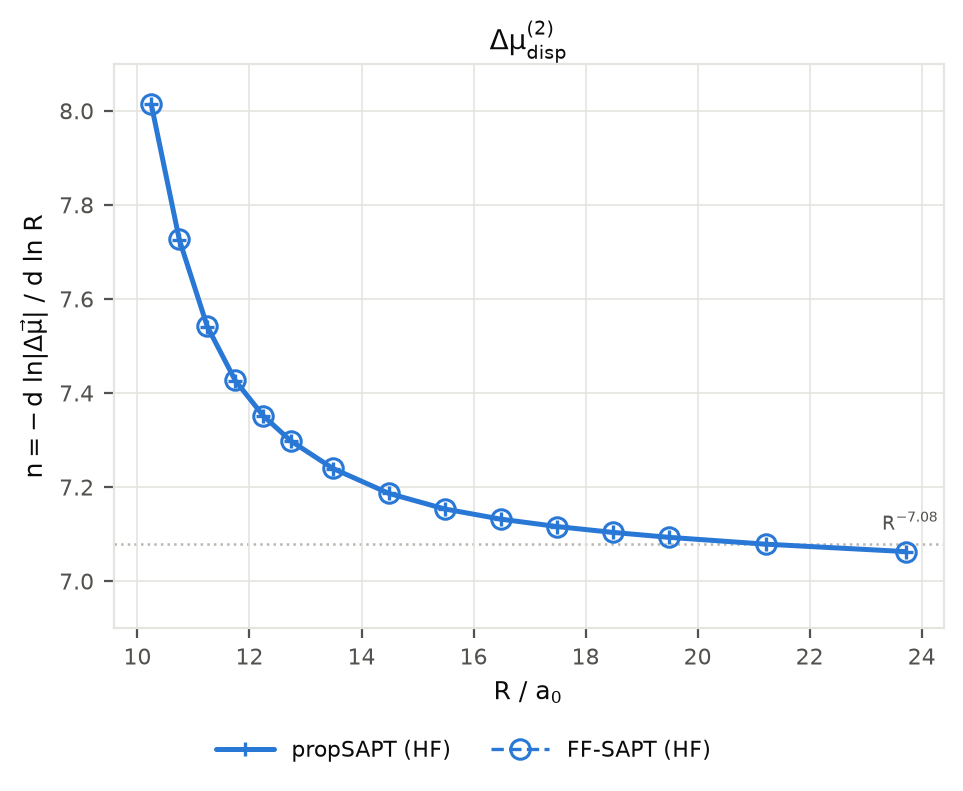}
    \caption{Asymptotic behaviour of the propSAPT(HF) and FF-SAPT(HF) at long range, presented as the local (effective) decay exponent calculated as in Eq.~(\ref{eq:exponent}) for the \ce{He\bond{...}Ar} system. Only dispersion-induced dipole exhibits the inverse power-law behaviour for the \ce{atom\bond{...}atom} system.}
    \label{fig:he-ar-dipole-asymptotics}
\end{figure}

\begin{figure}
    \centering
    \begin{subfigure}{\linewidth}
        \centering
        \includegraphics[width=\linewidth]{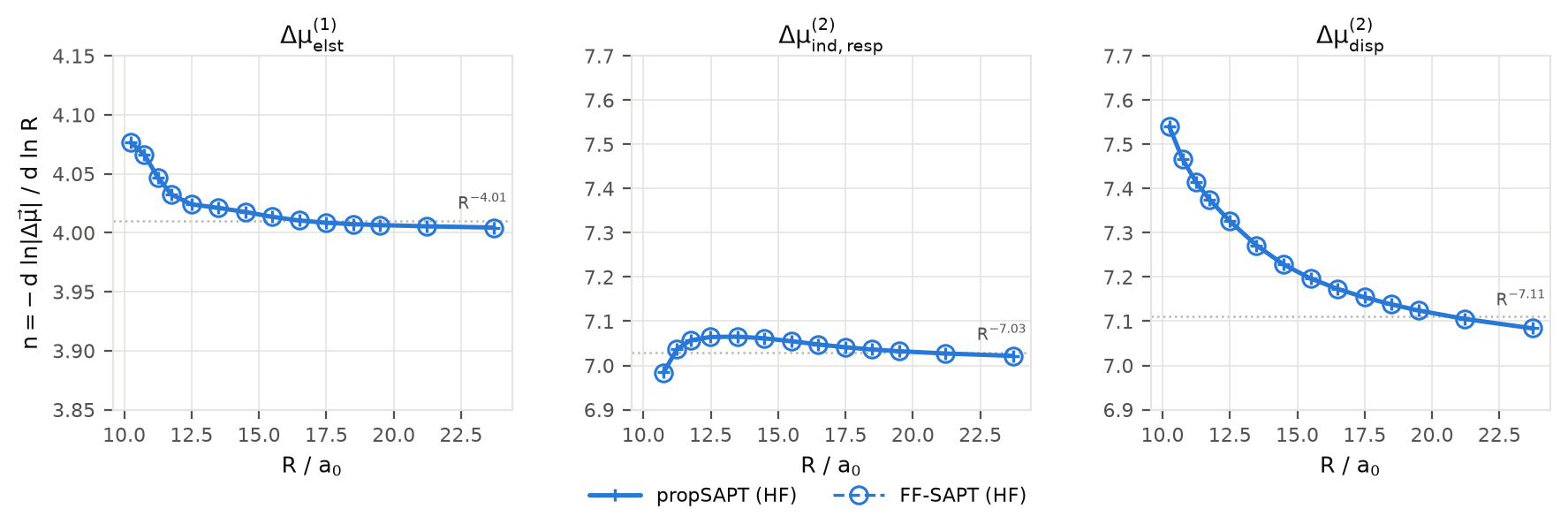}
        \caption{$\theta = 0\degree$ (\textit{linear})}
        \label{fig:he-h2-dipole-asymptotics-theta-0}
    \end{subfigure}
    \hfill
    \begin{subfigure}{\linewidth}
        \centering
        \includegraphics[width=\linewidth]{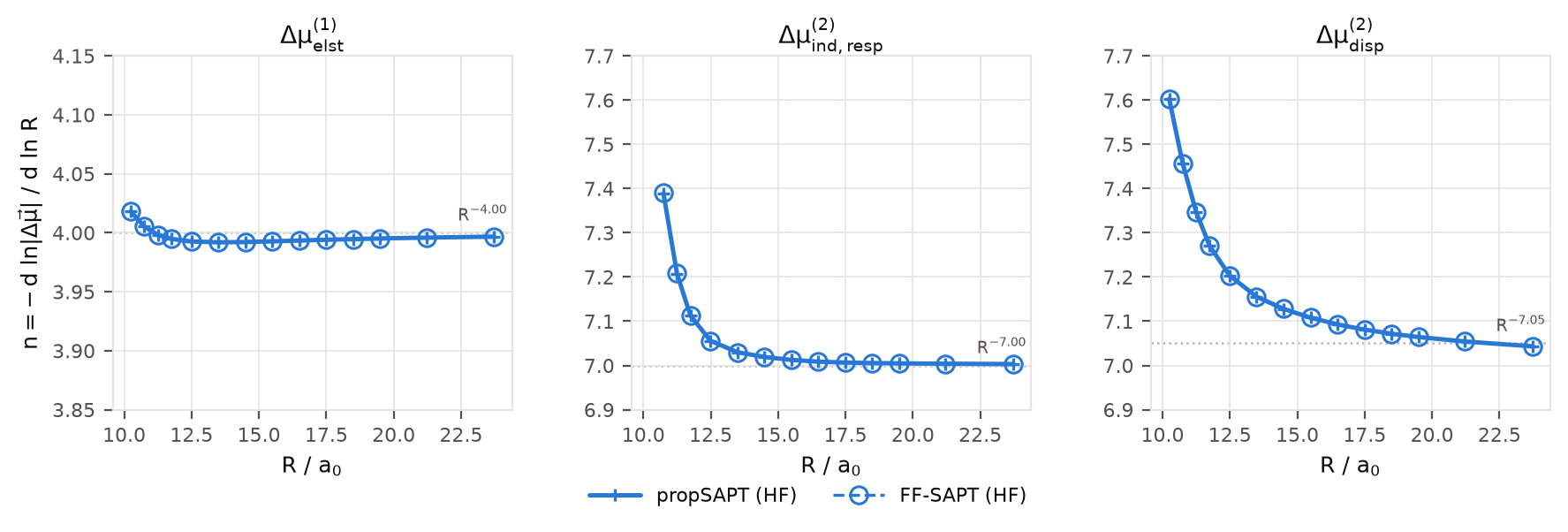}
        \caption{$\theta = 90\degree$ (\textit{T-shaped})}
        \label{fig:he-h2-dipole-asymptotics-theta-90}
    \end{subfigure}
    \caption{Asymptotic behaviour of the propSAPT(HF) and FF-SAPT(HF) contributions at long range, presented as the local (effective) decay exponent calculated as in Eq.~(\ref{eq:exponent}) for the \ce{He\bond{...}H2} system in the linear (panel a) and T-shaped (panel b) geometries.} 
    \label{fig:he-h2-dipole-asymptotics}
\end{figure}

In  Figs.~\ref{fig:he-ar-dipole-asymptotics} and~\ref{fig:he-h2-dipole-asymptotics} we show how the individual corrections approach their asymptotic power-law behavior at long range, as characterized by Eq.~(\ref{eq:exponent}). In all cases, the expected asymptotic exponents are recovered: for \ce{He\bond{...}Ar}, the relevant contribution decays as $R^{-7}$, whereas for \ce{He\bond{...}H2} the first-order electrostatic contribution follows the $R^{-4}$ law and the second-order induction-response and dispersion contributions decay as $R^{-7}$. Interestingly, the asymptotic $R^{-7}$ behaviour of dispersion appears at considerably larger separations than for the induction and electrostatic terms. The propSAPT and FF-SAPT results are indistinguishable over the whole range of distances and yield the same asymptotic exponents, providing an additional validation of the analytical propSAPT formulation.


\subsection{Accuracy of propSAPT to the second order}
\label{subsec:accuracy}

From a practical point of view, propSAPT offers an attractive alternative to supermolecular finite-field methods for the evaluation of interaction-induced properties. In this section, we assess its accuracy by comparison with finite-field CCSD(T) reference values. We additionally consider propSAPT(DFT), which represents the most straightforward and computationally efficient route for incorporating intramonomer electron correlation within the propSAPT framework.

\begin{figure}
    \centering
    \includegraphics[width=0.6\linewidth]{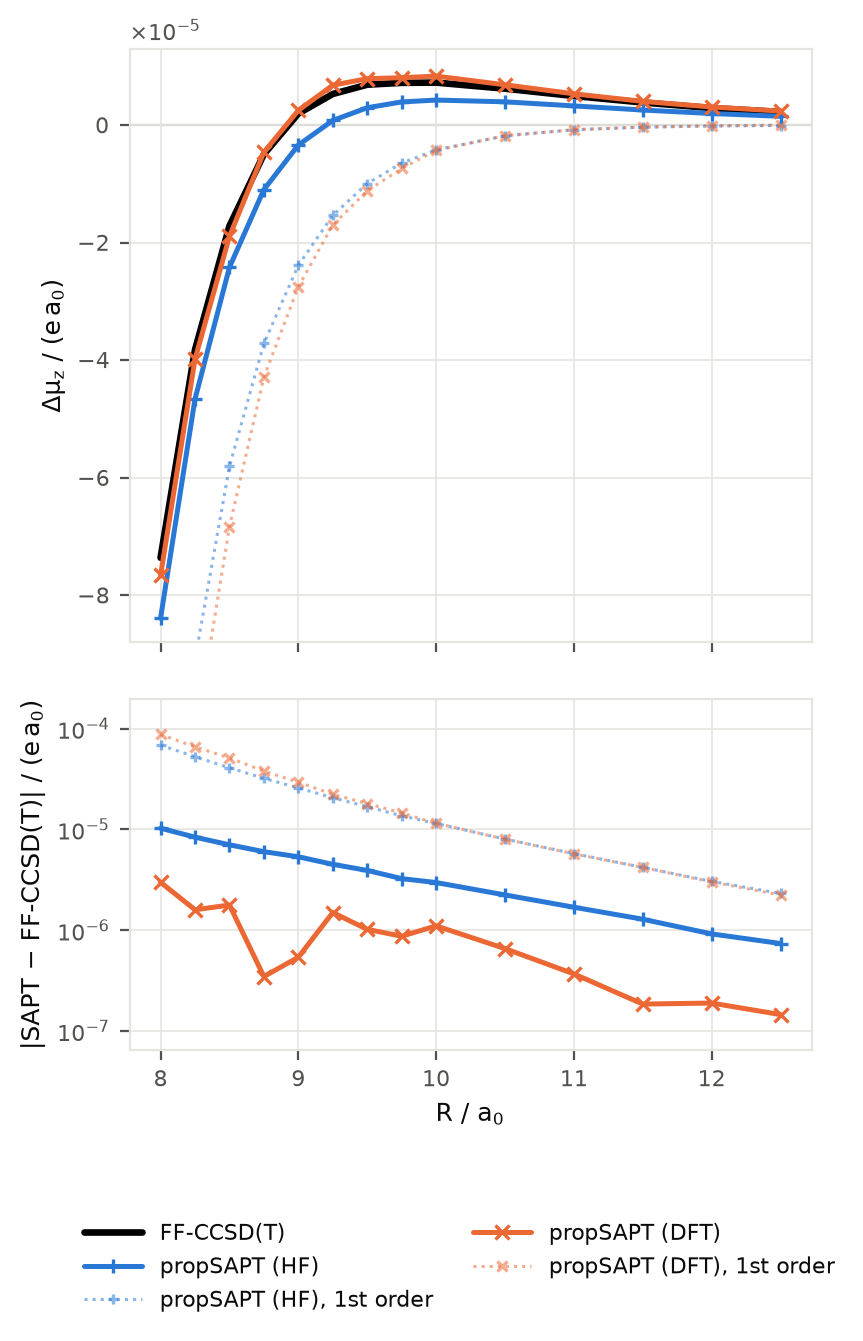}
    \caption{The comparison of the first- and second-order propSAPT interaction-induced dipole moments with the finite-field supermolecular CCSD(T) reference for the \ce{He\bond{...}Ar} complex. The lower panel shows the absolute error. (Note the logarithmic scale of the error panels' y-axis).}
    \label{fig:he-ar-dipole-totals}
\end{figure}

\begin{figure}
    \centering
    \includegraphics[width=0.6\linewidth]{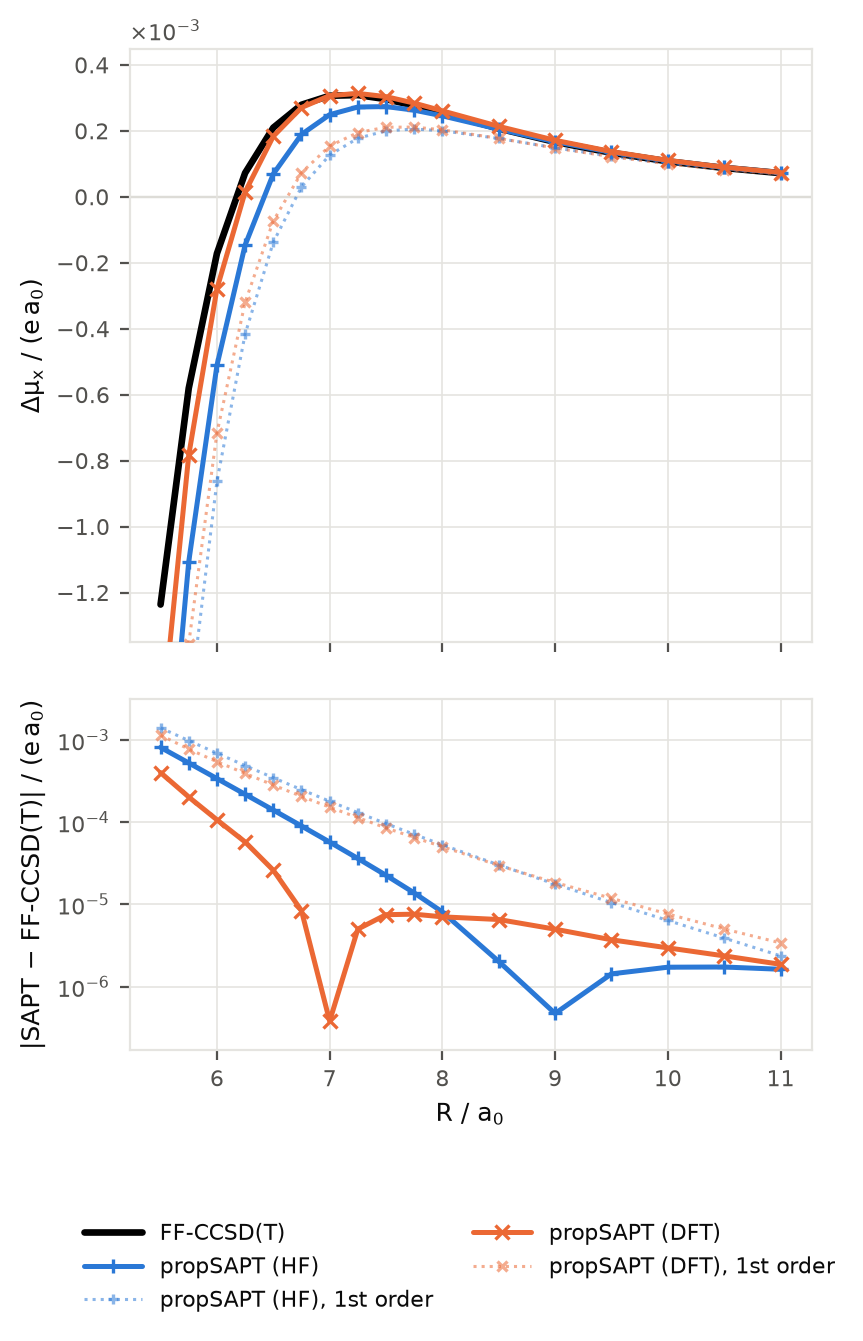}
    \caption{The single nonvanishing component of the interaction-induced dipole moment for the \ce{He\bond{...}H2} system at the T-shaped configuration. The lower panel shows the absolute error with respect to the finite-field supermolecular CCSD(T) calculation. (Note the sign change and the logarithmic y-axis on the error panel).}
    \label{fig:he-h2-dipole-totals}
\end{figure}

Let us first examine the importance of the second-order corrections for the overall performance of propSAPT for \ce{He\bond{...}Ar} and \ce{He\bond{...}H2}. Figures~\ref{fig:he-ar-dipole-totals} and \ref{fig:he-h2-dipole-totals} compare finite-field CCSD(T) results with first- and second-order propSAPT(HF) and propSAPT(DFT). For \ce{He\bond{...}Ar}, the second-order terms qualitatively change the shape of the induced-dipole moment curve: the first-order propSAPT result does not reproduce the turnover of the dipole moment and also exhibits visibly incorrect asymptotic behaviour. In contrast, for \ce{He\bond{...}H2} the qualitatively correct behaviour is already reproduced at first order. Nonetheless, the inclusion of the second-order contributions substantially improves the agreement with the reference, especially at the DFT/PBE0 level, where the propSAPT results are very close to the FF-CCSD(T) data.

\begin{table}
  \centering
  \small
  \caption{
    Interaction-induced dipole moments $\Delta\mu$ (in $e\,a_0$) of the \ce{He\bond{...}Ar}, \ce{He\bond{...}H2} and \ce{H2\bond{...}H2} dimers, from propSAPT and the FF-SAPT methods. The intermolecular vector $\vec{R}$ is directed along the $z$ axis, pointing from monomer A to monomer B. In case of \ce{H2\bond{...}H2} dimer, the $\theta_1$ and $\theta_2$ angles rotate monomers A and B, respectively, around their centre of mass in the $xz$ plane.
    $R$ in $a_0$, angles in degrees. The first-order exchange is $S^\infty$ and the second-order exchange corrections utilise the $S^2$ approximation. All calculations performed in the aug-cc-pVTZ basis set.
  }
  \label{tab:induced-dipoles-cartesian}
  \vspace{0.8em}
  \setlength{\tabcolsep}{4pt}
  \begin{tabular}{r rrr @{\hspace{1.5em}} l @{\hspace{1.5em}} rrrrr}
    \toprule
    \multicolumn{4}{c}{Jacobi coordinates} &  & \multicolumn{2}{c}{SAPT(HF)} & \multicolumn{2}{c}{SAPT(DFT)} & CCSD(T) \\
    \cmidrule(lr){1-4} \cmidrule(l{0pt}r{0.5em}){6-7} \cmidrule(l{0.5em}r{0.5em}){8-9} \cmidrule(l{0.5em}r{0pt}){10-10}
    $R$ & $\theta_1$ & $\theta_2$ & $\phi$ &  & propSAPT & FF & propSAPT & FF & FF \\
    \midrule
    \multicolumn{10}{l}{\textbf{He$\cdots$Ar}} \\
    \addlinespace[2pt]
    $4.00$ & -- & -- & -- & $\mu_{z}$ & $-7$\rlap{$.948{\times}10^{-2}$}\phantom{$.000{\times}10^{-0}$} & $-7$\rlap{$.948{\times}10^{-2}$}\phantom{$.000{\times}10^{-0}$} & $-8$\rlap{$.578{\times}10^{-2}$}\phantom{$.000{\times}10^{-0}$} & $-8$\rlap{$.347{\times}10^{-2}$}\phantom{$.000{\times}10^{-0}$} & $-5$\rlap{$.489{\times}10^{-2}$}\phantom{$.000{\times}10^{-0}$} \\
    $6.50$ & -- & -- & -- & $\mu_{z}$ & $-1$\rlap{$.609{\times}10^{-3}$}\phantom{$.000{\times}10^{-0}$} & $-1$\rlap{$.610{\times}10^{-3}$}\phantom{$.000{\times}10^{-0}$} & $-1$\rlap{$.714{\times}10^{-3}$}\phantom{$.000{\times}10^{-0}$} & $-1$\rlap{$.643{\times}10^{-3}$}\phantom{$.000{\times}10^{-0}$} & $-1$\rlap{$.483{\times}10^{-3}$}\phantom{$.000{\times}10^{-0}$} \\
    $10.00$ & -- & -- & -- & $\mu_{z}$ & $4$\rlap{$.289{\times}10^{-6}$}\phantom{$.000{\times}10^{-0}$} & $4$\rlap{$.295{\times}10^{-6}$}\phantom{$.000{\times}10^{-0}$} & $8$\rlap{$.365{\times}10^{-6}$}\phantom{$.000{\times}10^{-0}$} & $8$\rlap{$.107{\times}10^{-6}$}\phantom{$.000{\times}10^{-0}$} & $7$\rlap{$.261{\times}10^{-6}$}\phantom{$.000{\times}10^{-0}$} \\
    $15.00$ & -- & -- & -- & $\mu_{z}$ & $4$\rlap{$.243{\times}10^{-7}$}\phantom{$.000{\times}10^{-0}$} & $4$\rlap{$.264{\times}10^{-7}$}\phantom{$.000{\times}10^{-0}$} & $6$\rlap{$.576{\times}10^{-7}$}\phantom{$.000{\times}10^{-0}$} & $6$\rlap{$.394{\times}10^{-7}$}\phantom{$.000{\times}10^{-0}$} & $6$\rlap{$.290{\times}10^{-7}$}\phantom{$.000{\times}10^{-0}$} \\
    \midrule
    \multicolumn{10}{l}{\textbf{He$\cdots$H$_2$}} \\
    \addlinespace[2pt]
    $4.50$ & -- & $0$ & -- & $\mu_{z}$ & $-3$\rlap{$.234{\times}10^{-2}$}\phantom{$.000{\times}10^{-0}$} & $-3$\rlap{$.235{\times}10^{-2}$}\phantom{$.000{\times}10^{-0}$} & $-3$\rlap{$.177{\times}10^{-2}$}\phantom{$.000{\times}10^{-0}$} & $-3$\rlap{$.306{\times}10^{-2}$}\phantom{$.000{\times}10^{-0}$} & $-2$\rlap{$.661{\times}10^{-2}$}\phantom{$.000{\times}10^{-0}$} \\
    $6.50$ & -- & $0$ & -- & $\mu_{z}$ & $-2$\rlap{$.037{\times}10^{-3}$}\phantom{$.000{\times}10^{-0}$} & $-2$\rlap{$.037{\times}10^{-3}$}\phantom{$.000{\times}10^{-0}$} & $-1$\rlap{$.851{\times}10^{-3}$}\phantom{$.000{\times}10^{-0}$} & $-1$\rlap{$.867{\times}10^{-3}$}\phantom{$.000{\times}10^{-0}$} & $-1$\rlap{$.785{\times}10^{-3}$}\phantom{$.000{\times}10^{-0}$} \\
    $10.00$ & -- & $0$ & -- & $\mu_{z}$ & $-2$\rlap{$.071{\times}10^{-4}$}\phantom{$.000{\times}10^{-0}$} & $-2$\rlap{$.071{\times}10^{-4}$}\phantom{$.000{\times}10^{-0}$} & $-1$\rlap{$.999{\times}10^{-4}$}\phantom{$.000{\times}10^{-0}$} & $-2$\rlap{$.045{\times}10^{-4}$}\phantom{$.000{\times}10^{-0}$} & $-1$\rlap{$.960{\times}10^{-4}$}\phantom{$.000{\times}10^{-0}$} \\
    $15.00$ & -- & $0$ & -- & $\mu_{z}$ & $-4$\rlap{$.114{\times}10^{-5}$}\phantom{$.000{\times}10^{-0}$} & $-4$\rlap{$.114{\times}10^{-5}$}\phantom{$.000{\times}10^{-0}$} & $-4$\rlap{$.035{\times}10^{-5}$}\phantom{$.000{\times}10^{-0}$} & $-4$\rlap{$.141{\times}10^{-5}$}\phantom{$.000{\times}10^{-0}$} & $-3$\rlap{$.939{\times}10^{-5}$}\phantom{$.000{\times}10^{-0}$} \\
    \addlinespace
    $4.50$ & -- & $90$ & -- & $\mu_{z}$ & $-1$\rlap{$.399{\times}10^{-2}$}\phantom{$.000{\times}10^{-0}$} & $-1$\rlap{$.399{\times}10^{-2}$}\phantom{$.000{\times}10^{-0}$} & $-1$\rlap{$.296{\times}10^{-2}$}\phantom{$.000{\times}10^{-0}$} & $-1$\rlap{$.357{\times}10^{-2}$}\phantom{$.000{\times}10^{-0}$} & $-0$\rlap{$.950{\times}10^{-2}$}\phantom{$.000{\times}10^{-0}$} \\
    $6.50$$^{\mathrm{a}}$ & -- & $90$ & -- & $\mu_{z}$ & $0$\rlap{$.706{\times}10^{-4}$}\phantom{$.000{\times}10^{-0}$} & $0$\rlap{$.704{\times}10^{-4}$}\phantom{$.000{\times}10^{-0}$} & $1$\rlap{$.850{\times}10^{-4}$}\phantom{$.000{\times}10^{-0}$} & $2$\rlap{$.037{\times}10^{-4}$}\phantom{$.000{\times}10^{-0}$} & $2$\rlap{$.111{\times}10^{-4}$}\phantom{$.000{\times}10^{-0}$} \\
    $10.00$ & -- & $90$ & -- & $\mu_{z}$ & $1$\rlap{$.107{\times}10^{-4}$}\phantom{$.000{\times}10^{-0}$} & $1$\rlap{$.107{\times}10^{-4}$}\phantom{$.000{\times}10^{-0}$} & $1$\rlap{$.120{\times}10^{-4}$}\phantom{$.000{\times}10^{-0}$} & $1$\rlap{$.154{\times}10^{-4}$}\phantom{$.000{\times}10^{-0}$} & $1$\rlap{$.090{\times}10^{-4}$}\phantom{$.000{\times}10^{-0}$} \\
    $15.00$ & -- & $90$ & -- & $\mu_{z}$ & $2$\rlap{$.102{\times}10^{-5}$}\phantom{$.000{\times}10^{-0}$} & $2$\rlap{$.102{\times}10^{-5}$}\phantom{$.000{\times}10^{-0}$} & $2$\rlap{$.092{\times}10^{-5}$}\phantom{$.000{\times}10^{-0}$} & $2$\rlap{$.150{\times}10^{-5}$}\phantom{$.000{\times}10^{-0}$} & $2$\rlap{$.037{\times}10^{-5}$}\phantom{$.000{\times}10^{-0}$} \\
    \midrule
    \multicolumn{10}{l}{\textbf{H$_2\cdots$H$_2$}} \\
    \addlinespace[2pt]
    $5.00$ & $0$ & $90$ & $0$ & $\mu_{z}$ & $2$\rlap{$.503{\times}10^{-2}$}\phantom{$.000{\times}10^{-0}$} & $2$\rlap{$.503{\times}10^{-2}$}\phantom{$.000{\times}10^{-0}$} & $2$\rlap{$.342{\times}10^{-2}$}\phantom{$.000{\times}10^{-0}$} & $2$\rlap{$.387{\times}10^{-2}$}\phantom{$.000{\times}10^{-0}$} & $2$\rlap{$.611{\times}10^{-2}$}\phantom{$.000{\times}10^{-0}$} \\
    $6.50$ & $0$ & $90$ & $0$ & $\mu_{z}$ & $7$\rlap{$.843{\times}10^{-3}$}\phantom{$.000{\times}10^{-0}$} & $7$\rlap{$.844{\times}10^{-3}$}\phantom{$.000{\times}10^{-0}$} & $7$\rlap{$.128{\times}10^{-3}$}\phantom{$.000{\times}10^{-0}$} & $7$\rlap{$.339{\times}10^{-3}$}\phantom{$.000{\times}10^{-0}$} & $7$\rlap{$.346{\times}10^{-3}$}\phantom{$.000{\times}10^{-0}$} \\
    $10.00$ & $0$ & $90$ & $0$ & $\mu_{z}$ & $1$\rlap{$.323{\times}10^{-3}$}\phantom{$.000{\times}10^{-0}$} & $1$\rlap{$.323{\times}10^{-3}$}\phantom{$.000{\times}10^{-0}$} & $1$\rlap{$.216{\times}10^{-3}$}\phantom{$.000{\times}10^{-0}$} & $1$\rlap{$.246{\times}10^{-3}$}\phantom{$.000{\times}10^{-0}$} & $1$\rlap{$.199{\times}10^{-3}$}\phantom{$.000{\times}10^{-0}$} \\
    $16.00$ & $0$ & $90$ & $0$ & $\mu_{z}$ & $1$\rlap{$.999{\times}10^{-4}$}\phantom{$.000{\times}10^{-0}$} & $1$\rlap{$.999{\times}10^{-4}$}\phantom{$.000{\times}10^{-0}$} & $1$\rlap{$.845{\times}10^{-4}$}\phantom{$.000{\times}10^{-0}$} & $1$\rlap{$.893{\times}10^{-4}$}\phantom{$.000{\times}10^{-0}$} & $1$\rlap{$.815{\times}10^{-4}$}\phantom{$.000{\times}10^{-0}$} \\
    \addlinespace
    $5.00$ & $0$ & $45$ & $0$ & $\mu_{z}$ & $1$\rlap{$.352{\times}10^{-2}$}\phantom{$.000{\times}10^{-0}$} & $1$\rlap{$.352{\times}10^{-2}$}\phantom{$.000{\times}10^{-0}$} & $1$\rlap{$.271{\times}10^{-2}$}\phantom{$.000{\times}10^{-0}$} & $1$\rlap{$.293{\times}10^{-2}$}\phantom{$.000{\times}10^{-0}$} & $1$\rlap{$.427{\times}10^{-2}$}\phantom{$.000{\times}10^{-0}$} \\
     &  &  &  & $\mu_{x}$ & $6$\rlap{$.800{\times}10^{-3}$}\phantom{$.000{\times}10^{-0}$} & $6$\rlap{$.800{\times}10^{-3}$}\phantom{$.000{\times}10^{-0}$} & $6$\rlap{$.466{\times}10^{-3}$}\phantom{$.000{\times}10^{-0}$} & $6$\rlap{$.594{\times}10^{-3}$}\phantom{$.000{\times}10^{-0}$} & $6$\rlap{$.012{\times}10^{-3}$}\phantom{$.000{\times}10^{-0}$} \\
    $6.50$ & $0$ & $45$ & $0$ & $\mu_{z}$ & $4$\rlap{$.142{\times}10^{-3}$}\phantom{$.000{\times}10^{-0}$} & $4$\rlap{$.142{\times}10^{-3}$}\phantom{$.000{\times}10^{-0}$} & $3$\rlap{$.756{\times}10^{-3}$}\phantom{$.000{\times}10^{-0}$} & $3$\rlap{$.855{\times}10^{-3}$}\phantom{$.000{\times}10^{-0}$} & $3$\rlap{$.904{\times}10^{-3}$}\phantom{$.000{\times}10^{-0}$} \\
     &  &  &  & $\mu_{x}$ & $2$\rlap{$.947{\times}10^{-3}$}\phantom{$.000{\times}10^{-0}$} & $2$\rlap{$.947{\times}10^{-3}$}\phantom{$.000{\times}10^{-0}$} & $2$\rlap{$.782{\times}10^{-3}$}\phantom{$.000{\times}10^{-0}$} & $2$\rlap{$.845{\times}10^{-3}$}\phantom{$.000{\times}10^{-0}$} & $2$\rlap{$.685{\times}10^{-3}$}\phantom{$.000{\times}10^{-0}$} \\
    $10.00$ & $0$ & $45$ & $0$ & $\mu_{z}$ & $6$\rlap{$.764{\times}10^{-4}$}\phantom{$.000{\times}10^{-0}$} & $6$\rlap{$.765{\times}10^{-4}$}\phantom{$.000{\times}10^{-0}$} & $6$\rlap{$.216{\times}10^{-4}$}\phantom{$.000{\times}10^{-0}$} & $6$\rlap{$.371{\times}10^{-4}$}\phantom{$.000{\times}10^{-0}$} & $6$\rlap{$.136{\times}10^{-4}$}\phantom{$.000{\times}10^{-0}$} \\
     &  &  &  & $\mu_{x}$ & $5$\rlap{$.459{\times}10^{-4}$}\phantom{$.000{\times}10^{-0}$} & $5$\rlap{$.459{\times}10^{-4}$}\phantom{$.000{\times}10^{-0}$} & $5$\rlap{$.046{\times}10^{-4}$}\phantom{$.000{\times}10^{-0}$} & $5$\rlap{$.156{\times}10^{-4}$}\phantom{$.000{\times}10^{-0}$} & $4$\rlap{$.978{\times}10^{-4}$}\phantom{$.000{\times}10^{-0}$} \\
    $16.00$ & $0$ & $45$ & $0$ & $\mu_{z}$ & $1$\rlap{$.008{\times}10^{-4}$}\phantom{$.000{\times}10^{-0}$} & $1$\rlap{$.008{\times}10^{-4}$}\phantom{$.000{\times}10^{-0}$} & $0$\rlap{$.930{\times}10^{-4}$}\phantom{$.000{\times}10^{-0}$} & $0$\rlap{$.954{\times}10^{-4}$}\phantom{$.000{\times}10^{-0}$} & $0$\rlap{$.916{\times}10^{-4}$}\phantom{$.000{\times}10^{-0}$} \\
     &  &  &  & $\mu_{x}$ & $8$\rlap{$.280{\times}10^{-5}$}\phantom{$.000{\times}10^{-0}$} & $8$\rlap{$.280{\times}10^{-5}$}\phantom{$.000{\times}10^{-0}$} & $7$\rlap{$.615{\times}10^{-5}$}\phantom{$.000{\times}10^{-0}$} & $7$\rlap{$.772{\times}10^{-5}$}\phantom{$.000{\times}10^{-0}$} & $7$\rlap{$.525{\times}10^{-5}$}\phantom{$.000{\times}10^{-0}$} \\
    \addlinespace
    $5.00$ & $45$ & $90$ & $0$ & $\mu_{z}$ & $1$\rlap{$.166{\times}10^{-2}$}\phantom{$.000{\times}10^{-0}$} & $1$\rlap{$.166{\times}10^{-2}$}\phantom{$.000{\times}10^{-0}$} & $1$\rlap{$.086{\times}10^{-2}$}\phantom{$.000{\times}10^{-0}$} & $1$\rlap{$.117{\times}10^{-2}$}\phantom{$.000{\times}10^{-0}$} & $1$\rlap{$.203{\times}10^{-2}$}\phantom{$.000{\times}10^{-0}$} \\
     &  &  &  & $\mu_{x}$ & $-5$\rlap{$.946{\times}10^{-3}$}\phantom{$.000{\times}10^{-0}$} & $-5$\rlap{$.946{\times}10^{-3}$}\phantom{$.000{\times}10^{-0}$} & $-5$\rlap{$.754{\times}10^{-3}$}\phantom{$.000{\times}10^{-0}$} & $-5$\rlap{$.895{\times}10^{-3}$}\phantom{$.000{\times}10^{-0}$} & $-5$\rlap{$.296{\times}10^{-3}$}\phantom{$.000{\times}10^{-0}$} \\
    $6.50$ & $45$ & $90$ & $0$ & $\mu_{z}$ & $3$\rlap{$.734{\times}10^{-3}$}\phantom{$.000{\times}10^{-0}$} & $3$\rlap{$.734{\times}10^{-3}$}\phantom{$.000{\times}10^{-0}$} & $3$\rlap{$.429{\times}10^{-3}$}\phantom{$.000{\times}10^{-0}$} & $3$\rlap{$.498{\times}10^{-3}$}\phantom{$.000{\times}10^{-0}$} & $3$\rlap{$.477{\times}10^{-3}$}\phantom{$.000{\times}10^{-0}$} \\
     &  &  &  & $\mu_{x}$ & $-2$\rlap{$.526{\times}10^{-3}$}\phantom{$.000{\times}10^{-0}$} & $-2$\rlap{$.526{\times}10^{-3}$}\phantom{$.000{\times}10^{-0}$} & $-2$\rlap{$.397{\times}10^{-3}$}\phantom{$.000{\times}10^{-0}$} & $-2$\rlap{$.457{\times}10^{-3}$}\phantom{$.000{\times}10^{-0}$} & $-2$\rlap{$.310{\times}10^{-3}$}\phantom{$.000{\times}10^{-0}$} \\
    $10.00$ & $45$ & $90$ & $0$ & $\mu_{z}$ & $6$\rlap{$.502{\times}10^{-4}$}\phantom{$.000{\times}10^{-0}$} & $6$\rlap{$.502{\times}10^{-4}$}\phantom{$.000{\times}10^{-0}$} & $5$\rlap{$.983{\times}10^{-4}$}\phantom{$.000{\times}10^{-0}$} & $6$\rlap{$.137{\times}10^{-4}$}\phantom{$.000{\times}10^{-0}$} & $5$\rlap{$.889{\times}10^{-4}$}\phantom{$.000{\times}10^{-0}$} \\
     &  &  &  & $\mu_{x}$ & $-4$\rlap{$.661{\times}10^{-4}$}\phantom{$.000{\times}10^{-0}$} & $-4$\rlap{$.661{\times}10^{-4}$}\phantom{$.000{\times}10^{-0}$} & $-4$\rlap{$.327{\times}10^{-4}$}\phantom{$.000{\times}10^{-0}$} & $-4$\rlap{$.433{\times}10^{-4}$}\phantom{$.000{\times}10^{-0}$} & $-4$\rlap{$.253{\times}10^{-4}$}\phantom{$.000{\times}10^{-0}$} \\
    $16.00$ & $45$ & $90$ & $0$ & $\mu_{z}$ & $9$\rlap{$.921{\times}10^{-5}$}\phantom{$.000{\times}10^{-0}$} & $9$\rlap{$.921{\times}10^{-5}$}\phantom{$.000{\times}10^{-0}$} & $9$\rlap{$.160{\times}10^{-5}$}\phantom{$.000{\times}10^{-0}$} & $9$\rlap{$.396{\times}10^{-5}$}\phantom{$.000{\times}10^{-0}$} & $9$\rlap{$.009{\times}10^{-5}$}\phantom{$.000{\times}10^{-0}$} \\
     &  &  &  & $\mu_{x}$ & $-7$\rlap{$.062{\times}10^{-5}$}\phantom{$.000{\times}10^{-0}$} & $-7$\rlap{$.062{\times}10^{-5}$}\phantom{$.000{\times}10^{-0}$} & $-6$\rlap{$.521{\times}10^{-5}$}\phantom{$.000{\times}10^{-0}$} & $-6$\rlap{$.681{\times}10^{-5}$}\phantom{$.000{\times}10^{-0}$} & $-6$\rlap{$.423{\times}10^{-5}$}\phantom{$.000{\times}10^{-0}$} \\
    \bottomrule
  \end{tabular}
  \par\medskip
  \begin{minipage}{\linewidth}\footnotesize
    $^{\mathrm{a}}$~The total induced dipole for the T-shaped \ce{He\bond{...}H2} changes sign (direction) at $R \approx 6.5\ a_0$. Therefore, this entry is a residual of near-cancelling contributions; the difference between methods here is large in relative terms and small in absolute ones.
  \end{minipage}
\end{table}

In Table \ref{tab:induced-dipoles-cartesian}, we compare the performance of propSAPT based on HF and DFT/PBE0 orbitals. 
As discussed in the previous section, we see that at the SAPT(HF) level, the analytical propSAPT results are essentially indistinguishable from the corresponding finite-field values. This is not the case for SAPT(DFT); however, the agreement with FF-SAPT(DFT) remains within a few per cent, with a median relative deviation of approximately 2.5\%.
The largest relative discrepancy occurs close to the zero crossing of the induced dipole for the T-shaped \ce{He\bond{...}H2} complex, where the total moment results from a cancellation of larger contributions and relative errors are therefore strongly amplified. More importantly, the comparison with FF-CCSD(T) demonstrates that SAPT(DFT) provides a significantly improved description of the total induced dipole compared with SAPT(HF), primarily due to the inclusion of intramonomer correlation. 
The median relative error decreases from approximately 10\% for SAPT(HF) to about 4.5\% for FF-SAPT(DFT), and to about 3\% for the corresponding propSAPT(DFT) values. Larger deviations remain primarily in the short-range region, where perturbation theory lacks higher-order terms. This is particularly pronounced for \ce{He\bond{...}Ar}, where both SAPT(HF) and SAPT(DFT) overestimate the magnitude of the CCSD(T) induced dipole by about 40\%.


The \ce{H2\bond{...}H2} dimer is the most complex of the three systems considered here. Both monomers are anisotropic. While the leading long-range mechanism is, again, the quadrupole-induced dipole interaction, the angular dependence of $\Delta \vec{\mu}$ is very significant, so that the interaction-induced dipole must be resolved into several spherical components (\textit{vide infra}). 
The overall picture from Table~\ref{tab:induced-dipoles-cartesian} looks promising: as in previous cases, the finite-field SAPT data are reproduced either nearly exactly (for the HF-based case), or quite accurately (for the DFT-based one).

At long range ($R = 10\ a_0$), the behaviour is uniform across all three orientations and both non-zero Cartesian components: propSAPT(HF) overestimates the CCSD(T) dipole by
$10.0\%$, and propSAPT(DFT) by about $1.5\%$. At $R = 5\ a_0$, the
picture is less uniform, and the two components behave differently: the
component along $\vec{R}$ is underestimated by $3$--$5\%$ at the Hartree--Fock
level and by about $10\%$ at the Kohn--Sham one, while the perpendicular
component is overestimated by $13\%$ and $9\%$, respectively. Note that the
long-range deviation is the same to within $0.1\%$ for every geometry and both components.

\begin{figure}
    \centering
    \includegraphics[width=0.6\linewidth]{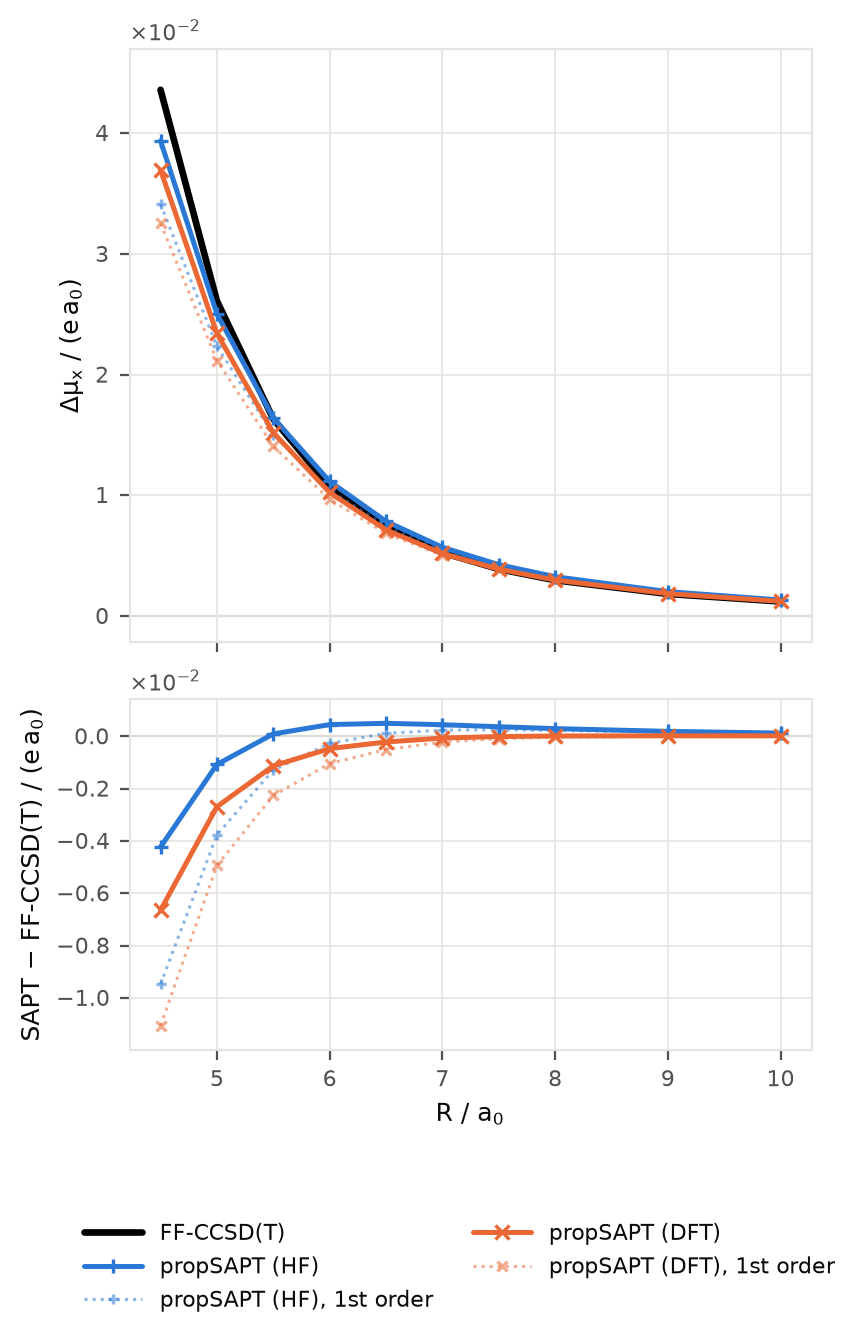}
    \caption{The single nonvanishing component of the interaction-induced dipole moment for the \ce{H2\bond{...}H2} system at a T-shaped configuration. The lower panel shows the error with respect to the finite-field CCSD(T) calculation.}
    \label{fig:h2-h2-comparison}
\end{figure}

As can be seen in the lower panel of Fig.~\ref{fig:h2-h2-comparison}, the
absolute deviation from CCSD(T) decreases monotonically above
$6\ a_0$, reaching $2\times10^{-5}\ e\,a_0$ for propSAPT(HF) and
$3 \times 10^{-6}\ e\,a_0$ for propSAPT(DFT) at the largest separation. 
To conclude this part, second-order propSAPT reproduces the CCSD(T) \ce{H2\bond{...}H2} interaction-induced
dipole to within a few per cent over the whole range of separations studied, and
the Kohn--Sham reference is clearly the better choice in the long-range region.

\subsection{Expansion of DMS into spherical components for molecular hydrogen complexes }
\label{subsec:expansion}

For molecular systems expressed in Jacobi coordinates, and for the interpretation of collision-induced absorption spectroscopy and bound-state complexes line intensities, it is informative to represent the DMS expansion in terms of spherical functions that depend on the coordinates of monomers. 
Following Ref.~\cite{vanderavoird1994Intermolecular}, the space-fixed spherical components of the dipole moment of a dimer composed of arbitrary (rigid) monomers can be expanded as
\begin{align}
    \mu^{\rm{SF}}_{q}(\vec{R}, \zeta^{\rm{SF}}_{1}, \zeta^{\rm{SF}}_{2}) & =
    \sum_{ \alpha \lambda}
    d_{ \alpha \lambda}(R)
    B_{ \alpha \lambda q}(
        \zeta^{\rm{SF}}_{1}, \zeta^{\rm{SF}}_{2}, \hat{R}
    )\ ,
\label{eq:dipole_expansion}
\end{align}
where
$\alpha =
\left(
    L_1, K_1,
    L_2, K_2,
    L_{12}
\right)$
is a joined tuple of indices and
$\vec{R} = R \hat{R}$ is the intermolecular vector directed from the
centre of mass of monomer 1 to that of monomer 2.
The Euler angles $\zeta_i^{\rm SF}$ specify the orientation of monomer $i$
with respect to the space-fixed frame, while $q=-1,0,1$ labels the spherical
components of the dipole moment.
The orthogonal angular basis functions are given by
\begin{align}
    B_{ \alpha \lambda q}(
        \zeta^{\rm{SF}}_{1}, \zeta^{\rm{SF}}_{2}, \hat{R}
    ) = &
    \sum_{M_1 M_2}
    \sum_{M_{12} m} 
    D^{(L_1)}_{M_1 K_1}(\zeta^{\rm{SF}}_1)^*
    D^{(L_2)}_{M_2 K_2}(\zeta^{\rm{SF}}_2)^*
    C^{\lambda}_{m}(\hat{R}) \\ \nonumber \times &
    \langle
        L_1 M_1, L_2 M_2 | L_{12} M_{12}
    \rangle
    \langle
        L_{12} M_{12}, \lambda m | 1q
    \rangle\ ,
\label{eq:dipole_angular_general}
\end{align}
where $D^{(L)}_{MK}$ denotes a Wigner $D$-matrix element,
$C^\lambda_m$ is a Racah-normalised spherical harmonic,
and $\langle j_1m_1,j_2m_2|jm\rangle$ denotes a Clebsch--Gordan coefficient.

For a linear diatomic molecule, the Wigner $D$-matrix element reduces to
\begin{equation}
    D^{(L)}_{MK}( \zeta^{\rm{SF}})^* =
    \delta_{K0} C^L_M(\hat{r})\ ,
\end{equation}
where $\hat{r}$ denotes the orientation of the molecular axis with respect
to the space-fixed frame.
Consequently, the tuple of indices required to describe the dipole moment simplifies to 
\begin{equation}
    \alpha =
    \left(
        L_1,L_2,L_{12}
    \right)\ ,
\end{equation}
and the angular basis function becomes
\begin{align}
    B_{ L_1 L_2 L_{12} \lambda q}(\hat{r}_1, \hat{r}_2,\hat{R}) =&
    \sum_{M_1 M_2}
    \sum_{M_{12} m} 
    C^{L_1}_{M_1}(\hat{r}_1) 
    C^{L_2}_{M_2}(\hat{r}_2) 
    C^{\lambda}_{m}(\hat{R}) \\ \nonumber \times &
    \langle
        L_1 M_1, L_2 M_2 | L_{12} M_{12}
    \rangle
    \langle
        L_{12} M_{12}, \lambda m | 1q
    \rangle\ .
\label{eq:dipole_angular_diatom}
\end{align}

For an atom--diatom complex with a spherically symmetric atomic fragment,
only $L_1=0$ contributes, with $M_1=0$ and $C^0_0=1$.
The first Clebsch--Gordan coefficient then becomes
\begin{equation}
    \langle 00, L_2 M_2 | L_{12} M_{12}\rangle =
    \delta_{L_2 L_{12}}
    \delta_{M_2 M_{12}}\ ,
\end{equation}
and the angular basis function reduces to
\begin{align}
    B_{ L \lambda q}(\hat{r},\hat{R}) = &
    \sum_{Mm} 
    C^{L}_{M}(\hat{r}) 
    C^{\lambda}_{m}(\hat{R}) 
    \langle
        L M, \lambda m | 1q
    \rangle\ .
\end{align}

The products of spherical harmonics coupled by the Clebsch--Gordan
coefficients can be written more compactly by introducing the irreducible
spherical tensor product,
\begin{equation}
    \Bigl[
        C^{j_1} \otimes C^{j_2}
    \Bigr]^k_q =
    \sum_{m_1 m_2}C^{j_1}_{m_1}C^{j_2}_{m_2}
    \langle
        j_1 m_1, j_2 m_2|kq
    \rangle\ .
\end{equation}
With this notation, the space-fixed dipole moment of a diatom--diatom complex
takes the form
\begin{align}
    \mu^{\rm{SF}}_q(\vec{R},\hat{r}_1,\hat{r}_2) =
    \sum_{L_1 L_2 L_{12} \lambda}
    d_{L_1 L_2 L_{12} \lambda}(R)
    \Bigl[
        \bigl[
            C^{L_1}(\hat{r}_1) \otimes
            C^{L_2}(\hat{r}_2)
        \bigr]^{L_{12}} \otimes
        C^{\lambda}(\hat{R})   
    \Bigr]^{1}_{q}\ ,
\end{align}
whereas for an atom--diatom complex,
\begin{align}
    \mu^{\rm{SF}}_q(\vec{R},\hat{r}) =
    \sum_{L \lambda}
    d_{L \lambda}(R)
    \Bigl[
        C^{L}(\hat{r}) \otimes
        C^{\lambda}(\hat{R})   
    \Bigr]^{1}_{q}\ .
\end{align}

For a homonuclear diatomic molecule, the orientations $\hat{r}_i$ and
$-\hat{r}_i$ are physically equivalent.
Since the Racah-normalised spherical harmonics satisfy
\begin{equation}
    C^{L_i}_{M_i}(-\hat{r}_i) =
    (-1)^{L_i}C^{L_i}_{M_i}(\hat{r}_i)\ ,
\end{equation}
the \ce{H2\bond{...}H2} dipole expansion contains only even values of $L_1$ and $L_2$.
In addition, the dipole moment has odd parity under spatial inversion,
$(\hat{R},\hat{r}_1,\hat{r}_2) \to $ $(-\hat{R},-\hat{r}_1,-\hat{r}_2)$,
which requires
\begin{equation}
    (-1)^{L_1+L_2+\lambda} = -1\ .
\end{equation}
Consequently, $\lambda$ is restricted to odd values.
Additionally, for the identical monomers, the dipole moment
should be invariant with respect to exchange of like monomers
$(\hat{r}_1 \leftrightarrow \hat{r}_2, \hat{R}\to-\hat{R})$:
\begin{equation}
    d_{L_1 L_2 L \lambda}(R)=
    (-1)^{1-L}
    d_{L_2 L_1 L \lambda}(R)\ .
    \label{eq:symm_d}
\end{equation}
For \ce{He\bond{...}H2}, the homonuclear symmetry of \ce{H2} similarly restricts $L$ to even values.
Odd parity under spatial inversion then requires $(-1)^{L+\lambda}=-1$, and
therefore $\lambda$ is again restricted to odd values. 
In the case of the \ce{H2} dimer, we will denote the tuple of indices that uniquely describes $d_{L_1 L_2 L_{12} \lambda}(R)$ as 
$\Lambda = \left(L_1, L_2, L_{12}, \lambda \right)$
and in case of the \ce{He\bond{...}H2} system as
$\Lambda = \left(L, \lambda \right)$.

The $d_{0221}$ and $d_{0223}$ coefficients form the lowest-rank anisotropic family ($02$):
the dipole is isotropic with respect to the orientation of one \ce{H2} monomer and
rank two with respect to the other.
This angular structure describes an induced dipole accompanying a rank-two rotational
excitation of one monomer, including $j=0\rightarrow2$, while leaving the rotational
state of its partner unchanged.
The values $\lambda=1$ and $3$ specify the two allowed end-over-end angular dependencies.
The contributions of the $02/20$ components to observable absorption have
been evaluated in calculations of \ce{H2\bond{...}H2} CIA
spectra~\cite{meyer1989Absorption,meyer1989Rototranslational,li2010CollisionInduced,abel2011CollisionInduced,karman2015Collisioninduced}
and in analyses of bound hydrogen-dimer spectra~\cite{mckellar1990Infrared,mckellar1991Farinfrared}.
Their established spectroscopic importance motivates the selection of $d_{0221}$
and $d_{0223}$ as representative coefficients in the present work.

\begin{figure}
    \centering
    \begin{subfigure}{\linewidth}
        \centering
        \includegraphics[width=0.6\linewidth]{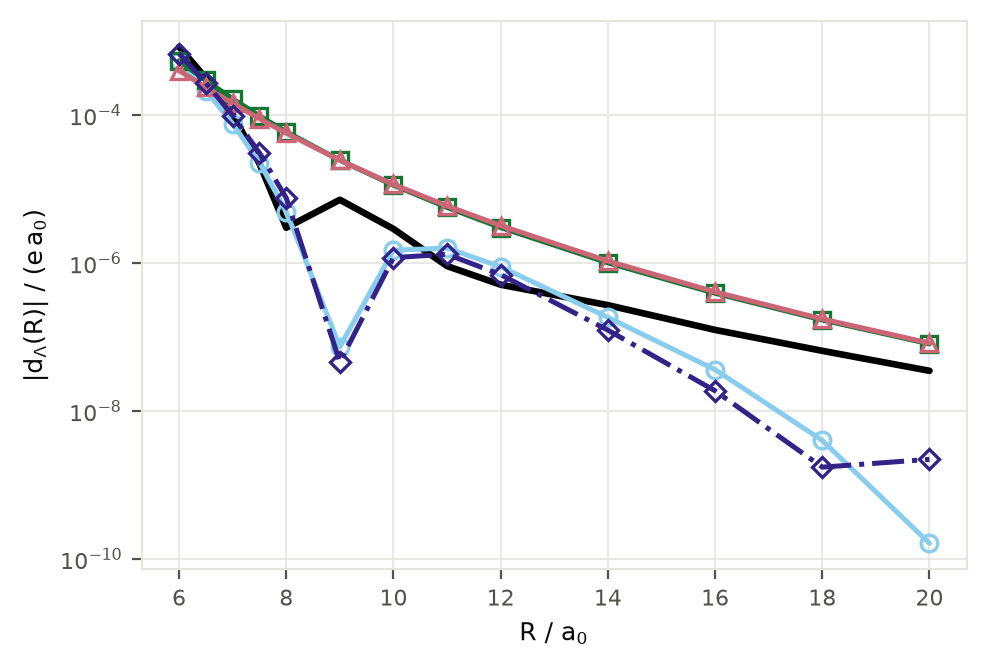}
        \caption{$\Lambda = (0,2,2,1)$}
        \label{fig:h2-h2-dms-fit-0221}
    \end{subfigure}
    \hfill
    \begin{subfigure}{\linewidth}
        \centering
        \includegraphics[width=0.6\linewidth]{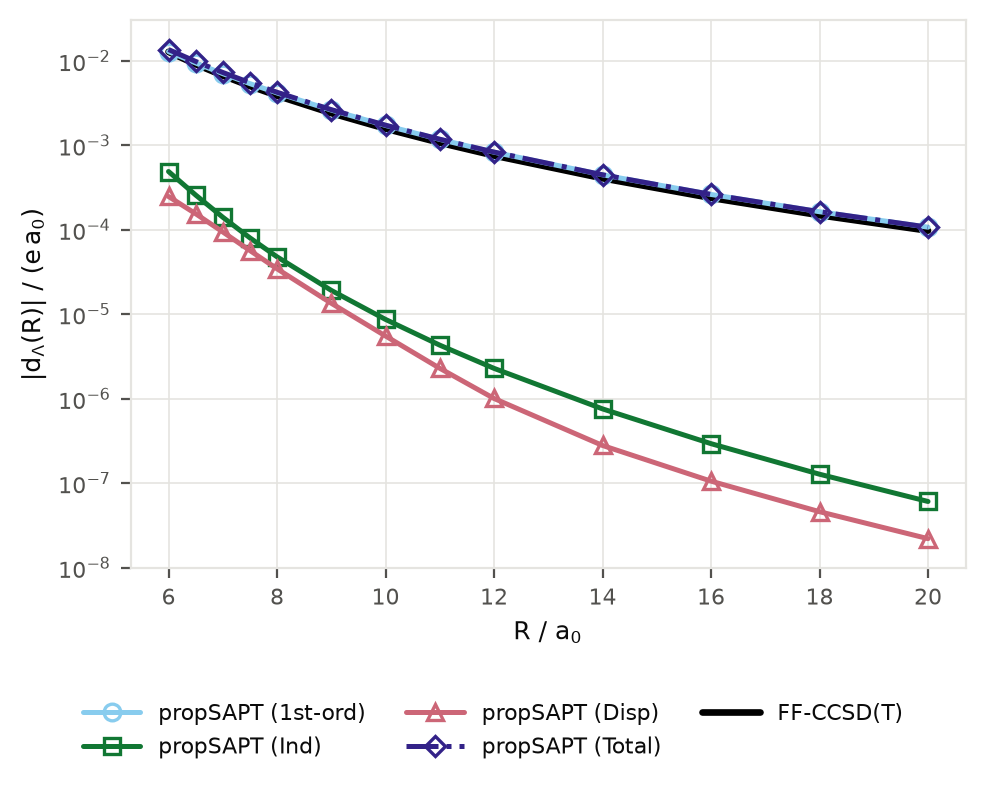}
        \caption{$\Lambda = (0,2,2,3)$}
        \label{fig:h2-h2-dms-fit-0223}
    \end{subfigure}
    \caption{The $d_{\Lambda}(R)$ coefficients of the \ce{H2\bond{...}H2} dipole moment surface fit for propSAPT(HF) and supermolecular FF-CCSD(T). 
    ``1st-ord'' denotes the sum of contributions to $d_{\Lambda}(R)$ arising from the sum
    $\vec{\mu}^{(1)}_\text{pol} + \vec{\mu}^{(1)}_\text{exch}$, 
    ``Ind'' from 
    $\vec{\mu}^{(2)}_\text{ind,r} + \vec{\mu}^{(2)}_\text{exch-ind,r}$ 
    and ``Disp'' from 
    $\vec{\mu}^{(2)}_\text{disp} + \vec{\mu}^{(2)}_\text{exch-disp}$.}
    \label{fig:h2-h2-dms-fit-coeffs}
\end{figure}

In Figures~\ref{fig:h2-h2-dms-fit-0221} and~\ref{fig:h2-h2-dms-fit-0223} we present the lowest-order dipole expansion terms for the \ce{H2\bond{...}H2} system, and Table~\ref{tab:dms-term-coefficients-h2-h2} shows a few more of them for several fixed distances.
According to the symmetry property~(\ref{eq:symm_d}), only one unique member of each symmetry-related pair is reported in Table~\ref{tab:dms-term-coefficients-h2-h2} for the molecular hydrogen dimer. 
The results show a strikingly different perturbative origin of the two coefficients. 
For $\Lambda = (0,2,2,1)$, the first-order terms decay exponentially, and the coefficient is determined by the second-order induction and dispersion terms, 
while the $\Lambda = (0,2,2,3)$ term is dominated by the first-order contribution at all separations, with induction and dispersion terms being more than two orders of magnitude smaller for separations above $10\ a_0$.
More specifically, these contributions to $d_{0221}$ have opposite signs and strongly cancel. This calls for particular attention, since dispersion interactions are very sensitive to the level of theory. 
In our example, the propSAPT(DFT) and propSAPT(HF) total moments at long range differ very significantly. Note also that since the supermolecular Hartree-Fock theory includes only the induction energy, despite the correct asymptotics in the first and second order, it will wrongly reproduce the $\Lambda = (0,2,2,1)$ term due to a lack of balance from dispersion.  
A comparable pattern is found for higher-order $\Lambda$: the $d_{0443}$ term is dominated by the second-order, $d_{0445}$ by the first-order contributions.

\begin{table}
  \centering
  \small
  \caption{%
    Radial expansion coefficients $d_{\Lambda}(R)$ (in units of $10^{-6}$ $e\,a_0$) of the \ce{H2\bond{...}H2} interaction-induced dipole moment surface in the Racah-normalized spherical-harmonic basis, resolved into the SAPT corrections, at three intermolecular separations $R$ (in $a_0$).
    $\Lambda = (L_1, L_2, L_{12}, \lambda)$ labels the basis function. 
    Each column is an independent fit to a given contribution.
  }
  \label{tab:dms-term-coefficients-h2-h2}
  \vspace{0.8em}
  \begin{tabular}{r l @{\hspace{2em}} rrrr @{\hspace{2em}} rrrr}
    \toprule
     &  & \multicolumn{4}{c}{propSAPT(HF)} & \multicolumn{4}{c}{propSAPT(DFT)} \\
    \cmidrule(l{0pt}r{0.5em}){3-6} \cmidrule(l{0.5em}r{0.5em}){7-10}
    $R$ & $\Lambda$ & 1st-order & Ind & Disp & total & 1st-order & Ind & Disp & total \\
    \midrule
    $6.00$ & $(0,2,2,1)$ & $535$ & $542$ & $-398$ & $679$ & $508$ & $500$ & $-545$ & $463$ \\
     & $(0,2,2,3)$ & $-12854$ & $-489$ & $-248$ & $-13591$ & $-11903$ & $-448$ & $-336$ & $-12686$ \\
     & $(0,4,4,3)$ & $71$ & $116$ & $-31$ & $156$ & $68$ & $107$ & $-50$ & $125$ \\
     & $(0,4,4,5)$ & $-550$ & $-56$ & $1$ & $-605$ & $-513$ & $-54$ & $1$ & $-566$ \\
     & $(2,2,1,1)$ & $-74$ & $-15$ & $19$ & $-71$ & $-80$ & $-17$ & $30$ & $-67$ \\
     & $(2,2,3,3)$ & $-4384$ & $172$ & $-125$ & $-4336$ & $-3877$ & $165$ & $-190$ & $-3901$ \\
    \addlinespace
    $8.00$ & $(0,2,2,1)$ & $5$ & $59$ & $-57$ & $8$ & $-2$ & $54$ & $-80$ & $-28$ \\
     & $(0,2,2,3)$ & $-4161$ & $-48$ & $-35$ & $-4244$ & $-3827$ & $-42$ & $-47$ & $-3917$ \\
     & $(0,4,4,3)$ & $-1$ & $14$ & $0$ & $13$ & $0$ & $12$ & $-1$ & $12$ \\
     & $(0,4,4,5)$ & $-103$ & $-5$ & $0$ & $-108$ & $-94$ & $-4$ & $0$ & $-99$ \\
     & $(2,2,1,1)$ & $-3$ & $-2$ & $5$ & $1$ & $-4$ & $-2$ & $8$ & $3$ \\
     & $(2,2,3,3)$ & $-1430$ & $22$ & $-15$ & $-1423$ & $-1277$ & $21$ & $-24$ & $-1279$ \\
    \addlinespace
    $10.00$ & $(0,2,2,1)$ & $1$ & $11$ & $-12$ & $1$ & $1$ & $10$ & $-16$ & $-5$ \\
     & $(0,2,2,3)$ & $-1716$ & $-9$ & $-6$ & $-1731$ & $-1583$ & $-8$ & $-8$ & $-1599$ \\
     & $(0,4,4,3)$ & $-1$ & $3$ & $1$ & $2$ & $-1$ & $2$ & $1$ & $2$ \\
     & $(0,4,4,5)$ & $-27$ & $-1$ & $0$ & $-28$ & $-24$ & $-1$ & $0$ & $-25$ \\
     & $(2,2,1,1)$ & $-1$ & $0$ & $1$ & $0$ & $-1$ & $0$ & $2$ & $1$ \\
     & $(2,2,3,3)$ & $-592$ & $4$ & $-2$ & $-589$ & $-529$ & $4$ & $-3$ & $-529$ \\
    \bottomrule
  \end{tabular}
\end{table}

\begin{figure}
    \centering
    \begin{subfigure}{\linewidth}
        \centering
        \includegraphics[width=0.6\linewidth]{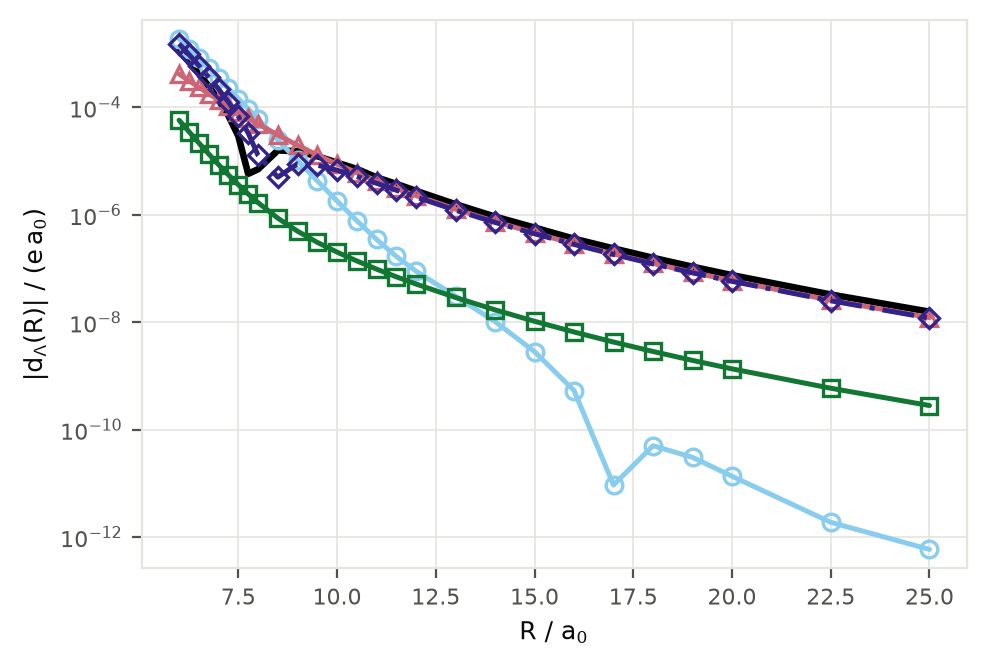}
        \caption{$\Lambda = (0,1)$}
        \label{fig:he-h2-dms-fit-0-1}
    \end{subfigure}
    \hfill
    \begin{subfigure}{\linewidth}
        \centering
        \includegraphics[width=0.6\linewidth]{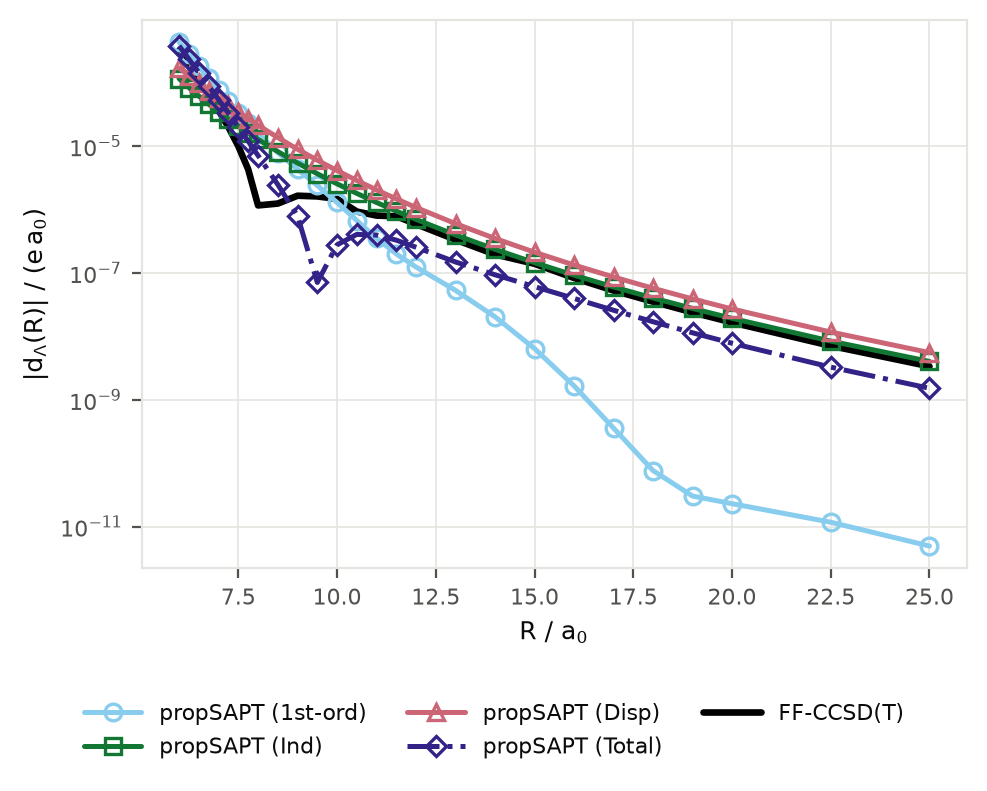}
        \caption{$\Lambda = (2,1)$}
        \label{fig:he-h2-dms-fit-2-1}
    \end{subfigure}
    \caption{The $d_{\Lambda}(R)$ coefficients of the \ce{He\bond{...}H2} dipole moment surface fit for propSAPT(HF) and  supermolecular FF-CCSD(T). 
    ``1st-ord'' denotes the sum of contributions to $d_{\Lambda}(R)$ arising from from the sum
    $\vec{\mu}^{(1)}_\text{pol} + \vec{\mu}^{(1)}_\text{exch}$, 
    ``Ind'' from 
    $\vec{\mu}^{(2)}_\text{ind,r} + \vec{\mu}^{(2)}_\text{exch-ind,r}$ 
    and ``Disp'' from 
    $\vec{\mu}^{(2)}_\text{disp} + \vec{\mu}^{(2)}_\text{exch-disp}$.}
    \label{fig:he-h2-dms-fit-coeffs}
\end{figure}

A similar contrast is found for \ce{He\bond{...}H2}. The $\Lambda=(2,3)$ coefficient (Table~\ref{tab:dms-term-coefficients-he-h2}) is dominated by the first-order contribution, whereas for the $d_{01}$ and $d_{21}$ terms, presented in Fig.~\ref{fig:he-h2-dms-fit-coeffs}, the second-order contributions become dominant. For $\Lambda=(0,1)$, the induction contribution is about an order of magnitude smaller than the dispersion-induced contribution, while for $\Lambda=(2,1)$ the induction and dispersion contributions are comparable in magnitude and largely cancel each other. In this case, the effect of intramonomer electron correlation is even more pronounced than for the \ce{H2\bond{...}H2} dimer. The higher-order coefficients, $d_{43}$ and $d_{45}$, are very small; the former has a mixed first- and second-order character, whereas the latter is again strongly dominated by the first-order contribution. 

\begin{table}
  \centering
  \small
  \caption{%
    Radial expansion coefficients $d_{\Lambda}(R)$ (in units of
    $10^{-6}$ $e\,a_0$) of the \ce{He\bond{...}H2} interaction-induced dipole
    moment surface in the Racah-normalized spherical-harmonic basis, resolved into
    the SAPT corrections, at three intermolecular separations $R$ (in $a_0$).
    $\Lambda = (L, \lambda)$ labels the basis function, with $L$ the rank on the
    \ce{H2} orientation and $\lambda$ the rank on $\vec{R}$. Each column is an
    independent fit to a given contribution.
  }
  \label{tab:dms-term-coefficients-he-h2}
  \vspace{0.8em}
  \begin{tabular}{r l @{\hspace{2em}} rrrr @{\hspace{2em}} rrrr}
    \toprule
     &  & \multicolumn{4}{c}{propSAPT(HF)} & \multicolumn{4}{c}{propSAPT(DFT)} \\
    \cmidrule(l{0pt}r{0.5em}){3-6} \cmidrule(l{0.5em}r{0.5em}){7-10}
    $R$ & $\Lambda$ & 1st-order & Ind & Disp & total & 1st-order & Ind & Disp & total \\
    \midrule
    $4.00$ & $(0,1)$ & $-43785$ & $-2803$ & $4277$ & $-42310$ & $-43836$ & $-2913$ & $5353$ & $-41397$ \\
     & $(2,1)$ & $13032$ & $2570$ & $-1809$ & $13793$ & $13707$ & $2692$ & $-2368$ & $14031$ \\
     & $(2,3)$ & $-26719$ & $-2571$ & $687$ & $-28604$ & $-28258$ & $-2830$ & $1005$ & $-30083$ \\
     & $(4,3)$ & $1745$ & $529$ & $-175$ & $2098$ & $1919$ & $546$ & $-233$ & $2233$ \\
     & $(4,5)$ & $-2082$ & $-309$ & $84$ & $-2307$ & $-2274$ & $-366$ & $123$ & $-2516$ \\
    \addlinespace
    $6.50$ & $(0,1)$ & $-820$ & $-22$ & $238$ & $-604$ & $-744$ & $-19$ & $300$ & $-463$ \\
     & $(2,1)$ & $181$ & $62$ & $-98$ & $145$ & $177$ & $60$ & $-134$ & $104$ \\
     & $(2,3)$ & $-2534$ & $-52$ & $36$ & $-2551$ & $-2498$ & $-51$ & $50$ & $-2499$ \\
     & $(4,3)$ & $13$ & $14$ & $-3$ & $24$ & $20$ & $14$ & $-4$ & $29$ \\
     & $(4,5)$ & $-96$ & $-4$ & $4$ & $-96$ & $-105$ & $-5$ & $5$ & $-105$ \\
    \addlinespace
    $10.00$ & $(0,1)$ & $-2$ & $0$ & $9$ & $7$ & $-2$ & $0$ & $12$ & $10$ \\
     & $(2,1)$ & $1$ & $3$ & $-4$ & $0$ & $2$ & $3$ & $-6$ & $-2$ \\
     & $(2,3)$ & $-415$ & $-2$ & $1$ & $-415$ & $-409$ & $-2$ & $2$ & $-409$ \\
     & $(4,3)$ & $1$ & $1$ & $0$ & $1$ & $0$ & $1$ & $0$ & $1$ \\
     & $(4,5)$ & $-7$ & $0$ & $0$ & $-7$ & $-7$ & $0$ & $0$ & $-7$ \\
    \bottomrule
  \end{tabular}
\end{table}

\section{Conclusions}
\label{sec:conclusions}

In this work, we have derived and implemented the second-order corrections to the
interaction-induced dipole moment within the propSAPT framework: the induction
and dispersion contributions together with their exchange counterparts, the
latter in the single-exchange approximation. The monomer properties enter through linear response functions, evaluated with either a Hartree--Fock or a Kohn--Sham reference. 
The molecular orbital contractions were automatically derived and implemented as a Python module that interfaces with the Psi4 package. Our theory is {\em fully relaxed}, which means that analytical derivatives reproduce the finite-field values of appropriate SAPT corrections to machine precision, as shown in Fig.~\ref{fig:correspondence_boxplot}. 

The local decay exponents confirm the expected leading powers, $R^{-4}$ for the electrostatic contribution and $R^{-7}$ for both induction and dispersion. Surprisingly, however, the dispersion-induced dipole approaches its leading $R^{-7}$ form at a significantly lower rate than the induction and first-order dipoles. Accurate long-range models of the dispersion dipole should therefore include higher inverse powers of $R$, together with the corresponding damping functions, rather than the leading term alone.

In all examples studied here, propSAPT reproduces the reference finite-field CCSD(T) data with good accuracy. In particular, propSAPT(DFT) yields results that are generally very close to the best reference values, showing the importance of intramonomer correlation for an accurate description of the interaction-induced dipole. 
The remaining discrepancies are largest for short-range cases, where higher-order polarisation is not present at the second-order theory.
The decomposition of the \ce{He\bond{...}H2} and \ce{H2\bond{...}H2} dipole-moment surfaces into spherical components reveals, however, that the role of dispersion is strongly component dependent and can become decisive for particular angular contributions. In some components, dispersion dominates the long-range dipole or strongly cancels the induction contribution, showing that an accurate treatment of dispersion is essential for a quantitative description of the full DMS. 

The spherical decomposition also has a direct spectroscopic consequence, since transitions between particular rotational states of the monomers and of the complex probe different angular components of the DMS. Some of these components have a simple classical interpretation: in \ce{H2\bond{...}H2}, $d_{0223}$, and in \ce{He\bond{...}H2}, $d_{23}$ are dominated by the first-order quadrupole-induced polarisation mechanism and exhibit the corresponding $R^{-4}$ behaviour at long range. Other components have a qualitatively different origin - the $d_{0221}$ of \ce{H2\bond{...}H2} is governed by second-order induction and dispersion contributions that strongly cancel, whereas the $d_{01}$ coefficient of \ce{He\bond{...}H2} is dominated by the dispersion-induced dipole. Consequently, the angular decomposition enables identification of classes of spectroscopic transitions whose intensities are particularly sensitive to dispersion-induced polarisation of the electronic density. Such transitions may therefore provide particularly direct spectroscopic signatures of the dispersion-induced distortion of the electronic density that drives Feynman's picture of dispersion forces. 

The current propSAPT method can still be improved in several directions. The recently introduced symmetry-adapted relaxation theory (SART) can in principle provide dimer properties to
infinite order in induction~\cite{dihingia2026SymmetryAdapted}, although without the separation into electrostatic and exchange contributions. At second order one may also dress the $\mathcal{R}_{11}$ operators with intramonomer
correlation, in the same way as in the CCD treatment of the dispersion energy of Williams et al.~\cite{williams_dispersion_1995} or in the CCPP dispersion~\cite{zuchowski_dispersion_2023}. Such a dressing would act precisely on the dispersion contribution, which is clearly very fragile for the intramonomer correlation description. 

The dipole moment is only one of the simplest quantities accessible in this way. 
What the theory actually delivers is the change of the one-electron density matrix upon the interaction, decomposed into electrostatic, exchange, induction, and dispersion contributions, obtained without subtraction and free of basis-set superposition error. 
Any one-electron observable follows from it: the interaction-induced density multipole moments of arbitrary rank, and, after the inclusion of the Pulay term~\cite{pulay1969initio}, also the Hellmann--Feynman forces on the nuclei, which can then be resolved into the same physical components as the interaction energy. 
It is worth noting that this closes the argument of Hunt, whose proof of Feynman's conjecture rests on exactly this object~\cite{hunt_dispersion_1990}: the interaction-induced distortion of the electronic charge density and the force it exerts on the nuclei of its own molecule. 
In future work, we would also like to explore measures of noncovalent interactions constructed from quantum information theory, which are likewise functionals of the interaction-induced density matrix. We expect that a direct and decomposable route to this quantity will be useful well beyond the collision-induced spectroscopy that motivated the present work.

\backmatter

\bmhead*{Acknowledgements}
This paper is dedicated to the memory of Professor Bogumi{\l} Jeziorski. Two of us (P.S.\.Z. and K.P.) had the privilege of working with him and remember him as an outstanding scientist and a generous and humble man.  

\bmhead*{Author contribution}
Conceptualisation: P.S.\.Z., K.P., B.T.; Methodology: B.T., K.P., P.S.\.Z.; Software: B.T.; Investigation: B.T., M.S.; Formal analysis: B.T., M.S., K.P., P.S.\.Z.; Supervision: P.S.\.Z.; Writing -- first draft: P.S.\.Z., B.T.; Writing -- review and editing: all authors.

\bmhead*{Funding}
P.S.\.Z. and B.T. are grateful for the support of the National Science Centre, Poland (NCN) through grant ``Sonata Bis 9'' No. 2019/34/E/ST4/00407 and ``Preludium 24'' No. 2025/57/N/ST4/04284, respectively.
This work supports the research activities of the COST Action COSY (CA21101).

\bmhead*{Data availability}
The data supporting the findings of this study, including the numerical data
used to generate the figures and tables, are available on the repository: \href{https://doi.org/10.5281/zenodo.23809047}{https://doi.org/10.5281/zenodo.23809047}.

\section*{Declarations}

\bmhead*{Ethics approval and consent to participate}
Not applicable.

\bmhead*{Consent for publication}
Not applicable.

\bmhead*{Conflict of interest}
We have no competing interests.

\bmhead*{Materials availability}
Not applicable.

\bmhead*{Code availability}
The propSAPT code is available at \href{https://doi.org/10.5281/zenodo.23951144}{https://doi.org/10.5281/zenodo.23951144}. \\
All input files required to reproduce the calculations reported in this work are available on the repository: \href{https://doi.org/10.5281/zenodo.23809047}{https://doi.org/10.5281/zenodo.23809047}.

\bibliography{references}

\begin{appendices}

\section{Molecular orbital expression for the second-order exchange-induction contribution to the induced property change}
\label{sec:appendix_ind}

\begin{equation}
\begin{split}
\Delta X^{(2)}_\text{exch-ind,resp}(S^2) &= 
- 4 S^{a}_{s} S^{b}_{a} (u_\text{ind,cp})^{a'}_{r} v^{rs}_{a'b} 
- 4 S^{a}_{b} S^{b}_{r} (u_\text{ind,cp})^{b'}_{s} v^{rs}_{ab'} 
- 2 S^{r}_{s} S^{b}_{r'} (u_\text{ind,cp})^{a}_{r} v^{r's}_{ab} 
- 2 S^{r}_{s} S^{b}_{a} (u_\text{ind,cp})^{a}_{r} (\omega_A)^{s}_{b} 
\\ &\hspace{-5em}
- 2 S^{r}_{b} S^{b}_{r'} (u_\text{ind,cp})^{a}_{r} (\omega_B)^{r'}_{a} 
- 2 S^{s}_{r} S^{a}_{s'} (u_\text{ind,cp})^{b}_{s} v^{rs'}_{ab} 
- 2 S^{s}_{r} S^{a}_{b} (u_\text{ind,cp})^{b}_{s} (\omega_B)^{r}_{a} 
- 2 S^{s}_{a} S^{a}_{s'} (u_\text{ind,cp})^{b}_{s} (\omega_A)^{s'}_{b} 
\\ &\hspace{-5em}
+ 2 S^{a}_{s} S^{b}_{a'} (u_\text{ind,cp})^{a'}_{r} v^{rs}_{ab} 
+ 2 S^{a}_{b} S^{b}_{a'} (u_\text{ind,cp})^{a'}_{r} (\omega_B)^{r}_{a} 
+ 2 S^{a}_{b} S^{b'}_{r} (u_\text{ind,cp})^{b}_{s} v^{rs}_{ab'} 
+ 2 S^{a}_{b} S^{b'}_{a} (u_\text{ind,cp})^{b}_{s} (\omega_A)^{s}_{b'} 
\\ &\hspace{-5em}
- 4 (C^{[1]}_A)^{r}_{a} S^{r'}_{s} S^{b}_{a'} (C^{(1)}_A)^{a'}_{r'} v^{as}_{rb} 
- 4 (C^{[1]}_A)^{r}_{a} S^{s}_{a'} S^{a'}_{s'} (C^{(1)}_B)^{b}_{s} v^{as'}_{rb} 
- 4 (C^{[1]}_A)^{r}_{a} S^{a}_{s} S^{b}_{r} (C^{(1)}_A)^{a'}_{r'} v^{r's}_{a'b} 
\\ &\hspace{-5em}
- 4 (C^{[1]}_A)^{r}_{a} S^{a}_{b} S^{b}_{r'} (C^{(1)}_B)^{b'}_{s} v^{r's}_{rb'} 
- 4 (C^{[1]}_A)^{r}_{a} S^{a'}_{s} S^{b}_{a'} (C^{(1)}_A)^{a}_{r'} v^{r's}_{rb} 
- 4 (C^{[1]}_B)^{s}_{b} S^{r}_{b'} S^{b'}_{r'} (C^{(1)}_A)^{a}_{r} v^{r'b}_{as} 
\\ &\hspace{-5em}
- 4 (C^{[1]}_B)^{s}_{b} S^{s'}_{r} S^{a}_{b'} (C^{(1)}_B)^{b'}_{s'} v^{rb}_{as} 
- 4 (C^{[1]}_B)^{s}_{b} S^{a}_{s} S^{b}_{r} (C^{(1)}_B)^{b'}_{s'} v^{rs'}_{ab'} 
- 4 (C^{[1]}_B)^{s}_{b} S^{a}_{s'} S^{b}_{a} (C^{(1)}_A)^{a'}_{r} v^{rs'}_{a's} 
\\ &\hspace{-5em}
- 4 (C^{[1]}_B)^{s}_{b} S^{a}_{b'} S^{b'}_{r} (C^{(1)}_B)^{b}_{s'} v^{rs'}_{as} 
- 4 (C^{[1]}_A)^{a}_{r} S^{r}_{s} S^{b}_{a} (C^{(1)}_A)^{a'}_{r'} v^{r's}_{a'b} 
- 4 (C^{[1]}_A)^{a}_{r} S^{r}_{b} S^{b}_{r'} (C^{(1)}_B)^{b'}_{s} v^{r's}_{ab'} 
\\ &\hspace{-5em}
- 4 (C^{[1]}_A)^{a}_{r} S^{r'}_{s} S^{b}_{a'} (C^{(1)}_A)^{a'}_{r'} v^{rs}_{ab} 
- 4 (C^{[1]}_A)^{a}_{r} S^{s}_{a'} S^{a'}_{s'} (C^{(1)}_B)^{b}_{s} v^{rs'}_{ab} 
- 4 (C^{[1]}_B)^{b}_{s} S^{r}_{b'} S^{b'}_{r'} (C^{(1)}_A)^{a}_{r} v^{r's}_{ab} 
\\ &\hspace{-5em}
- 4 (C^{[1]}_B)^{b}_{s} S^{s}_{r} S^{a}_{b} (C^{(1)}_B)^{b'}_{s'} v^{rs'}_{ab'} 
- 4 (C^{[1]}_B)^{b}_{s} S^{s}_{a} S^{a}_{s'} (C^{(1)}_A)^{a'}_{r} v^{rs'}_{a'b} 
- 4 (C^{[1]}_B)^{b}_{s} S^{s'}_{r} S^{a}_{b'} (C^{(1)}_B)^{b'}_{s'} v^{rs}_{ab} 
\\ &\hspace{-5em}
- 2 (C^{[1]}_A)^{r}_{a} S^{r'}_{s} S^{b}_{r} (C^{(1)}_A)^{a}_{r'} (\omega_A)^{s}_{b} 
- 2 (C^{[1]}_A)^{r}_{a} S^{r'}_{s} S^{b}_{r''} (C^{(1)}_A)^{a}_{r'} v^{r''s}_{rb} 
- 2 (C^{[1]}_A)^{r}_{a} S^{r'}_{b} S^{b}_{r''} (C^{(1)}_A)^{a}_{r'} (\omega_B)^{r''}_{r} 
\\ &\hspace{-5em}
- 2 (C^{[1]}_A)^{r}_{a} S^{s}_{r} S^{a}_{s'} (C^{(1)}_B)^{b}_{s} (\omega_A)^{s'}_{b} 
- 2 (C^{[1]}_A)^{r}_{a} S^{s}_{r'} S^{a}_{s'} (C^{(1)}_B)^{b}_{s} v^{r's'}_{rb} 
- 2 (C^{[1]}_A)^{r}_{a} S^{s}_{r'} S^{a}_{b} (C^{(1)}_B)^{b}_{s} (\omega_B)^{r'}_{r} 
\\ &\hspace{-5em}
- 2 (C^{[1]}_A)^{r}_{a} S^{a'}_{s} S^{b}_{a''} (C^{(1)}_A)^{a''}_{r} v^{as}_{a'b} 
- 2 (C^{[1]}_A)^{r}_{a} S^{a'}_{b} S^{b}_{a''} (C^{(1)}_A)^{a''}_{r} (\omega_B)^{a}_{a'} 
- 2 (C^{[1]}_A)^{r}_{a} S^{a'}_{b} S^{b'}_{r} (C^{(1)}_B)^{b}_{s} v^{as}_{a'b'} 
\\ &\hspace{-5em}
- 2 (C^{[1]}_B)^{s}_{b} S^{r}_{s} S^{b}_{r'} (C^{(1)}_A)^{a}_{r} (\omega_B)^{r'}_{a} 
- 2 (C^{[1]}_B)^{s}_{b} S^{r}_{s'} S^{b}_{r'} (C^{(1)}_A)^{a}_{r} v^{r's'}_{as} 
- 2 (C^{[1]}_B)^{s}_{b} S^{r}_{s'} S^{b}_{a} (C^{(1)}_A)^{a}_{r} (\omega_A)^{s'}_{s} 
\\ &\hspace{-5em}
- 2 (C^{[1]}_B)^{s}_{b} S^{s'}_{r} S^{a}_{s} (C^{(1)}_B)^{b}_{s'} (\omega_B)^{r}_{a} 
- 2 (C^{[1]}_B)^{s}_{b} S^{s'}_{r} S^{a}_{s''} (C^{(1)}_B)^{b}_{s'} v^{rs''}_{as} 
- 2 (C^{[1]}_B)^{s}_{b} S^{s'}_{a} S^{a}_{s''} (C^{(1)}_B)^{b}_{s'} (\omega_A)^{s''}_{s} 
\\ &\hspace{-5em}
- 2 (C^{[1]}_B)^{s}_{b} S^{a}_{s} S^{b'}_{a'} (C^{(1)}_A)^{a'}_{r} v^{rb}_{ab'} 
- 2 (C^{[1]}_B)^{s}_{b} S^{a}_{b'} S^{b''}_{r} (C^{(1)}_B)^{b'}_{s} v^{rb}_{ab''} 
- 2 (C^{[1]}_B)^{s}_{b} S^{a}_{b'} S^{b''}_{a} (C^{(1)}_B)^{b'}_{s} (\omega_A)^{b}_{b''} 
\\ &\hspace{-5em}
- 2 (C^{[1]}_A)^{a}_{r} S^{r}_{s} S^{s'}_{r'} (C^{(1)}_B)^{b}_{s'} v^{r's}_{ab} 
- 2 (C^{[1]}_A)^{a}_{r} S^{r}_{s} S^{s'}_{a} (C^{(1)}_B)^{b}_{s'} (\omega_A)^{s}_{b} 
- 2 (C^{[1]}_A)^{a}_{r} S^{r}_{b} S^{s}_{r'} (C^{(1)}_B)^{b}_{s} (\omega_B)^{r'}_{a} 
\\ &\hspace{-5em}
- 2 (C^{[1]}_A)^{a}_{r} S^{a'}_{b} S^{b'}_{a} (C^{(1)}_B)^{b}_{s} v^{rs}_{a'b'} 
- 2 (C^{[1]}_B)^{b}_{s} S^{r}_{s'} S^{s}_{r'} (C^{(1)}_A)^{a}_{r} v^{r's'}_{ab} 
- 2 (C^{[1]}_B)^{b}_{s} S^{r}_{s'} S^{s}_{a} (C^{(1)}_A)^{a}_{r} (\omega_A)^{s'}_{b} 
\\ &\hspace{-5em}
- 2 (C^{[1]}_B)^{b}_{s} S^{r}_{b} S^{s}_{r'} (C^{(1)}_A)^{a}_{r} (\omega_B)^{r'}_{a} 
- 2 (C^{[1]}_B)^{b}_{s} S^{a}_{b} S^{b'}_{a'} (C^{(1)}_A)^{a'}_{r} v^{rs}_{ab'} 
+ 2 (C^{[1]}_A)^{r}_{a} S^{r'}_{s} S^{b}_{r} (C^{(1)}_A)^{a'}_{r'} v^{as}_{a'b} 
\\ &\hspace{-5em}
+ 2 (C^{[1]}_A)^{r}_{a} S^{r'}_{b} S^{b}_{r} (C^{(1)}_A)^{a'}_{r'} (\omega_B)^{a}_{a'} 
+ 2 (C^{[1]}_A)^{r}_{a} S^{s}_{r} S^{a'}_{s'} (C^{(1)}_B)^{b}_{s} v^{as'}_{a'b} 
+ 2 (C^{[1]}_A)^{r}_{a} S^{s}_{r} S^{a'}_{b} (C^{(1)}_B)^{b}_{s} (\omega_B)^{a}_{a'} 
\\ &\hspace{-5em}
+ 2 (C^{[1]}_A)^{r}_{a} S^{a}_{s} S^{b}_{r'} (C^{(1)}_A)^{a'}_{r} v^{r's}_{a'b} 
+ 2 (C^{[1]}_A)^{r}_{a} S^{a}_{s} S^{b}_{a'} (C^{(1)}_A)^{a'}_{r} (\omega_A)^{s}_{b} 
+ 2 (C^{[1]}_A)^{r}_{a} S^{a}_{s} S^{b}_{a'} (C^{(1)}_A)^{a'}_{r'} v^{r's}_{rb} 
\\ &\hspace{-5em}
+ 2 (C^{[1]}_A)^{r}_{a} S^{a}_{b} S^{b}_{r'} (C^{(1)}_A)^{a'}_{r} (\omega_B)^{r'}_{a'} 
+ 2 (C^{[1]}_A)^{r}_{a} S^{a}_{b} S^{b}_{a'} (C^{(1)}_A)^{a'}_{r'} (\omega_B)^{r'}_{r} 
+ 2 (C^{[1]}_A)^{r}_{a} S^{a}_{b} S^{b'}_{r} (C^{(1)}_B)^{b}_{s} (\omega_A)^{s}_{b'} 
\\ &\hspace{-5em}
+ 2 (C^{[1]}_A)^{r}_{a} S^{a}_{b} S^{b'}_{r'} (C^{(1)}_B)^{b}_{s} v^{r's}_{rb'} 
+ 2 (C^{[1]}_A)^{r}_{a} S^{a'}_{s} S^{b}_{r} (C^{(1)}_A)^{a}_{r'} v^{r's}_{a'b} 
+ 2 (C^{[1]}_A)^{r}_{a} S^{a'}_{b} S^{b}_{r} (C^{(1)}_A)^{a}_{r'} (\omega_B)^{r'}_{a'} 
\\ &\hspace{-5em}
+ 2 (C^{[1]}_B)^{s}_{b} S^{r}_{s} S^{b'}_{r'} (C^{(1)}_A)^{a}_{r} v^{r'b}_{ab'} 
+ 2 (C^{[1]}_B)^{s}_{b} S^{r}_{s} S^{b'}_{a} (C^{(1)}_A)^{a}_{r} (\omega_A)^{b}_{b'} 
+ 2 (C^{[1]}_B)^{s}_{b} S^{s'}_{r} S^{a}_{s} (C^{(1)}_B)^{b'}_{s'} v^{rb}_{ab'} 
\\ &\hspace{-5em}
+ 2 (C^{[1]}_B)^{s}_{b} S^{s'}_{a} S^{a}_{s} (C^{(1)}_B)^{b'}_{s'} (\omega_A)^{b}_{b'} 
+ 2 (C^{[1]}_B)^{s}_{b} S^{a}_{s} S^{b}_{a'} (C^{(1)}_A)^{a'}_{r} (\omega_B)^{r}_{a} 
+ 2 (C^{[1]}_B)^{s}_{b} S^{a}_{s} S^{b'}_{r} (C^{(1)}_B)^{b}_{s'} v^{rs'}_{ab'} 
\\ &\hspace{-5em}
+ 2 (C^{[1]}_B)^{s}_{b} S^{a}_{s} S^{b'}_{a} (C^{(1)}_B)^{b}_{s'} (\omega_A)^{s'}_{b'} 
+ 2 (C^{[1]}_B)^{s}_{b} S^{a}_{s'} S^{b}_{r} (C^{(1)}_B)^{b'}_{s} v^{rs'}_{ab'} 
+ 2 (C^{[1]}_B)^{s}_{b} S^{a}_{s'} S^{b}_{a} (C^{(1)}_B)^{b'}_{s} (\omega_A)^{s'}_{b'} 
\\ &\hspace{-5em}
+ 2 (C^{[1]}_B)^{s}_{b} S^{a}_{s'} S^{b}_{a'} (C^{(1)}_A)^{a'}_{r} v^{rs'}_{as} 
+ 2 (C^{[1]}_B)^{s}_{b} S^{a}_{b'} S^{b}_{r} (C^{(1)}_B)^{b'}_{s} (\omega_B)^{r}_{a} 
+ 2 (C^{[1]}_B)^{s}_{b} S^{a}_{b'} S^{b}_{r} (C^{(1)}_B)^{b'}_{s'} v^{rs'}_{as} 
\\ &\hspace{-5em}
+ 2 (C^{[1]}_B)^{s}_{b} S^{a}_{b'} S^{b}_{a} (C^{(1)}_B)^{b'}_{s'} (\omega_A)^{s'}_{s} 
+ 2 (C^{[1]}_A)^{a}_{r} S^{r}_{s} S^{b}_{a'} (C^{(1)}_A)^{a'}_{r'} v^{r's}_{ab} 
+ 2 (C^{[1]}_A)^{a}_{r} S^{r}_{b} S^{b}_{a'} (C^{(1)}_A)^{a'}_{r'} (\omega_B)^{r'}_{a} 
\\ &\hspace{-5em}
+ 2 (C^{[1]}_A)^{a}_{r} S^{r}_{b} S^{b'}_{r'} (C^{(1)}_B)^{b}_{s} v^{r's}_{ab'} 
+ 2 (C^{[1]}_A)^{a}_{r} S^{r}_{b} S^{b'}_{a} (C^{(1)}_B)^{b}_{s} (\omega_A)^{s}_{b'} 
+ 2 (C^{[1]}_A)^{a}_{r} S^{r'}_{s} S^{b}_{a} (C^{(1)}_A)^{a'}_{r'} v^{rs}_{a'b} 
\\ &\hspace{-5em}
+ 2 (C^{[1]}_A)^{a}_{r} S^{r'}_{b} S^{b}_{a} (C^{(1)}_A)^{a'}_{r'} (\omega_B)^{r}_{a'} 
+ 2 (C^{[1]}_A)^{a}_{r} S^{s}_{a} S^{a'}_{s'} (C^{(1)}_B)^{b}_{s} v^{rs'}_{a'b} 
+ 2 (C^{[1]}_A)^{a}_{r} S^{s}_{a} S^{a'}_{b} (C^{(1)}_B)^{b}_{s} (\omega_B)^{r}_{a'} 
\\ &\hspace{-5em}
+ 2 (C^{[1]}_B)^{b}_{s} S^{r}_{b} S^{b'}_{r'} (C^{(1)}_A)^{a}_{r} v^{r's}_{ab'} 
+ 2 (C^{[1]}_B)^{b}_{s} S^{r}_{b} S^{b'}_{a} (C^{(1)}_A)^{a}_{r} (\omega_A)^{s}_{b'} 
+ 2 (C^{[1]}_B)^{b}_{s} S^{s}_{r} S^{a}_{b'} (C^{(1)}_B)^{b'}_{s'} v^{rs'}_{ab} 
\\ &\hspace{-5em}
+ 2 (C^{[1]}_B)^{b}_{s} S^{s}_{a} S^{a}_{b'} (C^{(1)}_B)^{b'}_{s'} (\omega_A)^{s'}_{b} 
+ 2 (C^{[1]}_B)^{b}_{s} S^{s}_{a} S^{a'}_{s'} (C^{(1)}_A)^{a}_{r} v^{rs'}_{a'b} 
+ 2 (C^{[1]}_B)^{b}_{s} S^{s}_{a} S^{a'}_{b} (C^{(1)}_A)^{a}_{r} (\omega_B)^{r}_{a'} 
\\ &\hspace{-5em}
+ 2 (C^{[1]}_B)^{b}_{s} S^{s'}_{r} S^{a}_{b} (C^{(1)}_B)^{b'}_{s'} v^{rs}_{ab'} 
+ 2 (C^{[1]}_B)^{b}_{s} S^{s'}_{a} S^{a}_{b} (C^{(1)}_B)^{b'}_{s'} (\omega_A)^{s}_{b'} 
+ 4 (C^{[1]}_A)^{r}_{a} S^{a'}_{s} S^{b}_{a'} (C^{(1)}_A)^{a''}_{r} v^{as}_{a''b} 
\\ &\hspace{-5em}
+ 4 (C^{[1]}_A)^{r}_{a} S^{a'}_{b} S^{b}_{r} (C^{(1)}_B)^{b'}_{s} v^{as}_{a'b'} 
+ 4 (C^{[1]}_A)^{r}_{a} S^{a'}_{b} S^{b'}_{a'} (C^{(1)}_B)^{b}_{s} v^{as}_{rb'} 
+ 4 (C^{[1]}_B)^{s}_{b} S^{a}_{s} S^{b'}_{a} (C^{(1)}_A)^{a'}_{r} v^{rb}_{a'b'} 
\\ &\hspace{-5em}
+ 4 (C^{[1]}_B)^{s}_{b} S^{a}_{b'} S^{b'}_{r} (C^{(1)}_B)^{b''}_{s} v^{rb}_{ab''} 
+ 4 (C^{[1]}_B)^{s}_{b} S^{a}_{b'} S^{b'}_{a'} (C^{(1)}_A)^{a'}_{r} v^{rb}_{as} 
+ 4 (C^{[1]}_A)^{a}_{r} S^{a'}_{b} S^{b}_{a} (C^{(1)}_B)^{b'}_{s} v^{rs}_{a'b'} 
\\ &\hspace{-5em}
+ 4 (C^{[1]}_A)^{a}_{r} S^{a'}_{b} S^{b'}_{a'} (C^{(1)}_B)^{b}_{s} v^{rs}_{ab'} 
+ 4 (C^{[1]}_B)^{b}_{s} S^{a}_{b} S^{b'}_{a} (C^{(1)}_A)^{a'}_{r} v^{rs}_{a'b'} 
+ 4 (C^{[1]}_B)^{b}_{s} S^{a}_{b'} S^{b'}_{a'} (C^{(1)}_A)^{a'}_{r} v^{rs}_{ab} 
\end{split}
\end{equation}

\section{Molecular orbital expression for the second-order exchange-dispersion contribution to the induced property change}
\label{sec:appendix_disp}

\begin{equation}
\begin{split}
\Delta X^{(2)}_\text{exch-disp}(S^2) &= 
- 4 S^{r}_{b} S^{b}_{r'} (u_\text{cp})^{ab'}_{rs} v^{r's}_{ab'} 
- 4 S^{r}_{b} S^{b}_{a} (u_\text{cp})^{ab'}_{rs} (\omega_A)^{s}_{b'} 
\\ &\hspace{-5em}
- 4 S^{s}_{a} S^{a}_{s'} (u_\text{cp})^{a'b}_{rs} v^{rs'}_{a'b} 
- 4 S^{s}_{a} S^{a}_{b} (u_\text{cp})^{a'b}_{rs} (\omega_B)^{r}_{a'} 
- 2 S^{r}_{s} S^{s'}_{r'} (u_\text{cp})^{ab}_{rs'} v^{r's}_{ab} 
- 2 S^{r}_{s} S^{s'}_{a} (u_\text{cp})^{ab}_{rs'} (\omega_A)^{s}_{b} 
\\ &\hspace{-5em}
- 2 S^{r}_{b} S^{s}_{r'} (u_\text{cp})^{ab}_{rs} (\omega_B)^{r'}_{a} 
- 2 S^{a}_{b} S^{b'}_{a'} (u_\text{cp})^{a'b}_{rs} v^{rs}_{ab'} 
+ 2 S^{r}_{b} S^{b'}_{r'} (u_\text{cp})^{ab}_{rs} v^{r's}_{ab'} 
+ 2 S^{r}_{b} S^{b'}_{a} (u_\text{cp})^{ab}_{rs} (\omega_A)^{s}_{b'} 
\\ &\hspace{-5em}
+ 2 S^{s}_{a} S^{a'}_{s'} (u_\text{cp})^{ab}_{rs} v^{rs'}_{a'b} 
+ 2 S^{s}_{a} S^{a'}_{b} (u_\text{cp})^{ab}_{rs} (\omega_B)^{r}_{a'} 
+ 4 S^{a}_{b} S^{b}_{a'} (u_\text{cp})^{a'b'}_{rs} v^{rs}_{ab'} 
+ 4 S^{a}_{b} S^{b'}_{a} (u_\text{cp})^{a'b}_{rs} v^{rs}_{a'b'} 
\\ &\hspace{-5em}
- 8 (C^{[1]}_A)^{r}_{a} S^{r'}_{b} S^{b}_{a'} t^{a'b'}_{r's} v^{as}_{rb'} 
- 8 (C^{[1]}_B)^{s}_{b} S^{s'}_{a} S^{a}_{b'} t^{a'b'}_{rs'} v^{rb}_{a's} 
- 8 (C^{[1]}_A)^{a}_{r} S^{r'}_{b} S^{b}_{a'} t^{a'b'}_{r's} v^{rs}_{ab'} 
- 8 (C^{[1]}_B)^{b}_{s} S^{s'}_{a} S^{a}_{b'} t^{a'b'}_{rs'} v^{rs}_{a'b} 
\\ &\hspace{-5em}
- 4 (C^{[1]}_A)^{r}_{a} S^{r'}_{s} S^{s'}_{a'} t^{a'b}_{r's'} v^{as}_{rb} 
- 4 (C^{[1]}_A)^{r}_{a} S^{r'}_{b} S^{b}_{r} t^{ab'}_{r's} (\omega_A)^{s}_{b'} 
- 4 (C^{[1]}_A)^{r}_{a} S^{r'}_{b} S^{b}_{r''} t^{ab'}_{r's} v^{r''s}_{rb'} 
- 4 (C^{[1]}_A)^{r}_{a} S^{s}_{r} S^{a}_{s'} t^{a'b}_{r's} v^{r's'}_{a'b} 
\\ &\hspace{-5em}
- 4 (C^{[1]}_A)^{r}_{a} S^{s}_{r} S^{a}_{b} t^{a'b}_{r's} (\omega_B)^{r'}_{a'} 
- 4 (C^{[1]}_A)^{r}_{a} S^{s}_{a'} S^{a'}_{s'} t^{ab}_{r's} v^{r's'}_{rb} 
- 4 (C^{[1]}_A)^{r}_{a} S^{s}_{a'} S^{a'}_{b} t^{ab}_{r's} (\omega_B)^{r'}_{r} 
- 4 (C^{[1]}_A)^{r}_{a} S^{a'}_{b} S^{b}_{a''} t^{a''b'}_{rs} v^{as}_{a'b'} 
\\ &\hspace{-5em}
- 4 (C^{[1]}_A)^{r}_{a} S^{a'}_{b} S^{b'}_{a'} t^{a''b}_{rs} v^{as}_{a''b'} 
- 4 (C^{[1]}_B)^{s}_{b} S^{r}_{s} S^{b}_{r'} t^{ab'}_{rs'} v^{r's'}_{ab'} 
- 4 (C^{[1]}_B)^{s}_{b} S^{r}_{s} S^{b}_{a} t^{ab'}_{rs'} (\omega_A)^{s'}_{b'} 
- 4 (C^{[1]}_B)^{s}_{b} S^{r}_{b'} S^{s'}_{r'} t^{ab'}_{rs'} v^{r'b}_{as} 
\\ &\hspace{-5em}
- 4 (C^{[1]}_B)^{s}_{b} S^{r}_{b'} S^{b'}_{r'} t^{ab}_{rs'} v^{r's'}_{as} 
- 4 (C^{[1]}_B)^{s}_{b} S^{r}_{b'} S^{b'}_{a} t^{ab}_{rs'} (\omega_A)^{s'}_{s} 
- 4 (C^{[1]}_B)^{s}_{b} S^{s'}_{a} S^{a}_{s} t^{a'b}_{rs'} (\omega_B)^{r}_{a'} 
- 4 (C^{[1]}_B)^{s}_{b} S^{s'}_{a} S^{a}_{s''} t^{a'b}_{rs'} v^{rs''}_{a's} 
\\ &\hspace{-5em}
- 4 (C^{[1]}_B)^{s}_{b} S^{a}_{b'} S^{b'}_{a'} t^{a'b''}_{rs} v^{rb}_{ab''} 
- 4 (C^{[1]}_B)^{s}_{b} S^{a}_{b'} S^{b''}_{a} t^{a'b'}_{rs} v^{rb}_{a'b''} 
- 4 (C^{[1]}_A)^{a}_{r} S^{r}_{s} S^{s'}_{a} t^{a'b}_{r's'} v^{r's}_{a'b} 
- 4 (C^{[1]}_A)^{a}_{r} S^{r}_{b} S^{s}_{a} t^{a'b}_{r's} (\omega_B)^{r'}_{a'} 
\\ &\hspace{-5em}
- 4 (C^{[1]}_A)^{a}_{r} S^{r'}_{s} S^{s'}_{a'} t^{a'b}_{r's'} v^{rs}_{ab} 
- 4 (C^{[1]}_B)^{b}_{s} S^{r}_{b} S^{s}_{r'} t^{ab'}_{rs'} v^{r's'}_{ab'} 
- 4 (C^{[1]}_B)^{b}_{s} S^{r}_{b} S^{s}_{a} t^{ab'}_{rs'} (\omega_A)^{s'}_{b'} 
- 4 (C^{[1]}_B)^{b}_{s} S^{r}_{b'} S^{s'}_{r'} t^{ab'}_{rs'} v^{r's}_{ab} 
\\ &\hspace{-5em}
- 2 (C^{[1]}_A)^{r}_{a} S^{r'}_{s} S^{s'}_{r} t^{ab}_{r's'} (\omega_A)^{s}_{b} 
- 2 (C^{[1]}_A)^{r}_{a} S^{r'}_{s} S^{s'}_{r''} t^{ab}_{r's'} v^{r''s}_{rb} 
- 2 (C^{[1]}_A)^{r}_{a} S^{r'}_{b} S^{s}_{r''} t^{ab}_{r's} (\omega_B)^{r''}_{r} 
- 2 (C^{[1]}_A)^{r}_{a} S^{r'}_{b} S^{b'}_{r} t^{a'b}_{r's} v^{as}_{a'b'} 
\\ &\hspace{-5em}
- 2 (C^{[1]}_A)^{r}_{a} S^{s}_{a'} S^{a''}_{s'} t^{a'b}_{rs} v^{as'}_{a''b} 
- 2 (C^{[1]}_A)^{r}_{a} S^{s}_{a'} S^{a''}_{b} t^{a'b}_{rs} (\omega_B)^{a}_{a''} 
- 2 (C^{[1]}_A)^{r}_{a} S^{a}_{b} S^{b'}_{r'} t^{a'b}_{rs} v^{r's}_{a'b'} 
- 2 (C^{[1]}_A)^{r}_{a} S^{a}_{b} S^{b'}_{a'} t^{a'b}_{rs} (\omega_A)^{s}_{b'} 
\\ &\hspace{-5em}
- 2 (C^{[1]}_A)^{r}_{a} S^{a}_{b} S^{b'}_{a'} t^{a'b}_{r's} v^{r's}_{rb'} 
- 2 (C^{[1]}_A)^{r}_{a} S^{a'}_{b} S^{b'}_{r} t^{ab}_{r's} v^{r's}_{a'b'} 
- 2 (C^{[1]}_B)^{s}_{b} S^{r}_{s} S^{s'}_{r'} t^{ab}_{rs'} (\omega_B)^{r'}_{a} 
- 2 (C^{[1]}_B)^{s}_{b} S^{r}_{s'} S^{s''}_{r'} t^{ab}_{rs''} v^{r's'}_{as} 
\\ &\hspace{-5em}
- 2 (C^{[1]}_B)^{s}_{b} S^{r}_{s'} S^{s''}_{a} t^{ab}_{rs''} (\omega_A)^{s'}_{s} 
- 2 (C^{[1]}_B)^{s}_{b} S^{r}_{b'} S^{b''}_{r'} t^{ab'}_{rs} v^{r'b}_{ab''} 
- 2 (C^{[1]}_B)^{s}_{b} S^{r}_{b'} S^{b''}_{a} t^{ab'}_{rs} (\omega_A)^{b}_{b''} 
- 2 (C^{[1]}_B)^{s}_{b} S^{s'}_{a} S^{a'}_{s} t^{ab'}_{rs'} v^{rb}_{a'b'} 
\\ &\hspace{-5em}
- 2 (C^{[1]}_B)^{s}_{b} S^{a}_{s} S^{b'}_{a'} t^{a'b}_{rs'} v^{rs'}_{ab'} 
- 2 (C^{[1]}_B)^{s}_{b} S^{a}_{s'} S^{b}_{a'} t^{a'b'}_{rs} v^{rs'}_{ab'} 
- 2 (C^{[1]}_B)^{s}_{b} S^{a}_{b'} S^{b}_{a'} t^{a'b'}_{rs} (\omega_B)^{r}_{a} 
- 2 (C^{[1]}_B)^{s}_{b} S^{a}_{b'} S^{b}_{a'} t^{a'b'}_{rs'} v^{rs'}_{as} 
\\ &\hspace{-5em}
- 2 (C^{[1]}_A)^{a}_{r} S^{r}_{b} S^{b'}_{a'} t^{a'b}_{r's} v^{r's}_{ab'} 
- 2 (C^{[1]}_A)^{a}_{r} S^{r'}_{b} S^{b'}_{a} t^{a'b}_{r's} v^{rs}_{a'b'} 
- 2 (C^{[1]}_B)^{b}_{s} S^{s}_{a} S^{a'}_{b'} t^{ab'}_{rs'} v^{rs'}_{a'b} 
- 2 (C^{[1]}_B)^{b}_{s} S^{s'}_{a} S^{a'}_{b} t^{ab'}_{rs'} v^{rs}_{a'b'} 
\\ &\hspace{-5em}
+ 2 (C^{[1]}_A)^{r}_{a} S^{r'}_{s} S^{s'}_{r} t^{a'b}_{r's'} v^{as}_{a'b} 
+ 2 (C^{[1]}_A)^{r}_{a} S^{r'}_{b} S^{s}_{r} t^{a'b}_{r's} (\omega_B)^{a}_{a'} 
+ 2 (C^{[1]}_A)^{r}_{a} S^{r'}_{b} S^{b'}_{r} t^{ab}_{r's} (\omega_A)^{s}_{b'} 
+ 2 (C^{[1]}_A)^{r}_{a} S^{r'}_{b} S^{b'}_{r''} t^{ab}_{r's} v^{r''s}_{rb'} 
\\ &\hspace{-5em}
+ 2 (C^{[1]}_A)^{r}_{a} S^{s}_{r} S^{a'}_{s'} t^{ab}_{r's} v^{r's'}_{a'b} 
+ 2 (C^{[1]}_A)^{r}_{a} S^{s}_{r} S^{a'}_{b} t^{ab}_{r's} (\omega_B)^{r'}_{a'} 
+ 2 (C^{[1]}_A)^{r}_{a} S^{s}_{r'} S^{a}_{s'} t^{a'b}_{rs} v^{r's'}_{a'b} 
+ 2 (C^{[1]}_A)^{r}_{a} S^{s}_{r'} S^{a}_{b} t^{a'b}_{rs} (\omega_B)^{r'}_{a'} 
\\ &\hspace{-5em}
+ 2 (C^{[1]}_A)^{r}_{a} S^{s}_{a'} S^{a}_{s'} t^{a'b}_{rs} (\omega_A)^{s'}_{b} 
+ 2 (C^{[1]}_A)^{r}_{a} S^{s}_{a'} S^{a}_{s'} t^{a'b}_{r's} v^{r's'}_{rb} 
+ 2 (C^{[1]}_A)^{r}_{a} S^{s}_{a'} S^{a}_{b} t^{a'b}_{r's} (\omega_B)^{r'}_{r} 
+ 2 (C^{[1]}_A)^{r}_{a} S^{a'}_{b} S^{b'}_{a''} t^{a''b}_{rs} v^{as}_{a'b'} 
\\ &\hspace{-5em}
+ 2 (C^{[1]}_B)^{s}_{b} S^{r}_{s} S^{s'}_{r'} t^{ab'}_{rs'} v^{r'b}_{ab'} 
+ 2 (C^{[1]}_B)^{s}_{b} S^{r}_{s} S^{s'}_{a} t^{ab'}_{rs'} (\omega_A)^{b}_{b'} 
+ 2 (C^{[1]}_B)^{s}_{b} S^{r}_{s} S^{b'}_{r'} t^{ab}_{rs'} v^{r's'}_{ab'} 
+ 2 (C^{[1]}_B)^{s}_{b} S^{r}_{s} S^{b'}_{a} t^{ab}_{rs'} (\omega_A)^{s'}_{b'} 
\\ &\hspace{-5em}
+ 2 (C^{[1]}_B)^{s}_{b} S^{r}_{s'} S^{b}_{r'} t^{ab'}_{rs} v^{r's'}_{ab'} 
+ 2 (C^{[1]}_B)^{s}_{b} S^{r}_{s'} S^{b}_{a} t^{ab'}_{rs} (\omega_A)^{s'}_{b'} 
+ 2 (C^{[1]}_B)^{s}_{b} S^{r}_{b'} S^{b}_{r'} t^{ab'}_{rs} (\omega_B)^{r'}_{a} 
+ 2 (C^{[1]}_B)^{s}_{b} S^{r}_{b'} S^{b}_{r'} t^{ab'}_{rs'} v^{r's'}_{as} 
\\ &\hspace{-5em}
+ 2 (C^{[1]}_B)^{s}_{b} S^{r}_{b'} S^{b}_{a} t^{ab'}_{rs'} (\omega_A)^{s'}_{s} 
+ 2 (C^{[1]}_B)^{s}_{b} S^{s'}_{a} S^{a'}_{s} t^{ab}_{rs'} (\omega_B)^{r}_{a'} 
+ 2 (C^{[1]}_B)^{s}_{b} S^{s'}_{a} S^{a'}_{s''} t^{ab}_{rs'} v^{rs''}_{a's} 
+ 2 (C^{[1]}_B)^{s}_{b} S^{a}_{b'} S^{b''}_{a'} t^{a'b'}_{rs} v^{rb}_{ab''} 
\\ &\hspace{-5em}
+ 2 (C^{[1]}_A)^{a}_{r} S^{r}_{s} S^{s'}_{a'} t^{a'b}_{r's'} v^{r's}_{ab} 
+ 2 (C^{[1]}_A)^{a}_{r} S^{r}_{b} S^{s}_{a'} t^{a'b}_{r's} (\omega_B)^{r'}_{a} 
+ 2 (C^{[1]}_A)^{a}_{r} S^{r'}_{s} S^{s'}_{a} t^{a'b}_{r's'} v^{rs}_{a'b} 
+ 2 (C^{[1]}_A)^{a}_{r} S^{r'}_{b} S^{s}_{a} t^{a'b}_{r's} (\omega_B)^{r}_{a'} 
\\ &\hspace{-5em}
+ 2 (C^{[1]}_B)^{b}_{s} S^{r}_{b} S^{s'}_{r'} t^{ab'}_{rs'} v^{r's}_{ab'} 
+ 2 (C^{[1]}_B)^{b}_{s} S^{r}_{b} S^{s'}_{a} t^{ab'}_{rs'} (\omega_A)^{s}_{b'} 
+ 2 (C^{[1]}_B)^{b}_{s} S^{r}_{b'} S^{s}_{r'} t^{ab'}_{rs'} v^{r's'}_{ab} 
+ 2 (C^{[1]}_B)^{b}_{s} S^{r}_{b'} S^{s}_{a} t^{ab'}_{rs'} (\omega_A)^{s'}_{b} 
\\ &\hspace{-5em}
+ 4 (C^{[1]}_A)^{r}_{a} S^{r'}_{b} S^{b}_{r} t^{a'b'}_{r's} v^{as}_{a'b'} 
+ 4 (C^{[1]}_A)^{r}_{a} S^{r'}_{b} S^{b'}_{a'} t^{a'b}_{r's} v^{as}_{rb'} 
+ 4 (C^{[1]}_A)^{r}_{a} S^{s}_{a'} S^{a'}_{s'} t^{a''b}_{rs} v^{as'}_{a''b} 
+ 4 (C^{[1]}_A)^{r}_{a} S^{s}_{a'} S^{a'}_{b} t^{a''b}_{rs} (\omega_B)^{a}_{a''} 
\\ &\hspace{-5em}
+ 4 (C^{[1]}_A)^{r}_{a} S^{a}_{b} S^{b}_{r'} t^{a'b'}_{rs} v^{r's}_{a'b'} 
+ 4 (C^{[1]}_A)^{r}_{a} S^{a}_{b} S^{b}_{a'} t^{a'b'}_{rs} (\omega_A)^{s}_{b'} 
+ 4 (C^{[1]}_A)^{r}_{a} S^{a}_{b} S^{b}_{a'} t^{a'b'}_{r's} v^{r's}_{rb'} 
+ 4 (C^{[1]}_A)^{r}_{a} S^{a}_{b} S^{b'}_{r} t^{a'b}_{r's} v^{r's}_{a'b'} 
\\ &\hspace{-5em}
+ 4 (C^{[1]}_A)^{r}_{a} S^{a'}_{b} S^{b}_{r} t^{ab'}_{r's} v^{r's}_{a'b'} 
+ 4 (C^{[1]}_A)^{r}_{a} S^{a'}_{b} S^{b'}_{a'} t^{ab}_{r's} v^{r's}_{rb'} 
+ 4 (C^{[1]}_B)^{s}_{b} S^{r}_{b'} S^{b'}_{r'} t^{ab''}_{rs} v^{r'b}_{ab''} 
+ 4 (C^{[1]}_B)^{s}_{b} S^{r}_{b'} S^{b'}_{a} t^{ab''}_{rs} (\omega_A)^{b}_{b''} 
\\ &\hspace{-5em}
+ 4 (C^{[1]}_B)^{s}_{b} S^{s'}_{a} S^{a}_{s} t^{a'b'}_{rs'} v^{rb}_{a'b'} 
+ 4 (C^{[1]}_B)^{s}_{b} S^{s'}_{a} S^{a'}_{b'} t^{ab'}_{rs'} v^{rb}_{a's} 
+ 4 (C^{[1]}_B)^{s}_{b} S^{a}_{s} S^{b}_{a'} t^{a'b'}_{rs'} v^{rs'}_{ab'} 
+ 4 (C^{[1]}_B)^{s}_{b} S^{a}_{s} S^{b'}_{a} t^{a'b}_{rs'} v^{rs'}_{a'b'} 
\\ &\hspace{-5em}
+ 4 (C^{[1]}_B)^{s}_{b} S^{a}_{s'} S^{b}_{a} t^{a'b'}_{rs} v^{rs'}_{a'b'} 
+ 4 (C^{[1]}_B)^{s}_{b} S^{a}_{b'} S^{b}_{a} t^{a'b'}_{rs} (\omega_B)^{r}_{a'} 
+ 4 (C^{[1]}_B)^{s}_{b} S^{a}_{b'} S^{b}_{a} t^{a'b'}_{rs'} v^{rs'}_{a's} 
+ 4 (C^{[1]}_B)^{s}_{b} S^{a}_{b'} S^{b'}_{a'} t^{a'b}_{rs'} v^{rs'}_{as} 
\\ &\hspace{-5em}
+ 4 (C^{[1]}_A)^{a}_{r} S^{r}_{b} S^{b}_{a'} t^{a'b'}_{r's} v^{r's}_{ab'} 
+ 4 (C^{[1]}_A)^{a}_{r} S^{r}_{b} S^{b'}_{a} t^{a'b}_{r's} v^{r's}_{a'b'} 
+ 4 (C^{[1]}_A)^{a}_{r} S^{r'}_{b} S^{b}_{a} t^{a'b'}_{r's} v^{rs}_{a'b'} 
+ 4 (C^{[1]}_A)^{a}_{r} S^{r'}_{b} S^{b'}_{a'} t^{a'b}_{r's} v^{rs}_{ab'} 
\\ &\hspace{-5em}
+ 4 (C^{[1]}_B)^{b}_{s} S^{s}_{a} S^{a}_{b'} t^{a'b'}_{rs'} v^{rs'}_{a'b} 
+ 4 (C^{[1]}_B)^{b}_{s} S^{s}_{a} S^{a'}_{b} t^{ab'}_{rs'} v^{rs'}_{a'b'} 
+ 4 (C^{[1]}_B)^{b}_{s} S^{s'}_{a} S^{a}_{b} t^{a'b'}_{rs'} v^{rs}_{a'b'} 
+ 4 (C^{[1]}_B)^{b}_{s} S^{s'}_{a} S^{a'}_{b'} t^{ab'}_{rs'} v^{rs}_{a'b}
\end{split}
\end{equation}

\end{appendices}

\end{document}